\pdfoutput=1
\documentclass[a4paper,12pt]{article}

\usepackage{amsfonts}
\usepackage{mathrsfs}
\usepackage{amsmath}
\usepackage{amssymb}
\usepackage{framed}

\usepackage[medium]{titlesec}
\usepackage{bm}
\usepackage{cite}

\usepackage[normalem]{ulem}
\usepackage{extarrows}
\usepackage{slashed}
\usepackage{isodateo}
\usepackage{graphicx}
\usepackage{xcolor}
\usepackage[bookmarksnumbered=true,bookmarksopen=true]{hyperref}
 \hypersetup{colorlinks,%
             linkcolor=[rgb]{0,0.3,0.6}, %
             citecolor=[rgb]{0,0.3,0.6}, %
             urlcolor=[rgb]{0,0.3,0.6}}
\usepackage[hmargin=.7in,vmargin=1.1in]{geometry}
\usepackage{indentfirst}
\usepackage{booktabs}

\usepackage{bbm}

\newcommand{\FR}[2]{\displaystyle\frac{\,{#1}\,}{#2}}
\newcommand{\fr}[2]{\mbox{$\frac{\,{#1}\,}{#2}$}}
\newcommand{\n}{\nonumber}

\graphicspath{{fig/}}

\def\bge{\begin{equation}}
\def\ede{\end{equation}}
\def\bga{\begin{aligned}}
\def\eda{\end{aligned}}
\def\bgb{\begin{bmatrix}}
\def\edb{\end{bmatrix}}
\def\bgp{\begin{pmatrix}}
\def\edp{\end{pmatrix}}
\def\bgm{\begin{matrix}}
\def\edm{\end{matrix}}
\def\bgs{\begin{subequations}}
\def\eds{\end{subequations}}
\newcommand{\order}[1]{\mathcal{O}({#1})}
\def\di{{\mathrm{d}}}

\def\mb{\mathbf}

\def\pd{\partial}
\def\ld{{\mathscr{L}}}

\def\la{\langle}\def\ra{\rangle}

\def\to{\rightarrow}
\def\To{\Rightarrow}
\def\ii{\mathrm{i}}

\def\al{\alpha}
\def\be{\beta}
\def\ga{\gamma}
\def\de{\delta}
\def\ep{\epsilon}

\def\lam{\lambda}

\def\si{\sigma}

\def\aa{\mathsf{a}}
\def\bb{\mathsf{b}}
\def\cc{\mathsf{c}}
\def\dd{\mathsf{d}}

\usepackage{mdframed}

\newmdenv[skipabove=0pt,%
          skipbelow=5pt,%
          leftmargin=0pt,%
          rightmargin=0pt,%
          innertopmargin=-5pt,%
          innerbottommargin=7pt,%
          innerleftmargin=2pt,%
          innerrightmargin=2pt,%
          splittopskip=0pt,%
          splitbottomskip=0pt,%
          linewidth=0pt,%
          nobreak=true]%
          {keyeqn2}

\newmdenv[backgroundcolor=gray!15,%
          skipabove=0pt,%
          skipbelow=5pt,%
          leftmargin=0pt,%
          rightmargin=0pt,%
          innertopmargin=-5pt,%
          innerbottommargin=7pt,%
          innerleftmargin=2pt,%
          innerrightmargin=2pt,%
          splittopskip=0pt,%
          splitbottomskip=0pt,%
          linewidth=0pt,%
          nobreak=true]%
          {keyeqn}

\usepackage{titlesec}          
\titleformat{\section}
{\normalfont\fontsize{15}{20}\bfseries}{\thesection}{1em}{}

\newcommand{\ob}[1]{\mkern 2mu \overline{\mkern -2mu #1 \mkern -2mu}\mkern 2mu}
\newcommand{\wt}[1]{\mkern 2mu \widetilde{\mkern -2mu #1 \mkern -2mu}\mkern 2mu}
\newcommand{\wh}[1]{\mkern 2mu \widehat{\mkern-2mu#1\mkern-2mu}\mkern 2mu}

\newcommand{\fnemail}[1]{\footnote{Email: \href{mailto:#1}{\nolinkurl{#1}}}}

\begin{document}

\title{\Large\textbf{Inflation Correlators at the One-Loop Order: Nonanalyticity, Factorization, Cutting Rule, and OPE\\[2mm]}}

\author{Zhehan Qin\fnemail{qzh21@mails.tsinghua.edu.cn}~~~~~ and ~~~~~Zhong-Zhi Xianyu\fnemail{zxianyu@tsinghua.edu.cn}\\[5mm]
\normalsize{\emph{Department of Physics, Tsinghua University, Beijing 100084, China}}}

\date{}
\maketitle

\vspace{20mm}

\begin{abstract}
\vspace{10mm}

Inflation correlators with one-loop massive exchange encode rich information about the dynamics of the massive loop particles. Their nonanalytic behavior in certain soft limits leads to characteristic oscillatory pattern, which is the leading signal of many particle models of cosmological collider physics. In this work, we investigate systematically such nonanalyticity for arbitrary one-particle-irreducible (1PI) one-loop correlators in various soft limits. With the partial Mellin-Barnes representation, we present and prove a factorization theorem and a cutting rule for arbitrary 1PI one-loop inflation correlators, which is reminiscent of the on-shell cutting rule for flat-space scattering amplitudes. We also show how to understand this factorization theorem from the viewpoint of operator product expansion on the future boundary. As an application of the one-loop factorization theorem, we derive new analytic and exact formulae for nonlocal cosmological collider signals for massive one-loop four-point inflation correlators of all possible 1PI topologies, including the bubble, the triangle, and the box graphs. Finally, we show how to push the computation of nonlocal signals to higher orders in the momentum ratio.

\end{abstract}

\newpage
\tableofcontents

\newpage
\section{Introduction}

It has been emphasized in recent years that the observation of large-scale inhomogeneity and anisotropy of our universe could be a unique chance to probe fundamental particle physics at very high energies, a program known as the cosmological collider (CC) physics. In typical scenarios, large-scale perturbations were generated by quantum fluctuations when the primordial universe was inflating at a Hubble rate $H$ up to $10^{14}$GeV \cite{Planck:2018jri}. Such a fast expansion, together with other field evolutions, can trigger spontaneous production of heavy particles with masses comparable to or much greater than $H$. The produced particles can then leave distinct signals in the scalar or tensor modes of primordial perturbations, called CC signals. See \cite{Chen:2009we,Chen:2009zp,Baumann:2011nk,Chen:2012ge,Pi:2012gf,Noumi:2012vr,Gong:2013sma} for earlier works, \cite{Arkani-Hamed:2015bza,Chen:2015lza,Chen:2016nrs,Chen:2016uwp,Chen:2016hrz,Lee:2016vti,An:2017hlx,An:2017rwo,Iyer:2017qzw,Kumar:2017ecc,Chen:2017ryl,Tong:2018tqf,Chen:2018sce,Chen:2018xck,Chen:2018cgg,Chua:2018dqh,Wu:2018lmx,Saito:2018omt,Li:2019ves,Lu:2019tjj,Liu:2019fag,Hook:2019zxa,Hook:2019vcn,Kumar:2018jxz,Kumar:2019ebj,Alexander:2019vtb,Wang:2019gbi,Wang:2019gok,Wang:2020uic,Li:2020xwr,Wang:2020ioa,Fan:2020xgh,Aoki:2020zbj,Bodas:2020yho,Maru:2021ezc,Lu:2021gso,Sou:2021juh,Lu:2021wxu,Pinol:2021aun,Cui:2021iie,Tong:2022cdz,Reece:2022soh,Qin:2022lva,Chen:2022vzh,Cabass:2022rhr,Cabass:2022oap,Niu:2022quw,Niu:2022fki,Aoki:2023tjm,Chen:2023txq,Tong:2023krn} for recent theoretical progresses, and \cite{Meerburg:2016zdz,MoradinezhadDizgah:2017szk,MoradinezhadDizgah:2018ssw,Kogai:2020vzz} for works related to observations.

The central objects in the study of CC physics are the correlators of primordial perturbations, which we call \emph{inflation correlators}. According to the inflation theory, the inflation correlators are expectation values of products of quantum field operators living in the inflationary universe. Since the inflationary spacetime is approximately de Sitter (dS), the inflation correlators can be thought of as correlation functions of bulk quantum fields in dS, but with their endpoints pinned at the future boundary. They form the so-called meta-observables, and are useful objects for the study of quantum field theories (QFTs) in dS \cite{Anninos:2012qw}. Compared to the studies of scattering amplitudes in Minkowski spacetime or boundary correlation functions in Anti de Sitter (AdS) spacetime, our understanding of inflation correlators, or dS correlators, is currently at a rather primitive stage. In recent years, partly inspired by the prospects of future cosmological observations, the study of inflation correlators has attracted lots of attentions, and many progresses have been made \cite{Arkani-Hamed:2018kmz,Baumann:2019oyu,Baumann:2020dch,Sleight:2019mgd,Sleight:2019hfp,Sleight:2020obc,Sleight:2021iix,Sleight:2021plv,Pajer:2020wnj,Pajer:2020wxk,Cabass:2021fnw,Pimentel:2022fsc,Jazayeri:2022kjy,Qin:2022lva,Qin:2022fbv,Wang:2021qez,Goodhew:2020hob,Melville:2021lst,Goodhew:2021oqg,Jazayeri:2021fvk,Bonifacio:2021azc,Hogervorst:2021uvp,Meltzer:2021zin,DiPietro:2021sjt,Tong:2021wai,Heckelbacher:2022hbq,Gomez:2021qfd,Gomez:2021ujt,Baumann:2021fxj,Baumann:2022jpr,Wang:2022eop,Lee:2022fgr,Qin:2023ejc}. 

Among all kinds of inflation correlators, the ones mediated by massive fields are of particular interest to CC physics. Heavy particles with mass $m\gtrsim H$ created during inflation quickly become nonrelativistic due to cosmic redshift, and their wavefunction thus oscillates like $e^{\pm\ii m t}$. Through interactions with nearly massless and long-lived scalar modes, the oscillation $e^{\pm\ii m t}$ is translated to the oscillations in the \emph{logarithm} of external momentum ratios in certain soft limits, which we shall call \emph{CC signals}. Therefore, by searching for CC signals from inflation correlators, we will have a chance to measure many properties of the intermediate massive particles, including their mass, spin, chemical potential, interaction type, etc.

As often happened, it is fruitful to view the inflation correlators as functions of \emph{complex} external momenta. By proper analytic continuation of these correlators, we can often develop useful techniques and gain new insights \cite{Goodhew:2020hob,Melville:2021lst,Goodhew:2021oqg,Meltzer:2021zin,DiPietro:2021sjt,Baumann:2021fxj}. Specialized to correlators with massive exchanges, the CC signals manifest themselves on the complex plane of intermediate momenta as branch cuts emanating from the origin \cite{Arkani-Hamed:2018kmz}. Therefore, by a careful study of nonanalyticity of the inflation correlators, we can develop new understanding and find new explicit results for CC signals. 

The analytic properties and CC signals of tree-level correlators with single massive exchange have been quite well understood \cite{Arkani-Hamed:2015bza,Arkani-Hamed:2018kmz}. In comparison, massive loop correlators are much less explored. Indeed, the first complete analytical result for arbitrary massive 1-loop exchange was obtained very recently in \cite{Xianyu:2022jwk}, where 1-loop bubble graphs of 3-point and 4-point correlators are computed. For more complex loop topologies, such as triangle and box graphs, currently there is no explicit result available. On the other hand, loop correlators are of great importance for CC phenomenology. In many particle models of CC physics, large CC signals are absent at the tree level, and show up firstly at the 1-loop level \cite{Chen:2016uwp,Chen:2016hrz,Chen:2018xck,Lu:2019tjj,Hook:2019zxa,Hook:2019vcn,Kumar:2018jxz,Wang:2019gbi,Wang:2020ioa,Lu:2021gso,Cui:2021iie,Tong:2022cdz}. In these models, the CC signals are typically generated in triangle or box diagrams. Therefore, there is a practical need for computing these diagrams, especially their CC signals. For this purpose, it is essential to have a good understanding of the nonanalyticity in generic massive 1-loop inflation correlators. 

In this work, we initiate a systematic study of the nonanalyticity of massive inflation correlators beyond the tree level. We begin with 1-loop processes, and consider the most general one-particle-irreducible (1PI) graphs with arbitrary couplings and arbitrary number of external points. In this work, by a 1PI graph, we mean that the graph cannot be broken into disconnected parts by removing one single \emph{bulk} propagator.\footnote{The reason for restricting ourselves to 1PI graphs is that we want to focus on the nonlocal signals from the loop propagators. For 1-loop graphs that are not 1PI, there must exist bulk tree propagators besides the loop propagators. These bulk tree propagators could contain additional CC signals which are unnecessary complications to us.} An example of such graphs is shown in Fig.\ \ref{fd_GenLoop}. To avoid unnecessary complications, we further take the external lines to be either massless scalars ($m^2=0$) or conformal scalars ($m^2=2H^2$), while the internal loop lines can be arbitrary massive scalars. For definiteness, we shall only consider principal scalars with $m>3H/2$. Generalization to lighter scalars is straightforward. It is also possible to generalize the result here to include spinning particles with dS boost breaking dispersions, which we shall consider in a subsequent work.

\paragraph{Outline of this work.}
Below we provide a summary of the main results of this work. 
 
\begin{description}
\item[1. Nonlocal signal (Sec.\ \ref{sec_cuttingrule})] We identify a particular nonanalytic behavior of a general inflation correlator, which we call the \emph{nonlocal signal}. In short, a nonlocal signal refers to the existence of a branch point at the origin of a complex momentum variable $P$ in the form of complex powers, $P^{\pm\ii\omega}$. Here $P\equiv |\mb P|$ is the magnitude of a 3-momentum $\mb P$, $\mb P$ is the sum of an arbitrary subset of external 3-momenta of a given graph, and $\omega$ is a positive real number, which determines the frequency of the (oscillatory) nonlocal signal. Very often, this frequency is related to the mass of the intermediate particle, and thus is a very useful quantity for CC phenomenology. 
\end{description}

\begin{figure}
\centering 
\includegraphics[width=0.4\textwidth]{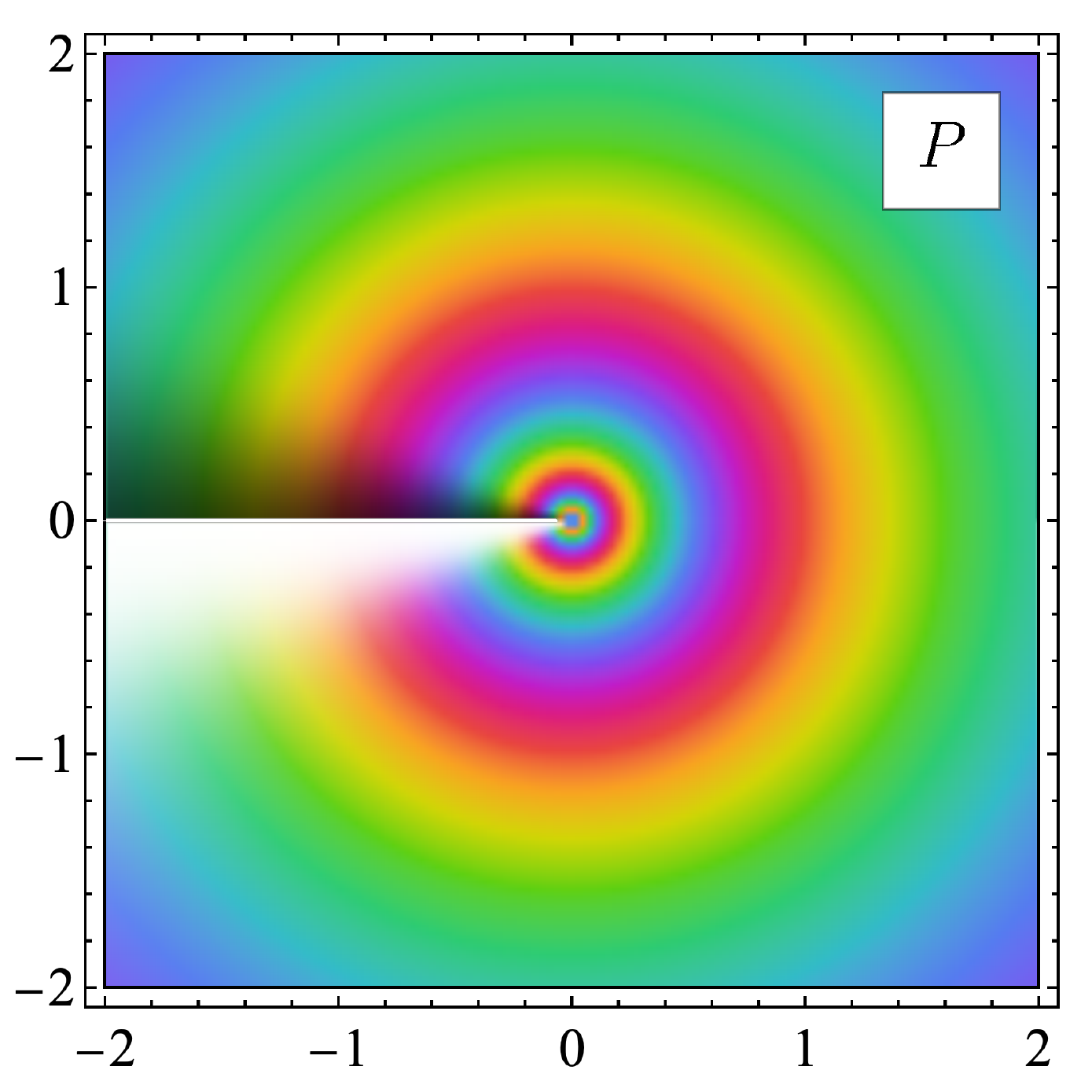} 
\caption{The function $P^{\ii\omega}$ with $\omega=4$ on the complex $P$-plane. The colors show the phase of the function while the brightness shows the magnitude. Both the logarithmic oscillation for $P>0$ and the branch cut for $P<0$ are clearly visible.}
  \label{fig_complexP}
\end{figure}

At this point, it is worth noting that the complex function $P^{\ii\omega}$ with $\omega>0$ gives rise to oscillatory signal on a logarithmic scale of $P$ for physical region $P>0$. On the other hand, when we take analytic continuation and consider the complex $P$-plane, the function $P^{\ii\omega}$ develops a branch cut on the negative real axis.\footnote{Of course, the exact location of the branch cut is of no invariant meaning. What is important here is that the function $P^{\ii\omega}$ does not return to its initial value when the argument $P$ goes a full circle around $P=0$.} We illustrate this behavior in Fig.\ \ref{fig_complexP}. Thus we see that the nonanalyticity of an inflation correlator is intimately tied to its oscillatory CC signals. Therefore, a study of nonanalyticity of inflation correlators is not only interesting from a purely theoretical point of view, but also important for phenomenological applications. We note in passing that there are other types of nonanalyticity of an inflation correlator, such as the so-called total energy pole (the pole that appears when the \emph{magnitude} sum of all external momenta goes to zero) and partial energy poles (the poles that appear when the \emph{magnitude} sum of all external momenta of a \emph{subgraph} goes to zero). These poles lie deeply in the unphysical region and thus cannot be directly probed by physically measured correlators. Finally, there is another class of nonanalyticity which we call \emph{local signals}. It corresponds to branch points on the complex plane of $E$ where $E$ is the magnitude sum of all external momenta at a vertex. A local signal can also show up in physical data as an oscillatory pattern, but it is more subtle to analyze than the nonlocal signal. We plan to consider the local signal in a subsequent work.\footnote{The terminology of nonlocal signal and local signal follows from previous works on tree-level correlators. See, e.g., \cite{Tong:2021wai,Qin:2022lva}.}

\begin{description}

\item[2. One-loop factorization theorem (Sec.\ \ref{sec_cuttingrule})] We state and prove a factorization theorem for an arbitrary 1PI 1-loop graph with massive exchanges. With this theorem, we can find all possible nonlocal signals in a given diagram, and get explicit and analytic expressions for all the nonlocal signals in the corresponding soft limits, without making any approximations. Technically, our proof of the factorization theorem is made possible by the technique of \emph{partial Mellin-Barnes (PMB) representation} we recently introduced in \cite{Qin:2022lva,Qin:2022fbv}.

\end{description}

\begin{figure}
\centering
\includegraphics[width=0.99\textwidth]{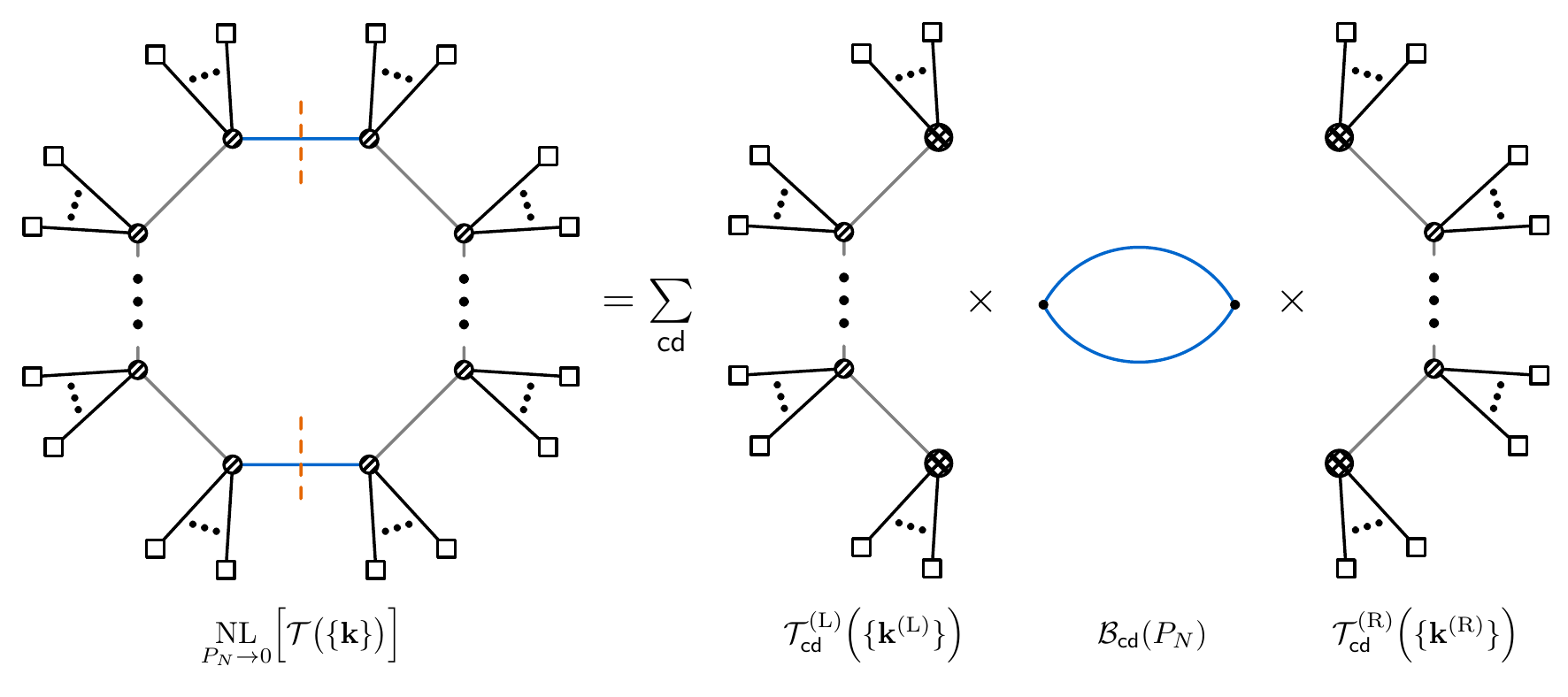}
\caption{An illustration of the 1-loop factorization theorem. On the left hand side, the two blue lines denote the soft lines, and the two cuts (dashed orange lines) show the cutting rule: one can replace the two soft lines by their real parts. On the right hand side, the three graphs denote the left tree graph, the bubble signal, and the right tree graph, respectively. The four hatched larger blobs on the right hand side denote the insertions with complex powers of conformal time variables.}
  \label{fig_theorem}
\end{figure}

Let us briefly explain the main content of this theorem with Fig.\ \ref{fd_GenLoop} and Fig.\ \ref{fig_theorem}. Essentially, the theorem says that a general 1PI 1-loop graph, such as Fig.\ \ref{fd_GenLoop}, exhibits a nonlocal signal, whenever the momentum injected into the graph through a \emph{connected} subset of vertices goes to zero, while all other external momenta remain fixed. (On the contrary, there is no nonlocal signal associated with the soft limit that the momentum injected through a \emph{disconnected} subset of vertices goes to zero.) For instance, Fig.\ \ref{fd_GenLoop} contains a nonlocal signal, when $\mb P_N\equiv \mb K_1+\cdots\mb +\mb K_N\to \mb 0$, in the form of $P_N^{3\pm2\ii(\wt\nu_1\pm\wt\nu_{N+1})}$. We shall show that this nonanalytic term is contributed by the parameter region of the loop momentum where the two blue lines in Fig.\ \ref{fd_GenLoop} are \emph{simultaneously} soft. Also, the explicit expression for this nonanalytic term factorizes into a product of three pieces: a left tree graph $\mathcal{T}_{\cc\dd}^\text{(L)}$, a right tree graph $\mathcal{T}_{\cc\dd}^\text{(R)}$, and a bubble signal $\mathcal{B}_{\cc\dd}$. See (\ref{eq_SoftThm}) for this explicit expression and Fig.\ \ref{fig_theorem} for an illustration. As shown in Fig.\ \ref{fig_theorem}, this factorization suggests that we can cut the graph through the two blue loop lines to construct the nonlocal signal. For this reason, we call the two blue lines in Fig.\ \ref{fig_theorem} \emph{soft lines} or \emph{cut lines}.

\begin{description}

\item[3. One-loop cutting rule (Sec.\ \ref{sec_cuttingrule})] The cut procedure described above can be formulated more precisely as a \emph{cutting rule} for computing nonlocal signals in arbitrary 1PI 1-loop graph. In short, the cutting rule states that we can replace the propagators for the two soft lines by their real parts. As we shall show, the real part of a bulk propagator is independent of its Schwinger-Keldysh indices, and thus is free from time ordering. As a result, in the computation of nonlocal signals, there is no integral that orders the two time variables of any cut line. We shall see that the cutting rule is a direct corollary of our factorization theorem.

\item[4. Boundary OPE (Sec.\ \ref{sec_ope})] It is possible to understand the factorization theorem from a boundary point of view in terms of OPEs. The key observation here is that the nonlocal signal in a 1-loop graph is generated when the two cut lines become soft simultaneously. For a bulk propagator in an inflating universe, the soft limit $k\to 0$ is often equivalent to the late-time limit $\tau\to 0$, since the momentum and the time are always combined in the form of $k\tau$. Again let us explain this point with Fig.\ \ref{fd_GenLoop}. When $\mb K_1+\cdots+\mb K_N\to\mb 0$, we can find the nonlocal signal by bringing all four modes associated with the two soft (blue) lines to the future boundary $\tau=0$. Then, we can perform two boundary OPE, pinching the two modes at $\tau_1$ and $\tau_{N}$, and pinching the other two modes at $\tau_{N+1}$ and $\tau_{V}$. This is exactly the origin of the bubble signal $\mathcal{B}_{\cc\dd}$ in the factorization theorem.

\item[5. Bubble, triangle, and box (Sec.\ \ref{sec_4pt})] We apply the factorization theorem to construct exact and analytical expressions for nonlocal signals in all 4-point inflaton correlators with 1PI 1-loop topologies, including the bubble, triangle, and box diagrams. According to the factorization theorem, we need three ingredients for this computation: the left tree graph, the right tree graph, and the bubble signal. The expression for the bubble signal is universal for all 1-loop processes and is explicitly given in Sec.\ \ref{sec_cuttingrule}, while the expressions for the left and right tree graphs were explicitly computed in our recent work \cite{Qin:2023ejc}. The nonlocal signal in the 1-loop bubble graph agrees with existing results, while the nonlocal signals in the triangle and box graphs are new. With these results, we define a \emph{pinched operator}, which shows that it is possible to pinch the hard line(s) in the triangle or box graphs to compute the nonlocal signal. See Fig.\ \ref{fig_pinched_triangle}.

\item[6. Subleading corrections (Sec.\ \ref{sec_full4pt})] We provide an algorithm that can in principle generate corrections to nonlocal signals to arbitrary given order in the soft momentum, and thereby generalize the factorization theorem to arbitrary momentum configurations. We apply this method to explicitly compute the $\order{k_s^2}$ corrections to the leading nonlocal signals in the $s$-channel soft limit $k_s\equiv|\mb k_1+\mb k_2|\to 0$ for all the three types of 1-loop 4-point functions, and thereby find the higher-order pinched operators. 

\end{description}

Above we have briefly introduced the main results of this paper. The details will be given in corresponding sections, and we will provide further discussions on these results in Sec.\ \ref{sec_conclusion}. There are three appendices following the main text. In App.\ \ref{app_math}, we collect several frequently used mathematical functions and shorthand notations. In App.\ \ref{app_loop}, we compute explicitly the loop momentum integrals for the bubble loop and the triangle loop with the PMB representation. Finally, in App.\ \ref{app_diffmass}, we compute the nonlocal signals at the leading order in the soft momentum for all 1PI 1-loop 4-point correlators with unequal loop masses.

\paragraph{Comparison with existing results.} 
The analytical properties of dS amplidutes have been under active studies in recent years. The discontinuity of dS amplitudes has been studied in, e.g., \cite{Goodhew:2020hob,Goodhew:2021oqg,Melville:2021lst,Jazayeri:2021fvk,Baumann:2021fxj,Sleight:2021plv}. Many of these results were obtained for wavefunction coefficients instead of correlation functions. In \cite{Tong:2021wai}, a cutting rule for extracting CC signals (including the nonlocal and local signals) from tree-level inflation correlators was proposed which was motivated by intuitive arguments in the bulk dS. This cutting rule was proved rigorously in \cite{Qin:2022lva,Qin:2022fbv} with PMB method. Compared with these results, our cutting rule applies to massive loop correlators, and can be used to compute loop signals. As all known cutting rules in both flat space and dS, our cutting rule cuts the integrand instead of the whole loop integral. Therefore, the cutting rules themselves are not enough to compute the whole signals in loop processes. In comparison, our factorization theorem provides an explicit expression for all nonlocal signals in an arbitrary 1PI 1-loop graph, and thus is more powerful than the cutting rules themselves. Indeed, as we have shown, the cutting rule shows up naturally as a corollary in our proof of the factorization theorem. 

The factorization found in this work is reminiscent of the factorization of scattering amplitudes in flat space. In the latter case, when the sum of a subset of external 4-momenta hits the physical mass of an intermediate state, the amplitude exhibits a pole due to the on-shell propagation of the intermediate massive particle. In our case, the existence of the branch cut associated with nonlocal signals can also be thought of as a consequence of on-shell production of intermediate massive particles. The difference here is that, due to the fast expansion, an on-shell massive particle does not have a stable wavefunction, and thus it does not produce a pole in the physical region of the momentum space. Instead, such on-shell particles produce softer singularities, namely branch cuts, in the unphysical regions. In the physically accessible regions, we see nonlocal (logarithmic) oscillations instead of discontinuities. 

In the study of dS amplitudes, there is another known factorization for tree-level correlators, which shows up when the sum of the \emph{magnitudes} of a subset of external momenta goes to zero, and is known as a partial energy pole \cite{Arkani-Hamed:2018kmz}. We emphasize that our factorization works in a physical limit and is physically accessible; It is not directly related to the partial-energy-pole factorization, which is deeply inside the unphysical region.

\paragraph{Notations and Conventions.} Most of notations and conventions of this work are in line with the ones adopted in our previous works on similar topics \cite{Qin:2022lva,Qin:2022fbv,Qin:2023ejc}. In particular, the spacetime metric is $\di s^2=a^2(\tau)(-\di\tau^2+\di\mb x^2)$ where $a(\tau)=-1/(H\tau)$ is the scale factor for an exponentially inflating universe, and $\tau\in(-\infty,0)$ is the conformal time. In the following, we always take $H=1$ for simplicity. We use bold letters such as $\mb k$ to denote 3-momenta and the corresponding italic letter $k\equiv|\mb k|$ to denote the magnitude of a 3-momentum. Following terminologies of the literature, we sometimes call the magnitude of a 3-momentum with the term \emph{energy}, although this terminology is not very precise. For many kinds of variables with indices, we shall use a compact notation to denote their sum. For instance, $k_{12}\equiv k_1+k_2$, $n_{1234}=n_{1}+n_{2}+n_{3}+n_{4}$, etc. We shall make heavy use of Mellin variables such as $s_i$ and $\bar s_i$. Unbarred and barred Mellin variables are independent; they have no relations whatsoever. When summing barred and unbarred Mellin variables, we shall use shorthand notation such as $s_{1\bar 2}\equiv s_{1}+\bar s_{2}$, $s_{i\bar jk\bar \ell}=s_{i}+s_{\bar j}+s_{k}+s_{\bar \ell}$. With this shorthand in mind, we shall also use $s_{\bar k}$ and $\bar s_k$ interchangeably.

\section{One-Loop Factorization Theorem and Cutting Rule}
 \label{sec_cuttingrule}
In this section, we study a general inflaton correlator mediated by arbitrary number of massive scalar fields at one-loop order. For the phenomenological applications in CC physics, one normally consider three-point or four-point inflaton correlators. In this section, however, we shall be slightly more general, and consider an arbitrary $B$-point correlator with $B\geq 4$, as shown in Fig. \ref{fd_GenLoop}. To be concrete, we shall assume that the external lines are either conformal scalars $\varphi_c$ with mass $m^2=2$ or (nearly) massless scalars such as the inflaton fluctuation $\varphi$. They are coupled via local interactions to a bunch of massive particles $\si_i$ ($i=1,\cdots,V$) of mass $m_i$. For definiteness and also for the direct application in CC physics, we shall assume that all $\si_i$ are principal scalars, namely, $m_i>3/2$. In this case, it is more convenient to define a real and positive \emph{mass parameter} $\wt\nu_i\equiv\sqrt{m_i^2-9/4}$. We stress that it is not necessary to stick to principal scalars. Indeed, our results can be directly generalized to either lighter scalars with mass $m\in(0,3H/2)$ or massive spinning particles with dS covariant dispersion relations. The couplings between the external modes ($\varphi_c$ or $\varphi$) and the internal massive modes $\si_i$ can be rather arbitrary, including arbitrary but finite integer number of spacetime derivatives. The only technical requirement is that these couplings should be infrared-safe, in the sense that the SK time integral does not contain any divergence in the late-time limit. 

Below, we shall first present a factorization theorem, which concerns the nonanalytic behavior of the one-loop graph $\mathcal{T}$ in Fig.\ \ref{fd_GenLoop} when the total momentum $\mb P_N$ flowing into the graph through a \emph{connected} subset of vertices becomes soft, see Fig. \ref{fig_theorem}. In such a limit, the factorization theorem states that the leading nonanalytic term of the graph factorizes into a product of three terms: the left tree graph, the right tree graph, and a bubble signal. Together with the factorization theorem, we shall also present a cutting rule, which states that one can freely replace all the cut loop propagators (the two blue lines in Fig.\ \ref{fig_theorem}) by their real parts when computing the nonanalytic term of $\mathcal{T}$.

\subsection{Factorization theorem}
\label{sec_theorem}

\begin{figure}
\centering
\vspace{-5mm}
\includegraphics[width=0.6\textwidth]{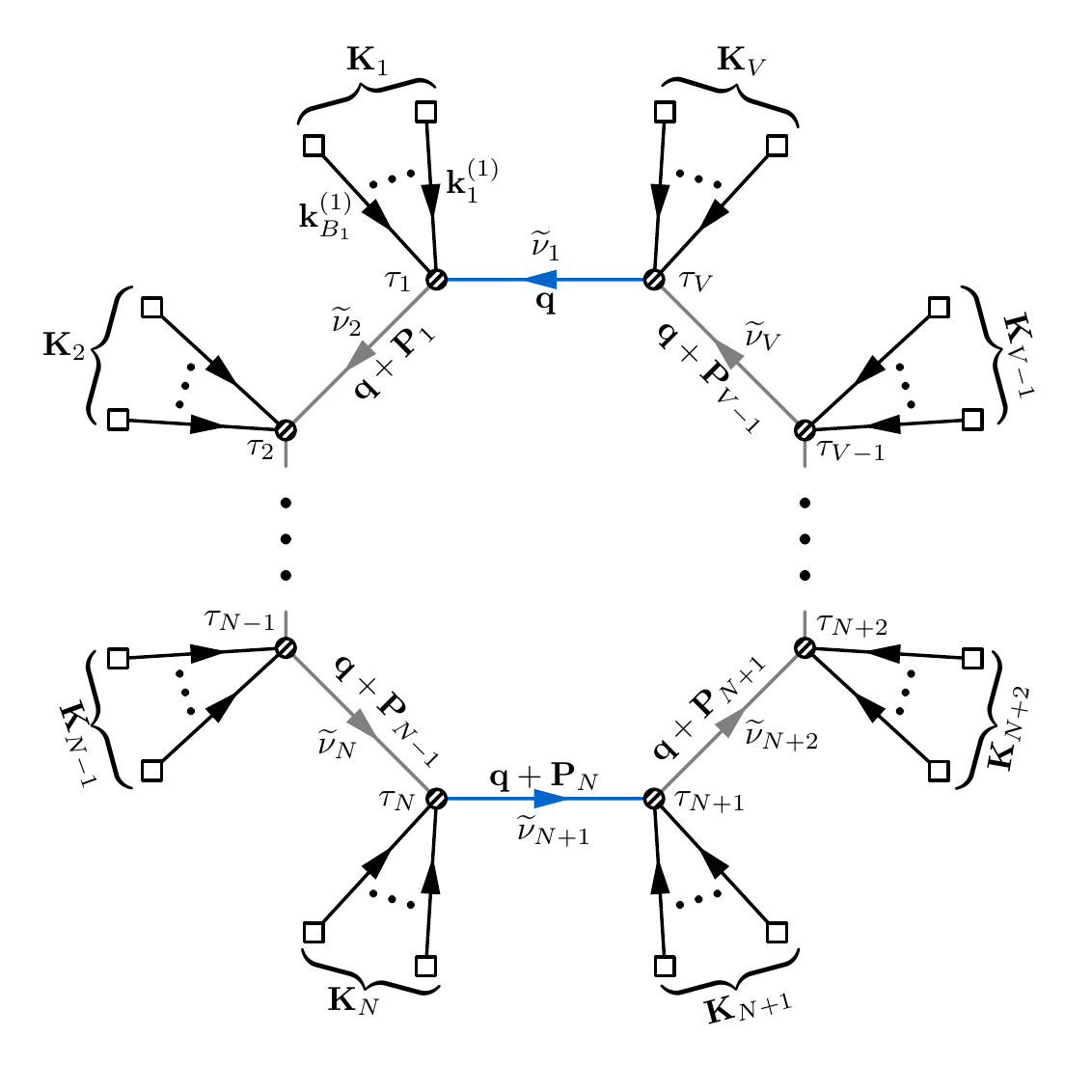}
\vspace{-5mm}
\caption{A general 1-loop inflation correlator. We follow the diagrammatic notation of \cite{Chen:2017ryl}. In particular, all the external (black) lines are either conformal scalars or massless scalars, while the internal (gray and blue) lines can be either identical or distinct massive scalar fields with $\wt\nu_i>0$ $(i=1,\cdots,V)$.  The two blue lines denote the soft lines that contribute to the nonlocal signal when $P_N\to 0$.}
  \label{fd_GenLoop}
\end{figure}

\paragraph{Specification of the 1-loop graph.}
The specific 1-loop graph we shall consider is shown in Fig.\ \ref{fd_GenLoop}. In this graph, we have a loop consisting of $V$ massive scalar lines with mass parameters $\wt\nu_i$ ($i=1,\cdots,V$). They are connected at $V$ vertices with external lines. We label each of the $V$ vertices by a number $i=1,\cdots,V$. (In this work, we do not consider two-point mixings.) At Vertex $i$, we have $B_i\geq 1$ external lines, each of which carries a 3-momentum $\mb k^{(i)}_j$ with $j=1,\cdots,B_i$. For convenience, we shall denote the total 3-momentum flowing into the Vertex $i$ by $\mb K_i$, and the sum of magnitude of 3-momenta $\mb k_j^{(i)}$ at the Vertex $i$ by $E_i$. Finally, we shall use $\mb P_n$ to denote the total external 3-momentum flowing into the graph through the first $n$ vertices (namely, from Vertex $i$ with $i=1,\cdots,n$), and we shall use $B$ to denote the total number of external legs. We summarize our kinematic speficification of the 1-loop diagram in the following equations:
\begin{align}
\label{eq_KEPB}
  &\mb K_i = \sum_{j=1}^{B_i}\mb k_{j}^{(i)}, &&E_i\equiv \sum_{j=1}^{B_i}\big|\mb k_{j}^{(i)}\big|,
  &&\mb P_n = \sum_{j=1}^n\mb K_j, 
  &&B=\sum_{j=1}^V B_j.
  &&(i=1,\cdots,V)
\end{align}
Clearly, $\mb P_V= \mb 0 $ as a consequence of momentum conservation. 

According to the diagrammatic rule in the Schwinger-Keldysh formalism, the graph can be written as a product of bulk propagators for all internal massive lines and bulk-to-boundary propagators for all external lines,
 integrated over all (conformal) time variables $\tau_i$ at all vertices, and also over the unconstrained loop momentum $\mb q$, and summed over all SK indices $\aa_i=\pm$. We refer the readers to \cite{Chen:2017ryl} for a pedagogical review of the diagrammatic rule in the SK formalism. We follow most of the diagrammatic notations and conventions in \cite{Chen:2017ryl}. In particular, the bulk-to-boundary propagator of a massless scalar reads:
\bge
  G_\aa(k;\tau)=\FR{1}{2k^3}(1-\ii\aa k\tau)e^{\aa\ii k\tau},
\ede
while the bulk-to-boundary propagator for a conformal scalar reads:
\bge
\label{eq_CSProp}
  C_\aa(k;\tau)=\FR{\tau\tau_f}{2k}e^{\aa\ii k\tau},
\ede
in which we have introduced a final time cutoff $\tau_f$, since a conformal scalar mode dies away in the late-time limit as $\tau$.  On the other hand, the bulk propagator $D_{\aa_1\aa_2}^{(\wt\nu)}(k;\tau_1,\tau_2)$ of a massive scalar with mass parameter $\wt\nu$ comes in two types. First, we have two opposite-sign propagators, also called the homogeneous propagators. These are solutions to homogeneous scalar equation of motion at each of the two time variables:\footnote{See Sec.\ 5 of \cite{Qin:2022fbv} for a discussion of SK propagators as solutions to scalar equations of motion. }
\begin{align}
\label{eq_Dmp}
  D_{-+}^{(\wt\nu)}(k;\tau_1,\tau_2)
  =&~\FR{\pi}{4}e^{-\pi\wt\nu}(\tau_1\tau_2)^{3/2}\mathrm{H}_{\ii\wt\nu}^{(1)}(-k\tau_1)\mathrm{H}_{-\ii\wt\nu}^{(2)}(-k\tau_2),\\
\label{eq_Dpm}
  D_{+-}^{(\wt\nu)}(k;\tau_1,\tau_2)
  =&~\Big[D_{-+}^{(\wt\nu)}(k;\tau_1,\tau_2)\Big]^*.
\end{align}
Second, we have two same-sign propagators, also called the inhomogeneous propagators, which are solutions to inhomogeneous scalar equation of motion with $\delta$-function source:
\begin{align}
\label{eq_Dpmpm}
  D_{\pm\pm}^{(\wt\nu)}(k;\tau_1,\tau_2)=&~D_{\mp\pm}^{(\wt\nu)}(k;\tau_1,\tau_2)\theta(\tau_1-\tau_2)+D_{\pm\mp}^{(\wt\nu)}(k;\tau_1,\tau_2)\theta(\tau_2-\tau_1).
\end{align}
As said above, we can consider arbitrary couplings for the Vertex $i$.  In general, this coupling has the schematic form $(\pd^\ell\varphi^{B_i})(\pd^m\si_i)(\pd^n\si_{i+1})$ $(\ell,m,n\geq 0)$, where $\si_i$ and $\si_{i+1}$ are the two massive scalars meeting at Vertex $i$, and $\varphi$ can also be replaced by $\varphi_c$. Here the derivatives can be either uncontracted temporal derivative $\pd_\tau$ or fully contracted spatial derivative $\pd_i$. As mentioned above, the only requirement is that these couplings do not lead to any late-time divergences in the SK time integrals over $\tau_i$. (On the other hand, the early-time limit of the SK time integral is always well behaved due to the Bunch-Davies initial condition, which is technically implemented by appropriate $\ii\ep$ prescriptions. See \cite{Chen:2017ryl}. Also, the loop momentum integral is also free of infrared divergences due to the finite masses in the propagators. The potential ultraviolet divergences of the loop momentum is irrelevant so long as we are only concerned with the nonanalytic part of the graph; More below.)
 
With all points clarified above, it is straightforward to write down an expression for the 1-loop graph in Fig.\ \ref{fd_GenLoop}:
\begin{align}
\label{eq_Tgeneral}
 \mathcal{T}(\{\mb k\})=&\sum_{\aa_1,\cdots,\aa_V=\pm} \int_{-\infty}^0\prod_{\ell=0}^V\bigg[\di\tau_\ell\,(\ii\aa_\ell)(-\tau_\ell)^{p_\ell}\prod_{i=1}^{B_\ell}C_{\aa_\ell}\Big(k_i^{(\ell)};\tau_\ell\Big)\bigg]\int\FR{\di^3\mb q}{(2\pi)^3}D_{\aa_V\aa_1}^{(\wt\nu_1)}\Big(q;\tau_V,\tau_1\Big)\n\\
  &\times 
  D_{\aa_1\aa_2}^{(\wt\nu_2)}\Big(|\mb q+\mb P_1|;\tau_1,\tau_2\Big) 
  \cdots D_{\aa_{V-1}\aa_V}^{(\wt\nu_V)}\Big(|\mb q+\mb P_{V-1}|;\tau_{V-1},\tau_V\Big).
\end{align}
Here, we use $\{\mb k\}$ to denote the collection of all external momenta. For simplicity, we assume the external modes to be conformal scalars, non-derivatively coupled at each vertex, with a time-dependent coupling that yields a factor $(-\tau_\ell)^{p_\ell}$ with a generic complex power $p_\ell$. We also omit all coupling constants. Generalizations to massless and/or derivatively coupled external modes are trivial. Furthermore, when some of the internal lines are identical particles, there could be nontrivial symmetry factors which we do not bother to include.

\paragraph{Nonlocal signal.} 
Recall that the question we want to address is in which soft kinematic configuration of the 1-loop graph in Fig.\ \ref{fd_GenLoop} can a nonlocal CC signal show up, and, what is the frequency and the size of the nonlocal signal. Recall that the nonlocal signal in an inflation correlator $\mathcal{T}$ means an oscillatory behavior in the logarithm of a momentum ratio: $\mathcal{T}\sim \cos[\omega\log(P/Q)+\varphi]$ in the limit $P\to 0$, where $P=|\mb P|$ and $\mb P$ is the sum of some external momenta $\mb k^{(i)}_j$. On the other hand, $Q$ is introduced only to make the argument of the logarithm dimensionless, and its specific form is inessential. The oscillatory behavior is originated from terms in $\mathcal{T}$ with noninteger power: $P^{\ii\omega}+\text{c.c.}$ where $\omega$ is a real number and is usually linked to the mass parameter $\wt\nu$. Therefore, searching for the nonlocal signals is equivalent to searching for external momentum sums $\mb P$ such that the correlator $\mathcal{T}$ contains a noninteger power term $\propto P^{\pm\ii\omega}$ in the limit $P\to 0$. 

With the above explanations, now, we are at the position to state the main theorem of this section, which we call the one-loop factorization theorem:

\paragraph{One-loop factorization theorem.} Consider the 1-loop graph shown in Fig.\ \ref{fd_GenLoop}, and let $\mb P_n$ be as defined in (\ref{eq_KEPB}). Then, for each of $N=1,\cdots,V-1$,\footnote{When $N=1$ or $N={V-1}$, it is required that $B_1>1$ or $B_V>1$, respectively, for our theorem to hold. More explanations below.} the 1-loop graph is nonanalytic in $P_N$ at $P_N=0$ on the complex $P_N$-plane, in the form of noninteger powers $P_N^{\pm 2\ii(\wt\nu_1\pm\wt\nu_{N+1})}$ (when $\wt\nu_1\neq\wt\nu_{N+1}$) or $P_N^{\pm 4\ii \wt\nu_1}$ (when $\wt\nu_1=\wt\nu_{N+1}$). In the limit $P_N\to 0$, each nonanalytic term with a given noninteger power factorizes into a product of a \emph{left tree correlator} $\mathcal{T}_{\cc\dd}^\text{(L)}$, a \emph{right tree correlator} $\mathcal{T}_{\cc\dd}^\text{(L)}$, and a \emph{bubble signal} $\mathcal{B}_{\cc\dd}(P_N)$:
\begin{keyeqn}
\begin{align}
\label{eq_SoftThm}
  \lim_{P_N\to 0}\mathcal{T}\big(\{\mb k\}\big)=\sum_{\cc,\dd=\pm}\mathcal{T}_{\cc\dd}^\text{(L)}\Big(\{\mb k^\text{(L)}\}\Big)\mathcal{T}_{\cc\dd}^\text{(R)}\Big(\{\mb k^\text{(R)}\}\Big)\mathcal{B}_{\cc\dd}(P_N)+\text{terms analytic in $P_N$}.
\end{align}
\end{keyeqn}
where the summation variables $\cc,\dd$ are subject to an additional constraint $\cc=\dd$ if $\wt\nu_1=\wt\nu_{N+1}$. In (\ref{eq_SoftThm}), the left tree graph $\mathcal{T}_{\cc\dd}^\text{(L)}$ and the right tree graph $\mathcal{T}_{\cc\dd}^\text{(R)}$ are respectively given by:
\begin{align}
\label{eq_Tleft}
  \mathcal{T}_{\cc\dd}^\text{(L)}\Big(\{\mb k^\text{(L)}\}\Big)=&\sum_{\aa_1,\cdots,\aa_N=\pm} \int_{-\infty}^0\prod_{\ell={1}}^N\bigg[\di\tau_\ell\,(\ii\aa_\ell)(-\tau_\ell)^{p_\ell} \prod_{i=1}^{B_\ell}C_{\aa_\ell}\Big(k_i^{(\ell)};\tau_\ell\Big)\bigg]\n\\
  & \times \prod_{n={1}}^{N-1}\Big[D_{\aa_n\aa_{n+1}}^{(\wt\nu_{n+1})}\big(P_n;\tau_n,\tau_{n+1}\big)\Big](-\tau_{1})^{3/2+\cc\ii\wt\nu_1}(-\tau_N)^{3/2+\dd\ii\wt\nu_{N+1}},\\
\label{eq_Tright}
  \mathcal{T}_{\cc\dd}^\text{(R)}\Big(\{\mb k^\text{(R)}\}\Big)=&\sum_{\aa_{N+1},\cdots,\aa_{V}=\pm} \int_{-\infty}^0\prod_{\ell={N+1}}^V\bigg[\di\tau_\ell\,(\ii\aa_\ell)(-\tau_\ell)^{p_\ell} \prod_{i=1}^{B_\ell}C_{\aa_\ell}\Big(k_i^{(\ell)};\tau_\ell\Big)\bigg]\n\\ 
  & \times \prod_{n={N+1}}^{V-1}\Big[D_{\aa_n\aa_{n+1}}^{(\wt\nu_{n+1})}\big(P_n;\tau_n,\tau_{n+1}\big)\Big](-\tau_{N+1})^{3/2+\dd\ii\wt\nu_{N+1}}(-\tau_V)^{3/2+\cc\ii\wt\nu_V},
\end{align}
where $\{\mb k^\text{(L)}\}$ denotes the set of external momenta $\mb k_j^{(i)}$ with $j=1,\cdots, B_i$ and $i=1,\cdots,N$, and $\{\mb k^\text{(R)}\}$ denotes the set of external momenta $\mb k_j^{(i)}$ with $j=1,\cdots, B_i$ and $i={N+1},\cdots,V$. Note that the two groups of momenta are separately conserved, i.e., $\sum\{\mb k^\text{(L)}\}=\sum\{\mb k^\text{(R)}\}=\mb 0$, as a consequence of the total momentum conservation plus the condition $P_N\to0$. The bubble signal $\mathcal{B}_{\cc\dd}(P_N)$ encodes all the nonanalytic behavior of $P_N$ as $P_N\to 0$, and is given by:  
\begin{align}
\label{eq_BubbleSignal}
  \mathcal{B}_{\cc\dd}(P_N) 
  =&~\FR{1}{(4\pi)^{7/2}} \Gamma\bgb -\fr32-\cc\ii\wt\nu_1-\dd\ii\wt\nu_{N+1},\fr32+\cc\ii\wt\nu_1,\fr32+\dd\ii\wt\nu_{N+1},-\cc\ii\wt\nu_1, -\dd\ii\wt\nu_{N+1} \\ 3+\cc\ii\wt\nu_1+\dd\ii\wt\nu_{N+1} \edb\n\\
  &\times P_N^3\Big(\FR{P_N}2\Big)^{2\ii(\cc\wt\nu_1+\dd\wt\nu_{N+1})} .
\end{align}
Here $\Gamma[\cdots]$ is our shorthand notation for products of Euler $\Gamma$ functions. See App.\ \ref{app_math} for the definition. 

\paragraph{One-loop nonlocal cutting rule.} As a side product of our derivation of the factorization theorem, we also get a cutting rule for obtaining the nonlocal signal in arbitrary 1-loop graph. That is, in the computation of the nonanalytic terms of the one-loop graph $\mathcal{T}$ to any orders in $P_N$ as $P_N\to 0$, one can cut the two ``soft'' loop propagators, $D_{\aa_V\aa_1}^{(\wt\nu_1)}(q;\tau_V,\tau_1)$ and $D_{\aa_{N}\aa_{N+1}}^{(\wt\nu_{N+1})}(q;\tau_N,\tau_{N+1})$, by replacing them with their real parts:
\begin{keyeqn}
\begin{align}
\label{eq_Cut}
   D_{\aa_V\aa_1}^{(\wt\nu_1)}\Big(q;\tau_V,\tau_1\Big)
   \To&~ \text{Re}\,D^{(\wt\nu_1)}\Big(q;\tau_V,\tau_1\Big); \\
   D_{\aa_{N}\aa_{N+1}}^{(\wt\nu_{N+1})}\Big(q;\tau_N,\tau_{N+1}\Big)
   \To&~ \text{Re}\,D^{(\wt\nu_{N+1})}\Big(q;\tau_N,\tau_{N+1}\Big).
\end{align}
\end{keyeqn}
Here, we deliberately remove the SK indices on the right hand sides of these equations, to highlight the fact that the real part of a bulk massive propagator is independent of SK labels. As a consequence, the time integrals nesting over the two cut lines are automatically removed in the computation of the nonanalytic terms, which justifies the name ``cutting rule.'' We emphasize that, as will be made explicit in the following proof, the cutting rule applies to the nonlocal CC signals to all orders in the expansion of the soft momenta $P_N$.

\subsection{Proof}
\label{sec_proof}

In this subsection, we shall prove the theorem stated above. To help the readers to keep track of what is going on, we first outline the main steps of the proof before going into details. Readers uninterested in technical details may only read the outline or skip the entire subsection. 

\paragraph{Outline.} The proof comes in the following several steps.
\begin{enumerate}
  \item \emph{Partial MB representation}: We use the MB representation for all massive loop lines \cite{Qin:2022lva,Qin:2022fbv}. The MB representation turns all the momentum and time dependences in the Hankel functions into powers, which greatly simplifies the time integrals and the loop momentum integral.

  \item \emph{Nested time integrals}: We then consider the time integrals. Without performing these integrations explicitly, we prove that all (nested) time integrals only produce right poles in all Mellin variables.\footnote{Right poles mean a series of poles at $s=an+b$ with $a>0$, $b\in\mathbb{C}$, and $n\in\mathbb{N}$. Our prescription is that all Mellin integral contours should be deformed to the left side of all right poles, even when $\text{Re}\,b<0$. Likewise, left poles mean a semi-infinite series of poles emanating to the left half of the complex $s$ plane, and the Mellin integral contour should be deformed to the right side of all left poles.}
   
  \item \emph{Loop momentum integral}: We then consider the loop integral over the momentum $\mb q$. After finishing the loop integral, the result is fully expressed in terms of external momenta $\mb P_\ell$. We show that any potential nonanalytic terms in $P_N\to 0$ must come from a region where both $\mb q$ and $\mb q+\mb P_N$ become soft, and this region gives rise to a term $\propto P_N^{3-2s_{V\bar 1N(\ob{N+1})}}$. Since we are looking at $P_N\to 0$ limit, this term tells us two things: 1) We should pick up ``left poles'' for the Mellin integrals over $s_V,\bar s_1,s_N,\bar s_{N+1}$; 2) The location of the left poles of these four Mellin variables should be such that $2s_{V\bar 1N(\ob{N+1})}$ takes noninteger values. Only in this case the term $P_N^{3-2s_{V\bar 1N(\ob{N+1})}}$ is truly nonanalytic in $P_N$ at $P_N=0$. Coupled with Step 2, we see that all poles from the time integrals can be discarded since they are right poles.

  \item \emph{Nonlocal signal poles}: We then collect all possible left poles with $2s_{V\bar 1N(\ob{N+1})}\notin \mathbb{Z}$. We call them nonlocal signal poles as explained in Step 3. The nonlocal signal poles are entirely from the MB representations of the Hankel functions in the two soft propagators $D_{\aa_V\aa_1}^{(\wt\nu_1)}(q;\tau_V,\tau_1)$ and $D_{\aa_{N}\aa_{N+1}}^{(\wt\nu_{N+1})}(q;\tau_N,\tau_{N+1})$. Both the loop momentum integral and the time integrals do not contribute to nonlocal signal poles.
  
  \item \emph{Cutting rule}: We then prove that, at the nonlocal signal poles identified in Step 4, the two soft propagators $D_{\aa_V\aa_1}^{(\wt\nu_1)}(q;\tau_V,\tau_1)$ and $D_{\aa_{N}\aa_{N+1}}^{(\wt\nu_{N+1})}(q;\tau_N,\tau_{N+1})$ are independent of any SK indices. They can be freely replaced by their real parts. In particular, the time-ordering $\theta$ functions become irrelevant in these real parts. Consequently, the nested time integrals over $\tau_V$ and $\tau_1$, and over $\tau_N$ and $\tau_{N+1}$, are automatically factorized. 
  
  \item \emph{Collecting residues}: We finally collect the residues of the Mellin integrand for the four ``soft Mellin variables'' $(s_V,\bar s_1,s_N,\bar s_{N+1})$ at the nonlocal signal poles at the leading order in $P_N$, and finish the Mellin integral over all other Mellin variables. The result just gives the leading nonanalytic contribution to the 1-loop graph $\mathcal{T}$ in the limit $P_N\to 0$, which has a factorized expression as given in (\ref{eq_SoftThm}).
\end{enumerate}

Below we carry out these steps in details.

\paragraph{Partial MB representation.} First, we use the partial MB representation of the 1-loop graph $\mathcal{T}(\{\mb k\})$ in (\ref{eq_Tgeneral}). See Sec.\ 3 of \cite{Qin:2022fbv} for a more comprehensive introduction to the partial MB representation. In short, we take the MB representation of all internal massive propagators, with all external lines unchanged. The MB representations for the two opposite-sign propagators $D_{\pm\mp}^{(\wt\nu)}(k;\tau_1,\tau_2)$ in (\ref{eq_Dmp}) and (\ref{eq_Dpm}) are given by:
\begin{align}
\label{eq_DScalarMB1}
    D_{\pm\mp}^{(\wt\nu)}(k;\tau_1,\tau_2) =&~ \FR{1}{4\pi}
    \int_{-\ii\infty}^{+\ii\infty}
    \FR{\di s_1}{2\pi\ii}\FR{\di s_2}{2\pi\ii}\,
    e^{\mp\ii\pi(s_1-s_2)}\Big(\FR{k}2\Big)^{-2s_{12}}
    (-\tau_1)^{-2s_1+3/2}(-\tau_2)^{-2s_2+3/2}\n\\
    &\times \Gamma\Big[s_1-\FR{\ii\wt\nu}2,s_1+\FR{\ii\wt\nu}2,s_2-\FR{\ii\wt\nu}2,s_2+\FR{\ii\wt\nu}2\Big],
\end{align} 
while the same-sign propagators are still given by (\ref{eq_Dpmpm}) with the opposite-sign propagators in (\ref{eq_Dpmpm}) replaced by their MB representation in (\ref{eq_DScalarMB1}).

\begin{figure}
\centering
\includegraphics[width=0.36\textwidth]{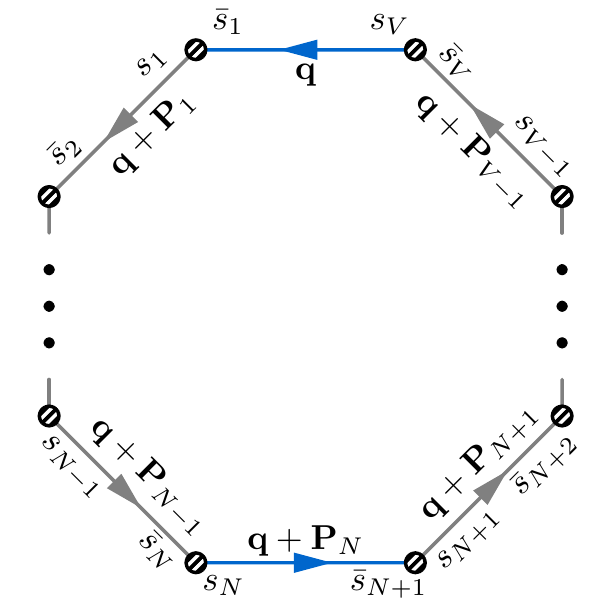}
\caption{The Mellin variables for the loop modes.} 
  \label{fig_mellin}
\end{figure}

In the 1-loop graph $\mathcal{T}(\{\mb k\})$ in (\ref{eq_Tgeneral}), for the massive propagator $D_{\aa_i\aa_j}^{(\wt\nu_j)}(p;\tau_i,\tau_j)$ connecting the two time variables $\tau_i$ and $\tau_j$, we designate two Mellin variables, $s_i$ for the mode at $\tau_i$, and $\bar s_j$ at $\tau_j$, as shown in Fig.\ \ref{fig_mellin}. With these Mellin variables, we rewrite all $V$ internal propagators as in (\ref{eq_DScalarMB1}). Therefore, we have a total of $2V$ Mellin integrals over $(s_1,\cdots, s_V,\bar s_1,\cdots,\bar s_V)$. Note that $s_\ell$ and $\bar s_\ell$ are independent variables. 

\paragraph{Poles of time integrals.} After taking the partial MB representation, let us look at the integrals over the time variables $\tau_1,\cdots,\tau_V$ in (\ref{eq_Tgeneral}). Using the explicit form the of the bulk-to-boundary propagator for conformal scalars (\ref{eq_CSProp}), together with the MB representation for the loop massive propagator (\ref{eq_DScalarMB1}), we observe that the time integrals have the following schematic form: 
\begin{align}
\label{eq_TimeInt}
&\int_{-\infty}^0 \di\tau_1\cdots\di\tau_V\, (-\tau_1)^{\wh p_1-2s_{1\bar1}}\cdots (-\tau_V)^{\wh p_V-2s_{V\ob V}}e^{\ii(\aa_1 E_1 \tau_1+\cdots+\aa_V E_V\tau_V)}\n\\
&\times \mathcal{N}(\tau_1,\tau_2)\mathcal{N}(\tau_2,\tau_3)\cdots \mathcal{N}(\tau_V-1,\tau_V)\mathcal{N}(\tau_V,\tau_1),
\end{align}
with $\wh p_\ell\equiv p_\ell+B_\ell+3/2$. Here $p_\ell$ comes from the explicit time dependence in the time integrals in (\ref{eq_Tgeneral}), $B_\ell$ counts the total number of external conformal scalar modes at Vertex $\ell$, and $3/2$ comes from the MB representation of the loop propagators. 
Here the ``nesting'' function $\mathcal{N}(\tau_i,\tau_j)$ can take any of the three possible choices: $1$, $\theta(\tau_i-\tau_j)$, and $\theta(\tau_j-\tau_i)$, depending on the choices of SK indices at all vertices. By using relations such as $\theta(\tau_i-\tau_j)+\theta(\tau_j-\tau_i)=1$, it is easy to see that the time integral (\ref{eq_TimeInt}) can be organized into a linear combination of many terms, and each term is a product of several monotonically ordered time integrals. For example, a typical term could look like:
\begin{align}
&\int_{-\infty}^0 \di\tau_1\cdots\di\tau_V\, (-\tau_1)^{\wh p_1-2s_{1\bar1}}\cdots (-\tau_V)^{\wh p_V-2s_{V\ob V}}e^{\ii(\aa_1 E_1 \tau_1+\cdots+\aa_V E_V\tau_V)}\n\\
&\times \Big[\theta(\tau_1-\tau_2)\cdots\theta(\tau_j-\tau_{j+1})\Big]\Big[\theta(\tau_{k}-\tau_{k+1})\cdots\theta(\tau_\ell-\tau_{\ell+1})\Big],
\end{align}
where $1\leq j$, $j+1<k$, $k+1<\ell$, and $\ell+1\leq V$. In this particular term, we identify two nested factors, one of which orders the time variables such that $\tau_1>\tau_2\cdots >\tau_j+1$, and the other gives $\tau_k>\cdots >\tau_{\ell+1}$. Since each nested factor can be integrated separately, we only need to consider one nested integral at a time. In particular, an integral over a time variable without any $\theta$ functions can be regarded as a singly nested time integral. 

So, without loss of generality,  let us consider a nested integral of the following form:
\begin{align}
\label{eq_NestedTInt}
  &\int_{-\infty}^0\di\tau_1\cdots\di\tau_\ell \, (-\tau_1)^{\wh p_1-2s_{1\bar1}}\cdots (-\tau_\ell)^{\wh p_\ell-2s_{\ell\bar\ell}}e^{\ii\aa(E_1 \tau_1+\cdots+E_\ell\tau_\ell)} 
 \theta(\tau_1-\tau_2)\cdots \theta(\tau_{\ell-1}-\tau_\ell).
\end{align}
Notice that we have set all SK indices equal: $\aa_1=\cdots=\aa_\ell\equiv \aa$, since the nested time integral arises only in this case. We would like to identify its pole structure in all Mellin variables $(s_1,\cdots,s_\ell,\bar s_1,\cdots,\bar s_\ell)$. Any pole of the above integral must come from a divergence of the integral, so we are led to ask when this integral diverges. First, the integrand is well behaved for any finite value of $\tau_i>0~(i=1,\cdots,\ell)$, and therefore, the potential divergences must come from two endpoints of the integral. Second, the integral is well behaved at $\tau_i\to -\infty$ for physical values of momenta, as a result of Bunch-Davies initial condition. Therefore, any pole of the above integral must come from the limit when one or several of $\tau_i$ goes to 0.

So, let us focus on the integral in the limit $\tau_i\to 0$ by introducing a small cutoff $0<\ep\ll \min\{E_1^{-1},\cdots, E_\ell^{-1}\}$:
\begin{align}
  &\int_{-\ep}^0\di\tau_1\cdots\di\tau_\ell \, (-\tau_1)^{\wh p_1-2s_{1\bar1}}\cdots (-\tau_\ell)^{\wh p_\ell-2s_{\ell\bar\ell}}e^{\ii\aa( E_1 \tau_1+\cdots+ E_\ell\tau_\ell)} 
 \theta(\tau_1-\tau_2)\cdots \theta(\tau_{\ell-1}-\tau_\ell).
\end{align}
In this restricted region, we can Taylor expand all exponential factors, and get
\begin{align}
  &\sum_{n_1,\cdots,n_\ell=0}^\infty\FR{(-\ii\aa)^{n_1+\cdots+n_\ell}E_1^{n_1}\cdots E_\ell^{n_\ell}}{n_1!\cdots n_\ell!}\int_{-\ep}^0\di\tau_1\cdots\di\tau_\ell \, (-\tau_1)^{\wh p_1-2s_{1\bar1}+n_1}\cdots (-\tau_\ell)^{\wh p_\ell-2s_{\ell\bar\ell}+n_\ell}\n\\ 
  &\times\theta(\tau_1-\tau_2)\cdots \theta(\tau_{\ell-1}-\tau_\ell).
\end{align}
Then the integral over $\tau_1$ can be done:
\begin{align}
  \int_{-\ep}^0\di\tau_1\,(-\tau_1)^{\wh p_1-2s_{1\bar 1}+n_1}\theta(\tau_1-\tau_2)
=\FR{(-\tau_2)^{\wh p_1-2s_{1\bar 1}+n_1+1}}{\wh p_1-2s_{1\bar 1}+n_1+1}. 
\end{align}
This is divergent when $\wh p_1-2s_{1\bar 1}+n_1=-1$. Therefore, we identify a set of poles for $s_{1\bar 1}$:
\bge
  s_{1\bar 1}=\FR{\wh p_1+n_1+1}{2}.~~~(n_1=0,1,2,\cdots)
\ede
These are right poles. Next, we insert the $\tau_2$ factor in the above result back to the integrand and finish the $\tau_2$ integral:
\begin{align}
  \int_{-\ep}^0\di\tau_2\,(-\tau_2)^{\wh p_{12}-2s_{1\bar 12\bar 2}+n_{12}+1}\theta(\tau_2-\tau_3)
=\FR{(-\tau_3)^{\wh p_{12}-2s_{1\bar 12\bar 2}+n_{12}+2}}{\wh p_{12}-2s_{1\bar 12\bar 2}+n_{12}+2}. 
\end{align}
Likewise, we identify a second set of poles:
\bge
  s_{1\bar 12\bar 2}=\FR{\wh p_{12}+n_{12}+2}{2}.~~~(n_1,n_2=0,1,2,\cdots)
\ede
These are again right poles. 

Now it becomes clear that the above procedure can be carried through all nested time variables, and the poles of the original nested time integral are given by:
\begin{align}
  s_{1\bar 1}=\FR{\wh p_1+n_1+1}{2}, ~~
  s_{1\bar 12\bar 2}=\FR{\wh p_{12}+n_{12}+2}{2},~~\cdots~~
  s_{1\bar 1\cdots \ell\bar \ell}=\FR{\wh p_{1\dots \ell}+n_{1\cdots \ell}+\ell}{2},
\end{align}
where $n_1,\cdots, n_\ell =0,1,2,\cdots$. Therefore we have proved that the nested time integrals only give rise to right poles.

It is useful to check the above result for special cases with $\ell = 1$ and $2$ in (\ref{eq_NestedTInt}). For $\ell =1$, 
\begin{align}
  &\int_{-\infty}^0\di\tau_1 (-\tau_1)^{\wh p_1-2s_{1\bar1}} e^{\ii \aa E_1 \tau_1}=\FR{1}{(\ii\aa E_1)^{\wh p_1-2s_{1\bar1}+1}}\Gamma(\wh p_1-2s_{1\bar1}+1).
\end{align}
Clearly, the poles are given by $s_{1\bar 1}=(\wh p_1+n_1+1)/2$, in agreement with the above analysis. 

Then, for the case of $\ell=2$,  
\begin{align}
&\int_{-\infty}^0 \di\tau_1\di\tau_2\,(-\tau_1)^{\wh p_1-2s_{1\bar1}}(-\tau_2)^{\wh p_2-2s_{2\bar 2}}e^{\ii\aa (E_1\tau_1+E_2\tau_2)}\theta(\tau_1-\tau_2) \n\\
=&~\FR{1}{(\ii\aa E_2)^{\wh p_{12}-2s_{1\bar 12\bar 2}+2}} 
\times{}_2\mathcal F_1\left[\bgm \wh p_1-2s_{1\bar 1}+1,\wh p_{12}-2s_{1\bar 12\bar 2}+2\\ \wh p_1-2s_{1\bar 1}+2\edm\middle| -\FR{E_1}{E_2}\right],
\end{align}
and the poles are entirely from the two upper $\Gamma$ factors in the dressed hypergeometric function ${}_2\mathcal F_1$ (defined in App.\ \ref{app_math}), and these poles are also identical to what we found above.

The upshot of this part is that the time integrals, however nested, only produce right poles for all Mellin variables. 

\paragraph{Momentum integral.} Next let us look at the loop momentum integral over $\mb q$ in (\ref{eq_Tgeneral}). After the partial MB representation, the loop momentum reads:
\begin{align}
\label{eq_LoopMomInt}
  &\mathcal{L}=\int\FR{\di^3\mb q}{(2\pi)^3}\mathcal{P}; 
  &&\mathcal{P}\equiv |\mb q|^{-2s_{V\bar1}}|\mb q+\mb P_1|^{-2s_{1\bar 2}}|\mb q+\mb P_2|^{-2s_{2\bar 3}}\cdots|\mb q+\mb P_{V-1}|^{-2s_{(V-1)\ob V}}.
\end{align}
Remember that we are interested in the limit $P_N\to 0$ with all other momenta appearing in (\ref{eq_Tgeneral}) kept finite, where $N$ can be any integer from 1 to $V-1$. In this limit, we can always pick up a value $Q$ such that $P_N<Q < \text{min}\{P_1,\cdots,P_{N-1},P_{N+1},\cdots,P_{V-1}\}$, and then separate the momentum integral into a soft-momentum integral $\mathcal{L}_\text{S}$ and a hard-momentum integral $\mathcal{L}_\text{H}$:
\begin{align}
  &\mathcal{L}=\mathcal{L}_\text{S}+\mathcal{L}_\text{H};
  &&\mathcal{L}_\text{S}=\int\FR{\di^3\mb q}{(2\pi)^3}\mathcal{P}\,\theta(Q-|\mb q|),
  &&\mathcal{L}_\text{H}=\int\FR{\di^3\mb q}{(2\pi)^3}\mathcal{P}\,\theta(|\mb q|-Q).
\end{align}
For the soft-momentum integral, we can Taylor expand all but the two factors $|\mb q|^{-2s_{V\bar 1}}$ and $|\mb q+\mb P_{N}|^{-2s_{N(\ob{N+1})}}$ around $\mb q=\mb 0$. The Taylor series is guaranteed to be convergent for the entire range of the soft-momentum integral:
\begin{align}
  |\mb q+\mb P_n|=\sum_{\ell=0}^\infty\FR{1}{\ell !} q^{i_1}\cdots q^{i_\ell}\bigg[\FR{\pd^\ell|\mb q+\mb P_n| }{\pd q^{i_1}\cdots\pd q^{i_\ell}}\bigg]_{q=0}. ~~~(n=1,\cdots,N-1,N+1,\cdots,{V-1})
\end{align}
In particular, when $P_N\to 0$, the leading term $|\mb q+\mb P_n|=P_n+\order{q}$ for all $n\neq N$ will contribute to the leading term in $P_N$, while higher order terms in $q$ lead to higher order terms in $P_N$. With these considerations, the soft-momentum integral $\mathcal{L}_\text{S}$ is reduced to:
\begin{align}
\label{eq_LowMomInt}
  \mathcal{L}_\text{S}
  =&~ \Big(P_1^{-2s_{1\bar 2}}\cdots P_{N-1}^{-2s_{(N-1)\ob N}}\Big)\Big(P_{N+1}^{-2s_{(N+1)(\ob{N+2})}}\cdots P_{V-1}^{-2s_{(V-1)\bar V}}\Big)\n\\
  &\times\int\FR{\di^3\mb q}{(2\pi)^3}|\mb q|^{-2s_{V\bar 1}}|\mb q+\mb P_N|^{-2s_{N(\ob{N+1})}}\theta\Big(Q-|\mb q|\Big)\Big[1+\order{P_N}\Big] .
\end{align}
On the other hand, the hard-momentum integral $\mathcal{L}_\text{H}$ is analytic in $P_N$, as one can see by Taylor expanding the integrand of $\mathcal{L}_\text{H}$ around $P_N=0$, which is guaranteed to be a convergent series since the integral variable $q>Q>P_N$. This tells us that any possible nonanalytic behavior in $P_N$ at $P_N=0$ must come from the soft-momentum integral (\ref{eq_LowMomInt}). In particular, so far as nonanalyticity in $P_N$ is concerned, we are free to extend the upper limit of the integral (\ref{eq_LowMomInt}) to $\infty$, since the additional terms introduced by this extension is again analytic in $P_N$. Therefore, we have:
\begin{align} 
  \mathcal{L} 
  =&~ \Big(P_1^{-2s_{1\bar 2}}\cdots P_{N-1}^{-2s_{(N-1)\ob N}}\Big)\Big(P_{N+1}^{-2s_{(N+1)(\ob{N+2})}}\cdots P_{V-1}^{-2s_{(V-1)\bar V}}\Big)\n\\
  &\times\int\FR{\di^3\mb q}{(2\pi)^3}|\mb q|^{-2s_{V\bar 1}}|\mb q+\mb P_N|^{-2s_{N(\ob{N+1})}}\Big[1+\order{P_N}\Big] +\text{terms analytic in $P_N$}.
\end{align}
The remaining integral is nothing but the bubble 1-loop integral, and can be directly done, as shown in App.\ \ref{app_bubble}:
\begin{align}
\label{eq_BubbleMomInt}
   &\int\FR{\di^3\mb q}{(2\pi)^3}|\mb q|^{-2s_{V\bar 1}}|\mb q+\mb P_N|^{-2s_{N(\ob{N+1})}} \n\\
   =&~\FR{P_N^{3-2s_{V\bar 1N(\ob{N+1})}}}{(4\pi)^{3/2}}\Gamma\bgb s_{V\bar 1N(\ob{N+1})}-\fr32,\fr32-s_{V\bar 1},\fr32-s_{N(\ob{N+1})} \\ 3-s_{V\bar 1N(\ob{N+1})}, s_{V\bar 1},s_{N(\ob{N+1})}\edb .
\end{align}
Therefore we finally get:
\begin{align}
\label{eq_LoopMomIntResult}
  \mathcal{L} 
  =&~\FR{1}{(4\pi)^{3/2}}\Gamma\bgb s_{V\bar 1N(\ob{N+1})}-\fr32,\fr32-s_{V\bar 1},\fr32-s_{N(\ob{N+1})} \\ 3-s_{V\bar 1N(\ob{N+1})}, s_{V\bar 1},s_{N(\ob{N+1})}\edb P_N^{3-2s_{V\bar 1N(\ob{N+1})}} \Big[1+\order{ P_N}\Big] \n\\
  &\times \Big(P_1^{-2s_{1\bar 2}}\cdots P_{N-1}^{-2s_{(N-1)\ob N}}\Big) \Big(P_{N+1}^{-2s_{(N+1)(\ob{N+2})}}\cdots P_{V-1}^{-2s_{(V-1)\bar V}}\Big)+\text{terms analytic in $ P_N$} 
\end{align}

We see that the nonanalytic behavior of $P_N$ is controlled by the power $P_N^{3-2s_{V\bar 1N(\ob{N+1})}}$. Nonanalyticity requires that $2s_{V\bar 1N(\ob{N+1})}$ takes noninteger values. This value is provided by the location of poles of the four variables $(s_V,\bar s_1,s_N,\bar s_{N+1})$. Also, this nonanalytic power must be combined into the power of a momentum ratio in the final result. Now, since we are taking $P_N\to 0$, the convergence of the resulting series requires that we pick up left poles for all these variables.

\paragraph{Left pole structure.} From the analysis of the momentum integral, we have seen that the nonanalyticity of the graph at $P_N=0$ is contributed by left poles of Mellin variables. Now we take a closer look at these poles. 

Above we have shown that the time integral does not yield any left poles. The momentum integral, on the other hand, has one set of left poles:
\bge
s_{V\bar 1N\ob{(N+1)}}=-n+\FR32. ~~~~~(n=0,1,\cdots)
\ede
But these poles only contribute to integer powers of $P_N$, which are analytic at $P_N=0$, and therefore we can discard them.

Then the remaining left poles are all contributed by the $\Gamma$ functions from the MB representation of Hankel functions in the massive propagators, as shown in (\ref{eq_DScalarMB1}). For the four relevant variables $(s_V,\bar s_1,s_N,\bar s_{N+1})$, these $\Gamma$ factors are:
\begin{align}
\label{eq_NonlocalGammas}
  &~\Gamma\bgb s_V-\FR{\ii\wt\nu_1}2,s_V+\FR{\ii\wt\nu_1}2,\bar s_1-\FR{\ii\wt\nu_1}2,\bar s_1+\FR{\ii\wt\nu_1}2\edb\n\\
  \times&~\Gamma\bgb s_N-\FR{\ii\wt\nu_{N+1}}2,s_N+\FR{\ii\wt\nu_{N+1}}2,\bar s_{N+1}-\FR{\ii\wt\nu_{N+1}}2,\bar s_{N+1}+\FR{\ii\wt\nu_{N+1}}2\edb.
\end{align}
We have 4 Mellin variables, each of which has 2 groups of left poles. Say, $s_V=-n_V\pm \ii\wt\nu_1/2$. For convenience, we call $s_V=-n_V+ \ii\wt\nu_1/2$ the negative poles, and $s_V=-n_V- \ii\wt\nu_1/2$ the positive poles. Now, in the result of loop momentum integral (\ref{eq_LoopMomIntResult}), the factor $1/\Gamma(s_{V\bar 1})$ implies that the poles of $s_V$ and $\bar s_1$ must be both positive or both negative. Similarly, the factor $1/\Gamma(s_{N(\ob{N+1})})$ in (\ref{eq_LoopMomIntResult}) implies that the poles of $s_N$ and $\bar s_{N+1}$ must be both positive or both negative. Then we have the following possibilities:
\begin{align}
\label{eq_NonlocalPoles}
   s_V=-n_V-\cc\FR{\ii\wt\nu_1}{2},~
   \bar s_1=-\bar n_1-\cc\FR{\ii\wt\nu_1}{2}, ~
   s_N=-n_N-\dd\FR{\ii\wt\nu_{N+1}}{2},~
   \bar s_{N+1}=-\bar n_{N+1}-\dd\FR{\ii\wt\nu_{N+1}}{2}, 
\end{align}
where $\cc=\pm 1$ and $\dd=\pm 1$. In the degenerate case $\wt\nu_1=\wt\nu_{N+1}$, we should also set $\cc=\dd$, in order that $2s_{V\bar 1N(\ob{N+1})}\notin\mathbb{Z}$. These are the only poles that can contribute to the nonlocal signal in $P_N$. We call them signal poles. 

\paragraph{Cutting the nested time integral.} Above we have identified all possible poles that could contribute to the nonlocal signal in the limit $P_N\to 0$, as summarized in (\ref{eq_NonlocalPoles}). An immediate yet important consequence of this result is that the nested time integrals across the cut are automatically factorized. That is, the time integrals nesting the two sides of the cut neither contribute left poles, nor contribute any finite residues at the signal poles in (\ref{eq_NonlocalPoles}). The first half of this statement (the absence of left poles) has been proved above. Now we show more explicitly that these nested time integrals vanish at all signal poles in (\ref{eq_NonlocalPoles}).

There are two possible nested time integrals across the cut, one nesting $\tau_V$ and $\tau_1$, the other nesting $\tau_N$ and $\tau_{N+1}$. The treatments for both integrals are identical so we only consider the former case.

The integrals over $\tau_V$ and $\tau_1$ are nested only when $\aa_V=\aa_1\equiv\aa$, due to the $\theta$ functions in the same-sign propagator $D_{\pm\pm}$, as shown in (\ref{eq_Dpmpm}). It proves convenient to rewrite $D_{\pm\pm}$ as:
\begin{align}
\label{eq_DppReIm}
  D_{\pm\pm}(q;\tau_V,\tau_1)=&~\text{Re}\,D_{\pm\pm}(q;\tau_V,\tau_1)+\ii\,\text{Im}\,D_{\pm\pm}(q;\tau_V,\tau_1);\\
  \label{eq_DppRe}
  \text{Re}\,D_{\pm\pm}(q;\tau_V,\tau_1)
  =&~\FR12\Big[D_{\mp\pm}(q;\tau_V,\tau_1)+D_{\pm\mp}(q;\tau_V,\tau_1)\Big]\\
  \label{eq_DppIm}
  \ii\,\text{Im}\,D_{\pm\pm}(q;\tau_V,\tau_1)
  =&~\FR12\Big[D_{\mp\pm}(q;\tau_V,\tau_1)-D_{\pm\mp}(q;\tau_V,\tau_1)\Big]\text{sgn}(\tau_V-\tau_1),
\end{align}
where $\text{sgn}(z)\equiv z/|z|$ extracts the sign of a real number $z\neq 0$. Thus we see that the real part of $D_{\pm\pm}$ is automatically factorized, while the imaginary part of $D_{\pm\pm}$ is nested. 

With the above representation for $D_{\pm\pm}$, we can express all four $D_{\aa_V\aa_1}$ in terms of opposite-sign propagators $D_{\mp\pm}$. Now let us see what $D_{\mp\pm}$ contributes inside the Mellin integrand. By examining the PMB representation of the propagator $D_{\aa_V\aa_1}$ in (\ref{eq_DScalarMB1}), we see that the SK indices affect the integrand only through the factor $e^{\mp\ii\pi(s_V-\bar s_1)}$. Now, at the poles given in (\ref{eq_NonlocalPoles}), this factor evaluates to $(-1)^{n_V-\bar n_1}$, which is independent of the two index choices of $D_{\mp\pm}$. In particular, this implies that:
\begin{enumerate}
  \item The real parts of all four propagators, namely $\text{Re}\,D_{\aa_V\aa_1}$ with $\aa_V=\pm 1$ and $\aa_1=\pm1$, all contribute the same factor $(-1)^{n_V-\bar n_1}$ to the Mellin integrand at the poles in (\ref{eq_NonlocalPoles}).
  \item  The imaginary parts of all four propagators, $\text{Im}\,D_{\aa_V\aa_1}$, vanish at the poles (\ref{eq_NonlocalPoles}). 
\end{enumerate}
The same argument also applies to all integrals nesting $\tau_N$ and $\tau_{N+1}$. Therefore, we conclude that we can freely replace the two cut propagators, $D_{\aa_V\aa_1}^{(\wt\nu_1)}(q;\tau_V,\tau_1)$ and $D_{\aa_{N}\aa_{N+1}}^{(\wt\nu_{N+1})}(q;\tau_N,\tau_{N+1})$, by their real parts, when computing the nonanalytic part of the correlator as $P_N\to 0$, as summarized in (\ref{eq_Cut}). In particular, this replacement means that the time integrals nesting over the two cut lines are automatically factorized at the signal poles. Therefore, we can think of this replacement as a cutting rule for extracting nonlocal signals of one-loop inflation correlators for a given soft momentum $P_N$. We stress that this analysis is done for all signal poles, and thus the cutting rule holds to all orders in the expansion of $P_N$.

\paragraph{Residues in the soft limit.}
Finally, we evaluate the whole Mellin integrals at the poles in (\ref{eq_NonlocalPoles}) with $n_V=\bar n_1=n_N=\bar n_{N+1}=0$, which we call the \emph{leading poles} since they give the leading terms of the nonlocal signal in the soft limit $P_N\to 0$. There are four types of contributions to these residues. First, we notice that, at the leading poles in \eqref{eq_NonlocalPoles}, four of the $\Gamma$ factors in \eqref{eq_NonlocalGammas} are singular with residue $1$, while the other four are regular. Together with two prefactors $1/(4\pi)$ and the factor $2^{2s_{V\bar 1 N(\ob{N+1})}}$ in (\ref{eq_DScalarMB1}), they give:
\begin{align}
  \FR{2^{-2\ii (\cc \wt\nu_1+\dd \wt\nu_{N+1})}}{(4\pi)^2}\Gamma\bgb -\cc\ii\wt\nu_1,-\cc\ii\wt\nu_1, -\dd\ii\wt\nu_{N+1}, -\dd\ii\wt\nu_{N+1}  \edb.
\end{align}
Second, we evaluate the loop momentum integral at the leading poles in (\ref{eq_NonlocalPoles}) and isolate the result in the first line of (\ref{eq_LoopMomIntResult}):
\begin{align}
  \mathcal{L} 
  \To&~\FR{1}{(4\pi)^{3/2}}\Gamma\bgb -\fr32-\cc\ii\wt\nu_1-\dd\ii\wt\nu_{N+1},\fr32+\cc\ii\wt\nu_1,\fr32+\dd\ii\wt\nu_{N+1} \\ 3+\cc\ii\wt\nu_1+\dd\ii\wt\nu_{N+1}, -\cc\ii\wt\nu_1 ,-\dd\ii\wt\nu_{N+1} \edb P_N^{3+2\ii(\cc\wt\nu_1+\dd\wt\nu_{N+1})}.
\end{align}
Let us put these two pieces together, and define the result to be $\mathcal{B}_{\cc\dd}(P_N)$ which we call the bubble signal:
\begin{align}
  \mathcal{B}_{\cc\dd}(P_N) 
  =&~\FR{1}{(4\pi)^{7/2}} \Gamma\bgb -\fr32 -\cc\ii\wt\nu_1-\dd\ii\wt\nu_{N+1},\fr32+\cc\ii\wt\nu_1,\fr32+\dd\ii\wt\nu_{N+1},-\cc\ii\wt\nu_1, -\dd\ii\wt\nu_{N+1} \\ 3+\cc\ii\wt\nu_1+\dd\ii\wt\nu_{N+1} \edb\n\\
  &\times P_N^3 \Big(\FR{P_N}2\Big)^{2\ii(\cc\wt\nu_1+\dd\wt\nu_{N+1})} .
\end{align}
It is called the bubble signal since this part is essentially the same as what we would get from computing a bubble loop; See Sec.\ \ref{sec_bubble}. The bubble signal encodes all the nonanalyticity of $P_N$ in the limit $P_N\to 0$. 

Now, we have two momentum factors remained, coming from the second line of the loop momentum integral in (\ref{eq_LoopMomIntResult}):
\bge
  \Big(P_1^{-2s_{1\bar 2}}\cdots P_{N-1}^{-2s_{(N-1)\ob N}}\Big) \Big(P_{N+1}^{-2s_{(N+1)(\ob{N+2})}}\cdots P_{V-1}^{-2s_{(V-1)\bar V}}\Big)
\ede
Comparing them with the original momentum integral (\ref{eq_LoopMomInt}), we see that this result amounts to replacing $|\mb q+\mb P_\ell|$ factors by $P_\ell$ for all $\ell=1,\cdots,N-1,N+1,\cdots,V-1$. Remember that we have only evaluated four Mellin integrals with $(s_V,\bar s_1,s_N,\bar s_{N+1})$. All other Mellin integrals are left undone at the moment. Now, we can put the factor $P_1^{-2s_{1\bar 2}}\cdots P_{N-1}^{-2s_{(N-1)\ob N}}$ back, and finish the Mellin integrals over $(s_1,\cdots,s_{N-1},\bar s_2,\cdots, \bar s_N)$. This gives us nothing but the expression for the ``left'' tree graph $\mathcal{T}_{\cc\dd}^\text{(L)}(\{\mb k^\text{(L)}\})$ after the cut, as given in (\ref{eq_Tleft}), and is the third piece of contribution to the residues. Likewise, finishing the Mellin integrals over $(s_{N+1},\cdots,s_{V-1},\bar s_{N+2},\cdots,\bar s_V)$, we get the ``right'' tree graph $\mathcal{T}_{\cc\dd}^\text{(R)}(\{\mb k^\text{(R)}\})$ after the cut, as given in (\ref{eq_Tright}), which makes the fourth contribution to the residues at the signal poles.

Putting all these pieces together, we finally reach a main result of this section, namely (\ref{eq_SoftThm}). This completes our proof of the factorization theorem. 

\subsection{Discussions}
\label{sec_theorem_discussions}

The factorization theorem stated and proved in this section not only tells us when do we expect to see a nonlocal CC signal in a generic 1-loop graph, but also gives the explicit expression for the leading nonlocal CC signal in terms of the two tree graphs $\mathcal{T}^{\text{L}}_{\cc\dd}$ and $\mathcal{T}^{\text{R}}_{\cc\dd}$. Therefore, this theorem can be directly put in use to compute more complicated 1-loop graphs than the simplest bubble graph. We shall pursue this direction in Sec.\ \ref{sec_4pt} and derive the most general nonlocal signals for all 1PI 1-loop graphs of the 4-point inflation correlators. On the other hand, there are several propositions one can directly extract from our proof of the factorization theorem. In the rest of this section, we shall briefly discuss these points, and we plan to expand on some of them in greater details in a future work.

\paragraph{Adjacent cut lines.}
First, in the figures and also in the proof, we have considered the cases where the two cut lines do not share any common vertex. That is, we assume $P_N\to 0$ with $1< N<V-1$. However, the theorem still holds for $N=1$ or $N=V-1$, namely, when the two cut lines share one common vertex, but with an important additional condition, namely that there are at least two external lines connected to the common vertex (Vertex 1 or Vertex $V$ for $N=1$ or $N={V-1}$, respectively). The reason for this additional condition is that we require enough freedom in the kinematics to independently take the $P_N\to 0$ limit. When there is only one external line connected to, say, Vertex 1, we have $B_1=1$. (Recall the definitions in (\ref{eq_KEPB}).) Then, the ``total external energy'' $E_1\equiv |\mb K_1|$ at this vertex is identical to the magnitude of $\mb P_1$. In this case it is impossible to take $P_1\to 0$ with $E_1$ fixed, and there could be additional nonanalytic behavior in $E_1$ which becomes degenerate with the nonanalyticity in $P_1$.

\paragraph{Subgraph.} Second, although we have presented our theorem for 1-loop boundary correlators assuming all bulk-to-boundary propagators being massive or conformal, our proof does not really rely on this assumption. The only occasion that we quote the explicit form of the bulk-to-boundary propagators is when we consider the pole structure of the time integrals. (See the discussion starting from (\ref{eq_TimeInt}).) Even at this place, what we really used is the regularity of the exponential factor $e^{\ii \aa_\ell E_\ell \tau_\ell}$ in the $\tau_\ell\to 0$ limit. This regularity is still true when we replace these conformal or massless bulk-to-boundary propagators by arbitrary massive bulk propagators. Therefore, our theorem also holds as a statement for a 1-loop bulk subgraph, \emph{provided that all the external momenta $\{\mb k\}$ in this subgraph can be held fixed}. That is, we require that those external momenta are not part of other loop integrals, so that it is possible to take $\mb P_N\to 0$ limit with all other momenta held at fixed finite values. 

\paragraph{No nonlocal signal in a disconnected soft limit.} 
In the factorization theorem and its proof, we have only considered the limit $P_N\to 0$ with $N=1,\cdots,V-1$. Recall that $\mb P_N=\mb K_1+\cdots+\mb K_N$ is the total energy flowing into the graph through the first $N$ vertices, that is, through a \emph{connected} subset of vertices. One may naturally ask what happens if we take the soft limit of the total momentum flowing into the graph through a \emph{disconnected} subset of vertices, with all other combinations of momenta fixed at finite values, which we call a ``disconnected soft limit.'' Is there any nonlocal CC signal associated with this disconnected soft limit? The answer to this question is no. To see this, let us have a second look at the integrand of the loop momentum integral (\ref{eq_LoopMomInt}): 
\bge
\label{eq_loopIntegrand}
  \mathcal{P}=|\mb q|^{-2s_{V\bar1}}|\mb q+\mb P_1|^{-2s_{1\bar 2}}|\mb q+\mb P_2|^{-2s_{2\bar 3}}\cdots|\mb q+\mb P_{V-1}|^{-2s_{(V-1)\ob V}}.
\ede
There are a total of $V$ factors here. When we take a connected soft limit (say $\mb P_N\to 0$), there are always a pair of factors in (\ref{eq_loopIntegrand}) sharing almost the same momentum. (In the case of $P_N\to 0$, the pair is given by $|\mb q|^{-2_{V\bar 1}}$ and $|\mb q+\mb P_N|^{-2s_{N(\ob{N+1})}}$.) As shown in the proof, in this case, we can ``pinch'' the rest $V-2$ loop lines by setting $\mb q\to 0$ in them, and move them out of the loop momentum integral. What remain in the integrand are just the two factors $|\mb q|^{-2_{V\bar 1}}$ and $|\mb q+\mb P_N|^{-2s_{N(\ob{N+1})}}$. As shown in (\ref{eq_BubbleMomInt}), integrating these two factors gives rise to the desired nonanalyticity when $P_N\to 0$. 

Now, if we take a disconnected soft limit, say, $\mb K_2+\mb K_V\to 0$ while keeping all other momentum combinations finite, then it is clear that no two momenta in the $V$ factors of (\ref{eq_loopIntegrand}) can become close to each other. Consequently, in all parameter space of the loop momentum $\mb q$, there can be at most one soft momentum at a time, and when a given momentum become soft (say $\mb q$), all the other $V-1$ lines can be ``pinched'' and pulled out of the integral. What remains inside the loop integral would be a single factor $|\mb q|^{-2s_{V\bar 1}}$. Clearly, this amounts to setting $s_{N(\ob{N+1})}=0$ in (\ref{eq_BubbleMomInt}), and it gives us a vanishing result due to the $\Gamma(s_{N(\ob{N+1})})$ factor in the denominator on the right hand side of (\ref{eq_BubbleMomInt}). So, we conclude that a disconnected soft limit does not generate any nonlocal CC signal. 

An immediate corollary of this discussion is that it is impossible to get nonlocal signal by ``cutting'' a single loop propagator; One has to cut at least two lines at a time. On the other hand, one may also ask if we could simultaneously see the nonlocal CC signals from more than two loop lines. The answer is yes, and one can achieve this by taking several successive cuts. We shall expand on all these points more systematically in a future work.

\section{A Boundary OPE Perspective}
\label{sec_ope}

It is usually helpful to look at inflation correlators from a boundary point of view. In particular, the bulk isometries of dS spacetime are mapped to the boundary conformal symmetries, which can be powerful tool to constrain the form of inflation correlators already at the purely kinematic level. In our current discussion, the 1-loop factorization theorem in \eqref{eq_SoftThm} from the last section states that we can cut the two soft lines and pinch all remaining hard lines to get the nonanalytic piece of a 1-loop graph in a soft limit. This is very similar to an OPE procedure on the future boundary: From the position-space point of view, the two tree graphs, $\mathcal{T}_{\cc\dd}^{(L)}$ in (\ref{eq_Tleft}) and $\mathcal{T}_{\cc\dd}^{(R)}$ in (\ref{eq_Tright}), are very far from each other, compared with the typical separation of points within each graph. Therefore, to compute the full loop graph at the leading order in the momentum transfer $P_N$, it seems reasonable to perform an OPE for operators within each subgraph, which simply pinches each subgraph to a common point in the position space at the leading order. 

However, this naive way of applying OPE is not quite correct. The reason is that all massive lines are attached to bulk vertices, and each bulk vertex carries a time variable $\tau_j$ that can take any value in $(-\infty,0)$; See (\ref{eq_Tgeneral}). \emph{A priori}, the SK integral receives contributions from all possible values of $\tau_j$, and it is not clear that one can simply treat these bulk vertices as boundary insertions. 

The way to make progress is to make the following two key observations: First, in the bulk propagators (and also in the bulk mode functions), the time variables $\tau$ always appear within the combination $k\tau$ where $k$ is the momenta carried by the propagator (or by the mode function). As a result, a bulk mode carrying very soft momenta $k\to 0$ will have $k\tau\to 0$, which means that this bulk mode is effectively living on the boundary $\tau\to 0$. Second, although the loop propagators can \emph{a priori} carry arbitrarily large (loop) momentum, the nonlocal signal is always contributed by the region of the loop integral where the two ``cut'' propagators (the two blue lines in Fig.\ \ref{fd_GenLoop}) carry very small momenta, as we have made clear in the proof of the factorization theorem. 

Combining the above two observations, we see that it is possible to treat the four bulk modes in the two cut propagators as boundary modes. We can then apply the boundary OPE to these four boundary modes, and this effectively pinches an arbitrary 1-loop configuration to a bubble. Thus it is indeed possible to reinterpret the 1-loop factorization theorem by a boundary OPE. Below, we shall expand on these points and show more explicitly how the factorization emerges from a boundary OPE viewpoint. We should emphasize at this point that the this OPE interpretation should not be treated as an independent proof of the factorization theorem. It is in any case less rigorous than the bulk proof in the last section. Still, this OPE interpretation may provide useful physical insights into what is happening in the generation of nonlocal CC signals.

\subsection{Pushing the soft lines to the future boundary}

The picture of the OPE is most transparent in the position space. Therefore, we shall begin by rewriting the 1-loop graph $\mathcal T(\{\mb k\})$, as shown in Fig.\ \ref{fd_GenLoop}, in terms of the expectation value of a string of bulk operator product in the position space. For this purpose, we introduce $V$ distinct massive scalar fields $\si^{(j)}(x)$ with mass parameter $\wt\nu_j$ $(j=1,\cdots,V)$. Then, we have:
\begin{align}
\label{eq_Tdeltageneral}
 \mathcal{T}(\{\mb k\}) (2\pi)^3\de(\mb P_V)=&\sum_{\aa_1,\cdots,\aa_V=\pm} \int_{-\infty}^0\prod_{\ell=1}^V\bigg[\di\tau_\ell\,\ii\aa_\ell(-\tau_\ell)^{p_\ell}\prod_{i=1}^{B_\ell}C_{\aa_\ell}\Big(k_i^{(\ell)};\tau_\ell\Big)\bigg]
  \int \prod_{n=1}^{V} \Big[\di^3\mb x_n\, e^{-\ii \mb K_n \cdot \mb x_n}\Big]\n\\ 
  &\times \Big\la \big[\si^{(1)}\si^{(2)}(x_1)\big]_{\aa_1} \big[\si^{(2)}\si^{(3)}(x_2)\big]_{\aa_2} \cdots \big[\si^{(V)}\si^{(1)}(x_V)\big]_{\aa_V}\Big\ra,
\end{align}
where $x_\ell\equiv(\tau_\ell,\mb x_\ell)$. The factor $(2\pi)^3\de(\mb P_V)$ on the left hand side comes from the total 3-momentum conservation produced by the Fourier transform of the position-space correlator $\la[\si^{(1)}\si^{(2)}]_{\aa_1}\cdots\ra$ in the last line. This position-space correlator can be thought of as the generating function $Z[J]$ of a bulk theory of $V$ free massive scalars $\si^{(\ell)}$, in the presence of external source terms $\int\di\tau\di^3\mb x\,J_{\ell}(\tau,\mb x)\si^{(\ell)}\si^{(\ell+1)}$, where $J_\ell(x)=(-\tau)^{p_\ell}\phi_c^{B_\ell}(x)$ and $\phi_c$ is the external conformal scalar. Furthermore, the subscript of the correlator $\la \cdots \ra_{\aa_1\cdots\aa_V}$ denotes the appropriate SK branch. More explicitly: 
\begin{align}
  \big\la \mathcal{O}_{\aa_1}(x_1)\cdots\mathcal{O}_{\aa_n}(x_n)\big\ra
  =\bigg\la \ob{\text{T}}\Big[\prod_{\{i|\aa_i=-\}}\mathcal{O}(x_i)\Big]\text{T}\Big[\prod_{\{j|\aa_j=+\}}\mathcal{O}(x_j)\Big]\bigg\ra,
\end{align}
where T denotes the time ordering and $\ob{\text{T}}$ denotes the anti-time ordering. More details of these SK correlators can be found in \cite{Chen:2017ryl}. 

It is straightforward to see that the expression (\ref{eq_Tdeltageneral}) is equivalent to the original expression (\ref{eq_Tgeneral}). In fact, at the leading order in perturbation theory, the bulk correlator is reduced to a product of propagators after Wick contractions. Since we are assuming $V$ distinct massive scalars in this process, the Wick contraction is unique, and the result is:
\bge
\label{eq_WickContraction}
\Big\la \big[\si^{(1)}\si^{(2)}(x_1)\big]_{\aa_1} \cdots \big[\si^{(V)}\si^{(1)}(x_V)\big]_{\aa_V}\Big\ra
\sim
D^{(\wt\nu_1)}_{\aa_V\aa 1}(x_V,x_1)\cdots D^{(\wt\nu_V)}_{\aa_{V-1}\aa_{V}}(x_{V-1},x_V). 
\ede
where $D_{\aa_i\aa_j}^{(\wt\nu_j)}(x_i,x_j)$ is the position-space bulk propagator for the field $\si^{(j)}$, and is related to the previously used momentum-space propagator via:
\bge
\label{eq_DabPosition}
D_{\aa_i\aa_j}^{(\wt\nu_j)}(x_i,x_j) = 
\int \FR{\di^3\mb k}{(2\pi)^3} D_{\aa_i\aa_j}^{(\wt\nu_j)}(k;\tau_i,\tau_j)e^{\ii \mb k\cdot(\mb x_i-\mb x_j)}.
\ede
If we insert \eqref{eq_WickContraction} and \eqref{eq_DabPosition} into the right hand side of \eqref{eq_Tdeltageneral}, and finish the Fourier transformation, we recover nothing but \eqref{eq_Tgeneral} with an extra factor $(2\pi)^3\de(\mb P_V)$ denoting the momentum conservation.

Now let us consider the soft limit $P_N\to 0$ and the corresponding nonlocal signal in \eqref{eq_Tdeltageneral}. Recall that 
in momentum space, the nonlocal signal associated with $P_N$ comes totally from the configuration where the $\si^{(1)}$ propagator and the $\si^{(N+1)}$ propagator carry soft momenta, while all other loop lines carry hard momenta. This particular configuration in the momentum space has two implications for the corresponding position-space picture:
\begin{enumerate}
\item On a fixed equal-time slice, the bulk operators on the right hand side of \eqref{eq_Tdeltageneral} are divided into two sets far away from each other:
$\{\si^{(j)}(x_i)|1\leq i \leq N\}$ and $\{\si^{(j)}(x_i)|N+1\leq i \leq V\}$.
These two sets of operators are connected by the two ``soft" propagators $D_{\aa_V\aa_1}^{(\wt\nu_1)}(x_V,x_1)$ and $D_{\aa_N\aa_{N+1}}^{(\wt\nu_{N+1})}(x_N,x_{N+1})$. The softness in momentum space implies large distance in position space.
On the other hand, operators among each set are close to each other in 3-dimension equal-time slice in the bulk, since all the other momenta are ``hard" compared to the two soft momenta.  

\item In the time direction, the four bulk modes associated with the two soft propagators are close to the late-time boundary. To see this, let us consider the time integral over $\tau_i$ ($i=V,1,N,N+1$) more carefully. Due to the exponential factor $e^{\pm \ii E_i\tau_i}$ in \eqref{eq_Tdeltageneral}, the time integral receives most contribution from  the region $-1/E_i<\tau_i<0$. For this range of $\tau_i$, we always have $|q\tau_1|\ll 1$, $|q\tau_V|\ll 1$, $|(\mb q+\mb P_N)\tau_{N}|\ll 1$, and $|(\mb q+\mb P_N)\tau_{N+1}|\ll 1$. The soft modes with momenta $|\mb q|, |\mb q+\mb P_N| \ll E_i$ crossed the horizon at the time $\tau_i^*\sim -1/|\mb q|$ or $-1/|\mb q+\mb P_N|$. These conditions allow us to expand the operator $\si^{(j)}(x_i)$ around $\tau_i\sim 0$. 
\end{enumerate}

\paragraph{More on the late-time expansion}

Above we have shown that it is legitimate to push the four soft modes, $\si^{(1)}(x_1)$, $\si^{(1)}(x_V)$, $\si^{(N+1)}(x_N)$, and $\si^{(N+1)}(x_{N+1})$, to the future boundary. That is, we can perform small $\tau$ expansion for them. Now, let us pause and have a closer look at the late-time expansion of a bulk massive field $\si (\tau,\mb x)$:
\bge
\label{eq_sigma(x)}
\si (x)
= \int \FR{\di^3\mb k}{(2\pi)^3}\, \Big[u(k,\tau)a_{\mb k} + u^*(k,\tau) a_{-\mb k}^\dagger\Big] e^{\ii\mb k\cdot \mb x}.
\ede
Here $k=|\mb k|$ denotes the magnitude of the 3-momentum.
$a_{\mb k}$ and $a^\dagger_{\mb k}$ are the annihilation and the creation operators, respectively, with standard commutation relation $[a_{\mb p},a_{\mb q}^\dagger] = (2\pi)^3\de(\mb p-\mb q)$. $u(k,\tau)$ is the mode function satisfying the equation of motion:
\bge
u(k,\tau) = \FR{\sqrt\pi}2 e^{-\pi\wt\nu/2}(-\tau)^{3/2} \mathrm H_{\ii\wt\nu}^{(1)}(-k\tau).
\ede
Now that the time dependence in the operator $\si(x)$ is all in the mode function, so we can expand the mode function around $\tau\sim0$:
\bge
\label{eq_ulatetime}
\lim_{\tau\to 0} u(k,\tau) = -\ii \sqrt{\FR{2}{\pi k^3}}\bigg[ e^{\pi\wt\nu/2}\Gamma(-\ii\wt\nu)\Big(-\FR{k\tau}2\Big)^{3/2+\ii\wt\nu}+e^{-\pi\wt\nu/2}\Gamma(\ii\wt\nu)\Big(-\FR{k\tau}2\Big)^{3/2-\ii\wt\nu}\bigg].
\ede
Insert \eqref{eq_ulatetime} into \eqref{eq_sigma(x)}, we obtain the late-time expansion for the scalar field:
\bge
\label{eq_sigmalatetime}
\lim_{\tau\to0}\si(\tau,\mb x) = \si_+(\mb x) (-\tau)^{3/2+\ii\wt\nu} + \si_-(\mb x) (-\tau)^{3/2-\ii\wt\nu},
\ede
where $\si_\pm$ are boundary operators:
\bge
\si_\pm(\mb x) = \FR{-\ii}{\sqrt{4\pi}}\Gamma(\mp\ii\wt\nu) \int \FR{\di^3\mb k}{(2\pi)^3}\,  \Big(\FR k2\Big)^{\pm\ii\wt\nu}
\Big[ e^{\pm\pi\wt\nu/2} a_{\mb k} - e^{\mp\pi\wt\nu/2} a^\dagger_{-\mb k}\Big] e^{\ii\mb k\cdot \mb x}.
\ede
As mentioned before, the bulk isometries of dS induce conformal symmetries on the future boundary, and it can be shown that the two boundary fields $\si_\pm(\mb x)$ thus defined transform as primary scalars with conformal weights $\Delta_\pm=3/2\pm\ii\wt\nu$. 

The fields $\si_\pm(\mb x)$ are the leading two boundary operators in the late-time expansion (\ref{eq_sigmalatetime}). One can also consider higher order terms:
\bge
\label{eq_sigmalatetimeFull}
\si (\tau,\mb x) = \sum_{\Delta \in \{\Delta_\pm\}} \si_{\Delta}(\mb x)(-\tau)^\Delta,
\qquad \{\Delta_\pm\} = \Big\{2n+\FR32 \pm \ii\wt\nu \Big| n=0,1,2\cdots\Big\},
\ede
The higher order operators $\si_{\Delta}$ with $n\neq 0$ in \eqref{eq_sigmalatetimeFull} are descendants of the corresponding primary fields $\si_\pm$, and their effects are suppressed by at least $\mathcal O(\tau^2)$ in the late-time limit, and thus can be neglected when we consider the leading nonlocal signal in the squeezed limit. We also point out that each term in the expansion \eqref{eq_sigmalatetimeFull} is in one-to-one correspondence with the contribution of residues at nonlocal poles \eqref{eq_NonlocalPoles} of the $\Gamma$ factors $\Gamma(s-\ii\wt\nu/2)\Gamma(s+\ii\wt\nu/2)$ in the MB representation \eqref{eq_DScalarMB1}.

\subsection{Boundary OPE and factorization theorem}

With all preparation made above, we can now perform the late-time expansion \eqref{eq_sigmalatetime} to the four soft modes in the position-space correlator $\la[\si^{(1)}\si^{(2)}]_{\aa_1}\cdots\ra$ in (\ref{eq_Tdeltageneral}):
\begin{align}
\label{eq_latetime1}
\si^{(1)}(x_V) \sim &~ \sum_{\cc_1=\pm} \si^{(1)}_{\cc_1}(\mb x_V)(-\tau_V)^{3/2+\cc_1\ii\wt\nu_1},\\
\label{eq_latetime2}
\si^{(1)}(x_1) \sim &~ \sum_{\cc_2=\pm} \si^{(1)}_{\cc_2}(\mb x_1)(-\tau_1)^{3/2+\cc_2\ii\wt\nu_1},\\ 
\label{eq_latetime3}
\si^{(N+1)}(x_N) \sim &~ \sum_{\dd_1=\pm} \si^{(N+1)}_{\dd_1}(\mb x_N)(-\tau_N)^{3/2+\dd_1\ii\wt\nu_{N+1}},\\ 
\label{eq_latetime4}
\si^{(N+1)}(x_{N+1}) \sim &~ \sum_{\dd_2=\pm} \si^{(N+1)}_{\dd_2}(\mb x_{N+1})(-\tau_{N+1})^{3/2+\dd_2\ii\wt\nu_{N+1}},
\end{align}
where we only keep the leading terms. 
Inserting \eqref{eq_latetime1}-\eqref{eq_latetime4} into the position-space correlator in (\ref{eq_Tdeltageneral}), we get:
\begin{align}
&~\Big\la \big[\si^{(1)}\si^{(2)}(x_1)\big]_{\aa_1} \cdots \big[\si^{(V)}\si^{(1)}(x_V)\big]_{\aa_V}\Big\ra
\sim \mathcal M_{\aa_1\cdots\aa_V}^{\text{Bulk}} \times \sum_{\cc_1,\cc_2,\dd_1,\dd_2}\mathcal M_{\cc_1\cc_2\dd_1\dd_2}^{\text{Boundary}}\n\\
&\times (-\tau_V)^{3/2+\cc_1\ii\wt\nu_1}(-\tau_1)^{3/2+\cc_2\ii\wt\nu_1}(-\tau_N)^{3/2+\dd_1\ii\wt\nu_{N+1}}(-\tau_{N+1})^{3/2+\dd_2\ii\wt\nu_{N+1}},
\end{align}
where
\begin{align}
\label{eq_BulkCorrelator}
\mathcal M_{\aa_1\cdots\aa_V}^{\text{Bulk}} 
\equiv &~ \prod_{i=1}^{N-1}D_{\aa_i\aa_{i+1}}^{(\wt\nu_{i+1})}(x_i,x_{i+1})
 \prod_{j=N+1}^{V-1}D_{\aa_j\aa_{j+1}}^{(\wt\nu_{j+1})}(x_i,x_{j+1}) \\
\label{eq_BoundaryCorrelator}
\mathcal M_{\cc_1\cc_2\dd_1\dd_2}^{\text{Boundary}}
\equiv&~ \Big\la \si^{(1)}_{\cc_1}(\mb x_V)\si^{(1)}_{\cc_2}(\mb x_1)\si^{({N+1})}_{\dd_1}(\mb x_N)\si^{({N+1})}_{\dd_2}(\mb x_{N+1})\Big\ra.
\end{align}
Now we focus on the boundary correlator \eqref{eq_BoundaryCorrelator}. The four boundary operators in the bra-ket are divided into two groups: $\{\si^{(1)}_{\cc_2}(\mb x_1), \si^{({N+1})}_{\dd_1}(\mb x_N)\}$ and $\{\si^{({N+1})}_{\dd_2}(\mb x_{N+1}),\si^{(1)}_{\cc_1}(\mb x_V)\}$. The two groups are far from each other, while the two operators in each group are close to each other. This configuration allows us to do the OPE at the boundary:\footnote{The OPEs in \eqref{eq_OPE1} and \eqref{eq_OPE2} correspond to a free theory. If we turn on interactions among these boundary fields, there will be other terms appearing on the right hand side, which may have lower scaling dimension than the operators shown here. In particular, for the product of two identical massive scalars $\si^{(i)}(\mb x)\si^{(i)}(\mb y)$, the right hand side also includes the identity operator. These operators contribute only when we include more interaction insertions. The identity operator does not contribute to the connected diagram. }
\begin{align}
\label{eq_OPE1}
\si_{\cc_2}^{(1)}(\mb x_1)\si_{\dd_1}^{({N+1})}(\mb x_N) =&~ \Big[ \si_{\cc_2}^{(1)}(\mb x_N) + (\mb x_1-\mb x_N) \cdot \nabla \si_{\cc_2}^{(1)}(\mb x_N)+ \cdots\Big] \si_{\dd_1}^{({N+1})}(\mb x_N) ,\\
\label{eq_OPE2}
\si_{\dd_2}^{({N+1})}(\mb x_{N+1})\si_{\cc_1}^{(1)}(\mb x_V) =&~ \Big[\si_{\dd_2}^{({N+1})}(\mb x_V) + (\mb x_{N+1}-\mb x_V)\cdot \nabla\si_{\dd_2}^{({N+1})}(\mb x_V) \cdots \Big]\si_{\cc_1}^{(1)}(\mb x_V).
\end{align}
For the soft configuration where the two groups of modes $\{\si^{(j)}(x_i)|1\leq i \leq N\}$ and $\{\si^{(j)}(x_i)|N+1\leq i \leq V\}$ are far from each other, the behavior of the boundary correlator \eqref{eq_BoundaryCorrelator} is dominated by the leading terms, namely:
\begin{align}
\label{eq_OPE3}
\si_{\cc_2}^{(1)}(\mb x_1)\si_{\dd_1}^{({N+1})}(\mb x_N) \sim &~ \Big[\si_{\cc_2}^{(1)}\si_{\dd_1}^{({N+1})}\Big](\mb x_N) ,\\
\label{eq_OPE4}
\si_{\dd_2}^{({N+1})}(\mb x_{N+1})\si_{\cc_1}^{(1)}(\mb x_V) \sim&~ \Big[\si_{\dd_2}^{({N+1})}\si_{\cc_1}^{(1)}\Big](\mb x_V).
\end{align}

Now we can calculate the right hand side of \eqref{eq_Tdeltageneral} and check that the leading nonlocal piece does give the factorization theorem \eqref{eq_SoftThm}.
We rewrite the integral \eqref{eq_Tdeltageneral} with the leading late-time expansion \eqref{eq_latetime1}-\eqref{eq_latetime4} and the leading OPE \eqref{eq_OPE3}, \eqref{eq_OPE4}, together with the following rearrangement of the Fourier kernel:
\begin{align}
\label{eq_FourierKernel}
&~\exp\Big[ -\ii\sum_{\ell=1}^V \mb K_\ell\cdot\mb x_\ell\Big]=\exp\Big[-\ii\mb P_V\cdot \mb x_V\Big]\exp\Big[-\ii\mb P_N\cdot (\mb x_N-\mb x_V)\Big]\n\\
&\times
\exp\Big[-\ii\sum_{\ell=1}^{N-1}\mb K_\ell\cdot \mb x_\ell+ \ii \mb P_{N-1}\cdot \mb x_N\Big] \exp\Big[-\ii\sum_{\ell=N+1}^{V-1}\mb K_\ell\cdot \mb x_\ell+\ii (\mb P_{V-1}-\mb P_N)\cdot\mb x_V\Big].
\end{align}
We can further change the measurement for later convenience:
\bge
\di^3 \mb x_1\cdots \di^3 \mb x_V = \di^3\mb x_1\cdots \di^3\mb x_{N-1} \di^3\mb y_N \di^3\mb x_{N+1}\cdots \di^3\mb x_V,\qquad \mb y_N\equiv \mb x_N-\mb x_V.
\ede
The integral \eqref{eq_Tdeltageneral} is thus divided into four parts:

\begin{enumerate}
\item The first part involves the boundary correlator \eqref{eq_BoundaryCorrelator}:
\begin{align}
 & \int \di^3\mb y_N\,e^{-\ii \mb P_N\cdot (\mb x_N-\mb x_V)}
  \Big\la \si^{(1)}_{\cc_1}(\mb x_V)\si^{(1)}_{\cc_2}(\mb x_N) \si^{({N+1})}_{\dd_1}(\mb x_V)\si^{({N+1})}_{\dd_2}(\mb x_N)\Big\ra\n\\
  =& \int \di^3\mb y_N\,e^{-\ii \mb P_N\cdot \mb y_N}
  \Big\la \si^{(1)}_{\cc_1}(\mb 0)\si^{(1)}_{\cc_2}(\mb y_N) \si^{({N+1})}_{\dd_1}(\mb 0)\si^{({N+1})}_{\dd_2}(\mb y_N)\Big\ra,
\end{align}
where we have made used of the spatial translation symmetry. Again, in the free-theory limit, this four-point correlation function breaks into a product of two two-point functions $\la\si^{(1)}_{\cc_1}(\mb 0)\si^{(1)}_{\cc_2}(\mb y_N)\ra$ and $\la\si^{({N+1})}_{\dd_1}(\mb 0)\si^{({N+1})}_{\dd_2}(\mb y_N)\ra$. Since $\si^{(1)}_\cc$ and $\si^{(N+1)}_\dd$ are primary scalars on the boundary, when $\cc_1=\cc_2\equiv\cc$ and $\dd_1=\dd_2\equiv\dd$, we have $\la\si^{(1)}_{\cc}(\mb 0)\si^{(1)}_{\cc}(\mb y_N)\ra\sim y_N^{-3\mp\ii\cc\wt\nu_1}$ and $\la\si^{({N+1})}_{\dd_1}(\mb 0)\si^{({N+1})}_{\dd_2}(\mb y_N)\ra\sim y_N^{-3\mp\ii\dd\wt\nu_{N+1}}$. We see that the equal-sign choices $\cc_1=\cc_2\equiv\cc$ and $\dd_1=\dd_2\equiv\dd$ just give rise to the bubble signal in \eqref{eq_BubbleSignal}.\footnote{Naively, we would expect that the opposite-sign choice $\cc_1=-\cc_2$ makes the two-point correlator $\la\si^{(1)}_{\cc_1}(\mb 0)\si^{(1)}_{\cc_2}(\mb y_N)\ra$ vanish, since the two fields $\si^{(1)}_{\cc_1}$ and $\si^{(1)}_{\cc_2}$ now have different scaling dimensions. However, $\si^{(1)}_{\cc_1}$ and $\si^{(1)}_{\cc_2}$ does not really vanish, but is proportional to $\de^{(3)}(\mb y_N)$, since the two representations with $\Delta_\pm=3/2\pm\ii\wt\nu$ are related by a shadow transformation, and thus are isomorphic to each other. With this said, the result $\la\si^{(1)}_{\cc_1}(\mb 0)\si^{(1)}_{\cc_2}(\mb y_N)\ra\sim \de^{(3)}(\mb y_N)$ shows that we should discard all opposite-sign combinations, since they do not produce any complex power dependences on the coordinate difference $\mb y_N$, and so they do not contribute to nonlocal signals. Similar analysis can also be done for $\dd_1=-\dd_2$. The net result is that we only need to keep the same-sign choice for both soft lines.}
\item
The second part is the integral of the first product $\prod_{i=1}^{N-1}D$ in the bulk correlator $\mathcal{M}^{\text{(Bulk)}}_{\aa_1\cdots \aa_V}$ in \eqref{eq_BulkCorrelator} with the third factor on the right hand side of \eqref{eq_FourierKernel}, over the spatial coordinates $\mb x_1\cdots\mb x_{N-1}$ and temporal coordinates $\tau_1\cdots\tau_N$, with SK indices $\aa_1,\cdots,\aa_N$ summed. This is nothing but the left tree graph  $\mathcal T_{\cc\dd}^{(\text{L})}$ \eqref{eq_Tleft}. We note that, here and also in the next entry, one should set $\mb P_N=\mb 0$ for the leading result in the soft limit.
\item
Similarly, the third part is the integral of the second product $\prod_{i={N+1}}^{V-1}D$ in the bulk correlator \eqref{eq_BulkCorrelator} with the fourth factor on the right hand side of \eqref{eq_FourierKernel}, over the spatial coordinates $\mb x_{N+1}\cdots\mb x_{V-1}$ and temporal coordinates $\tau_{N+1}\cdots\tau_V$, with SK indices $\aa_{N+1},\cdots,\aa_V$ summed over. This is nothing but the right tree graph  $\mathcal T_{\cc\dd}^{(\text{R})}$ \eqref{eq_Tright}.
\item
Finally, the last part is simply:
\bge
\int \di^3\mb x_V\, e^{-\ii\mb P_V\cdot \mb x_V} = (2\pi)^3\de(\mb P_V),
\ede
which is the extra momentum conservation constraint in \eqref{eq_Tdeltageneral}.
\end{enumerate}
Combining the four parts together, we recover the same result of our factorization theorem \eqref{eq_SoftThm}. Thus we have shown that the ``cut and pinch'' process of the factorization theorem does have a boundary interpretation in terms of OPE.

\section{Applications: Bubble, Triangle, and Box}
\label{sec_4pt}

\begin{figure}[t]
\centering 
\includegraphics[width=0.32\textwidth]{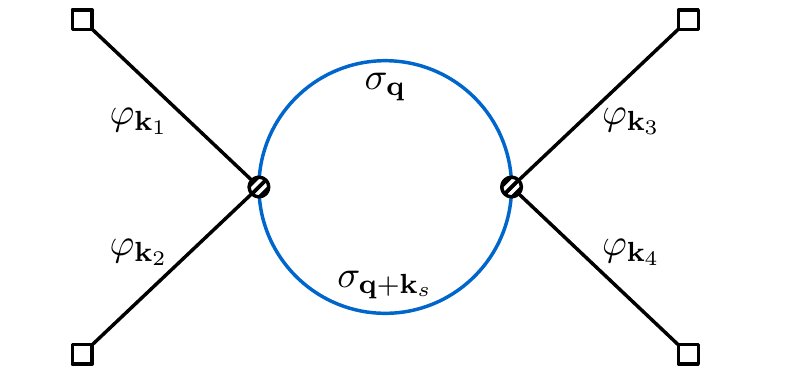}
\includegraphics[width=0.32\textwidth]{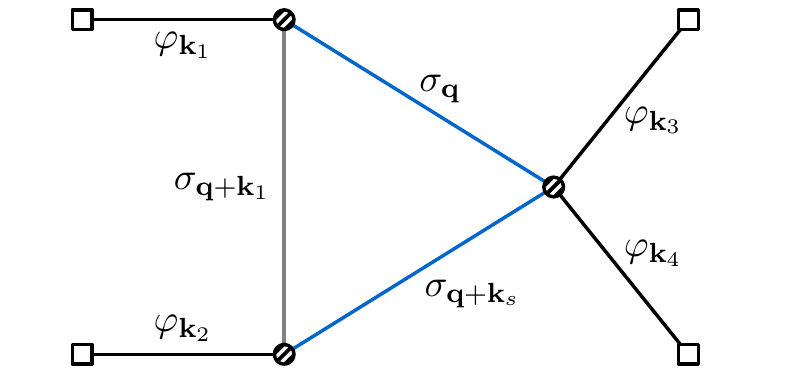}
\includegraphics[width=0.32\textwidth]{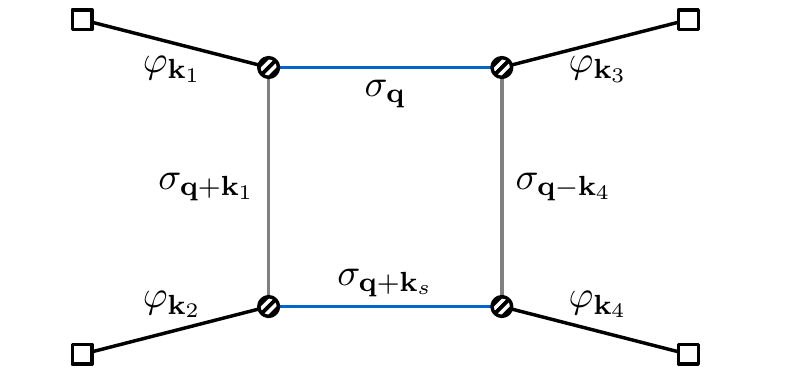} 
\caption{All possible 4-point inflation correlators with nontrivial 1PI 1-loop topologies. The external (black) lines denote bulk-to-boundary propagators of massless field $\varphi$, while the internal (gray and blue) lines denote bulk propagator of the massive scalar $\si$. The blue lines represent the soft lines that contribute to the nonlocal signal in the $s$-channel soft limit $k_s\to 0$.}
  \label{fig_4pt}
\end{figure}

In this section, we apply the factorization theorem \eqref{eq_SoftThm} in Sec.\ \ref{sec_cuttingrule} to the case of 4-point inflation correlators of all possible 1PI topologies, including the bubble, the triangle, and the box diagrams. See Fig.\ \ref{fig_4pt}. (The case of seagull diagram is also 1PI but is trivial.) These three 1-loop diagrams have been very well studied in flat spacetime, and it has long been known that they form a linearly independent basis for expressing more general 1-loop diagrams via  Passarino-Veltman reduction \cite{Passarino:1978jh,Elvang:2015rqa}. In dS, the boundary correlators corresponding to these three diagrams are much less understood. The complete analytical result for the 1-loop bubble was obtained only very recently \cite{Xianyu:2022jwk}, and the  result for the triangle and box diagrams are still unavailable at the moment. Our new analytical results in this section thus mark a step towards a full analytical understanding of these graphs in dS.

With the application to CC physics in mind, we consider the case where the four external lines are time-derivatively coupled massless scalars $\varphi$. The corresponding bulk-to-boundary propagator is:
\bge
G_\aa'(k;\tau) = \FR{1}{2k}\tau e^{\aa\ii k\tau}.
\ede
We shall consider the $s$-channel squeezed limit, i.e., we send $k_s\equiv |\mb k_1+\mb k_2|\to 0$ and keep all other independent magnitudes of momenta ($k_1,\cdots,k_4, k_t\equiv |\mb k_1+\mb k_4|$) finite. Also, for simplicity, we shall assume that all massive modes in the loop are identical, with a common mass parameter $\wt\nu$. This is the most commonly appeared case in CC models. The more general situation with different masses in the loop will be presented in App.\ \ref{app_diffmass}. 

Specialized to the 4-point 1-loop diagrams, the factorization theorem \eqref{eq_SoftThm} can be rewritten as:
\begin{align}
\label{eq_4ptFac}
    \lim_{k_s\to 0}\mathcal{T}\big(\{\mb k\}\big)=\sum_{\cc=\pm}\mathcal{T}_{\cc}^\text{(L)}\big(\mb k_1,\mb k_2\big)\mathcal{T}_{\cc}^\text{(R)}\big(\mb k_3,\mb k_4\big)\mathcal{B}_{\cc}(k_s)+\text{terms analytic in $k_s$},
\end{align}
where $\mathcal T_{\cc}^{(\text{L,R})}\equiv \mathcal T_{\cc\cc}^{(\text{L,R})}$.
The bubble signal \eqref{eq_BubbleSignal} is reduced to:
\begin{align}
\label{eq_bubbleNu}
    \mathcal{B}_{\cc}(k_s) 
    =&~\FR{k_s^3}{(4\pi)^{7/2}} \Gamma\bgb -2\cc\ii\wt\nu-\fr32 \\ 3+2\cc\ii\wt\nu\edb
    \Gamma^2\Big[\fr32+\cc\ii\wt\nu,-\cc\ii\wt\nu \Big]
    \Big(\FR{k_s}2\Big)^{4\cc\ii\wt\nu} .
\end{align}
The computation of the nonlocal signals in the 4-point functions are then reduced to a computation of the left and the right tree graphs, which we shall discuss for all three cases in turn.
  
\subsection{Bubble diagram}
\label{sec_bubble}

\begin{figure}
\centering
\includegraphics[width=0.95\textwidth]{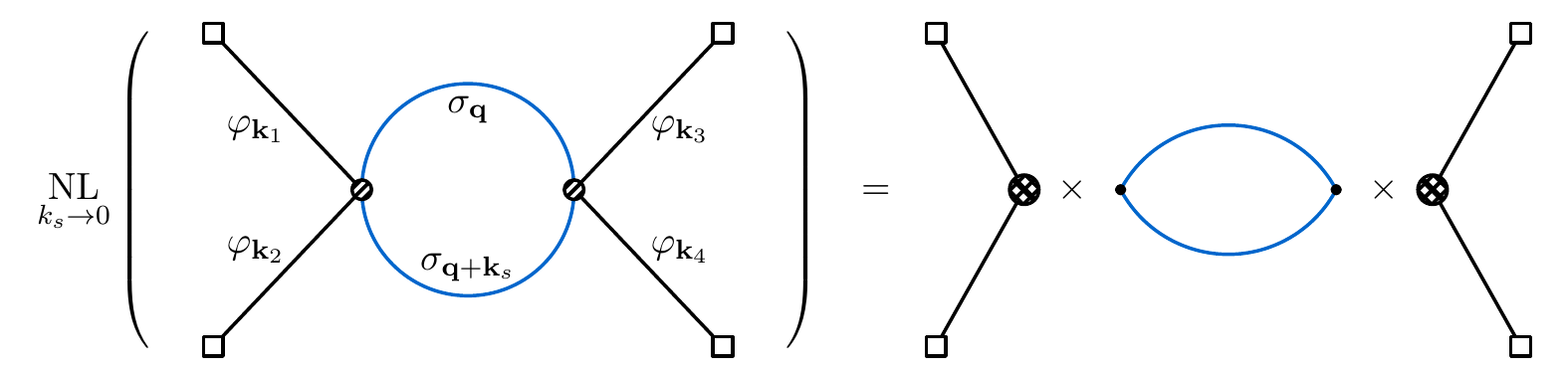} 
\caption{The nonlocal signal in the 1-loop bubble graph. On the right hand side, the two hatched larger blobs denote the insertions with complex powers of conformal time variables.} 
  \label{fig_bubble}
\end{figure}

We first consider the bubble diagram, for which we choose the following coupling between the external massless mode $\varphi$ and the internal massive mode $\si$:
\bge
\Delta \ld = \FR14 a^2  \varphi'^2 \si^2.
\ede
Again, the massive field $\si$ has the mass parameter $\wt\nu$.

In this simple case, the computation of the left and right tree graphs is pretty straightforward:
\begin{align}
\label{eq_TLbubble}
  \mathcal T_\cc^{(\text{L})} =&~ \FR{1}{4k_1k_2}\sum_{\aa_1=\pm}(\ii\aa_1)\int_{-\infty}^0 \di\tau_1\,e^{\aa_1\ii k_{12}\tau_1}
   (-\tau_1)^{3/2+\cc\ii \wt\nu}(-\tau_1)^{3/2+\cc\ii \wt\nu} 
  = \mathcal{P}^{(1)}_\cc(\wt\nu)\FR{k_{12}^{-4-2\cc\ii\wt\nu}}{4k_1k_2} ,\\
  \label{eq_TRbubble}
  \mathcal T_\cc^{(\text{R})} =&~ \FR{1}{4k_3k_4}\sum_{\aa_2=\pm}(\ii\aa_2)\int_{-\infty}^0 \di\tau_2\,e^{\aa_2\ii k_{34}\tau_2}
  (-\tau_2)^{3/2+\cc\ii \wt\nu}(-\tau_2)^{3/2+\cc\ii \wt\nu} 
  = \mathcal{P}^{(1)}_\cc(\wt\nu)\FR{k_{34}^{-4-2\cc\ii\wt\nu}}{4k_3k_4} ,
\end{align}
where we have introduced the \emph{pinch coefficient of first order} $\mathcal{P}_\cc^{(1)}(\wt\nu)$, which is defined by:
\bge
\label{eq_P1}
  \mathcal{P}^{(1)}_\cc(\wt\nu)\equiv 2\sin(\pi\ii \cc\wt\nu)\Gamma(4+2\cc\ii\wt\nu).
\ede
Now we can directly substitute the two amplitudes $\mathcal{T}^\text{(L)}_\cc$, $\mathcal{T}^\text{(R)}_\cc$ obtained here, and the bubble signal $\mathcal{B}_\cc$ in (\ref{eq_bubbleNu}), into the factorization equation (\ref{eq_4ptFac}), to obtain the leading nonlocal signal in the bubble diagram. In addition, the bubble diagram with identical particle has an extra symmetry factor $1/2$. Altogether, the final result is:
\begin{keyeqn}
\begin{align}
\label{eq_bubbleNL}
  \lim_{k_s\to 0} \Big[\mathcal T_\text{bubble}(\{\mb k\})\Big]_{\text{NL}} 
  = &- \FR{k_s^3}{8(4\pi)^{7/2}k_1k_2k_3k_4k_{12}^4k_{34}^4}\sinh^2(\pi\wt\nu) \Big(\FR{k_s^2}{4k_{12}k_{34}}\Big)^{2\ii\wt\nu}\n\\
  &\times (3+2\ii\wt\nu)\Gamma\Big[4+2\ii\wt\nu,-\FR32-2\ii\wt\nu\Big]\Gamma^2\Big[\FR32+\ii\wt\nu,-\ii\wt\nu\Big]+ \text{c.c.}.
\end{align}
\end{keyeqn}
We illustrate this result in Fig.\ \ref{fig_bubble}.

On the other hand, the complete analytical result for this diagram has been worked out in \cite{Xianyu:2022jwk}. In particular, there is a ``nonlocal signal'' piece in the full result. This nonlocal signal is the only piece nonanalytic in $k_s$ as $k_s\to 0$ and we quote it here:
\begin{align}
\label{eq_fullBubbleNL}
  &~\Big[\mathcal T_\text{bubble}(\{\mb k\})\Big]_{\text{NL}} 
  =\FR{(r_1r_2)^{3/2+2\ii\wt{\nu}}}{8k_1k_2k_3k_4(k_{12}k_{34})^{5/2}}\frac{}{}\sum_{n=0}^\infty\FR{(1+n)_{\frac{1}{2}} \big[(1+\ii\wt\nu+n)_{\frac{1}{2}}\big]^2(1+2\ii\wt\nu+n)_{\frac{1}{2}}}{\pi^{2}\cos(2\pi\ii\wt\nu)(1+2\ii\wt\nu+2n)_{2}}(\fr{3}{2}+2\ii\wt{\nu}+2n)\n\\
  &~\times{}_2\mathcal{F}_1\left[\bgm 2+\ii\wt\nu+n,\fr{5}{2}+\ii\wt\nu+n \\ \fr{5}{2}+2\ii\wt\nu+2n
  \edm\middle|r_1^2\right]
  {}_2\mathcal{F}_1\left[\bgm 2+\ii\wt\nu+n,\fr{5}{2}+\ii\wt\nu+n \\ \fr{5}{2}+2\ii\wt\nu+2n
  \edm\middle|r_2^2\right]
  (r_1r_2)^{2n}+\text{c.c.}.
\end{align}
Here $r_1=k_s/k_{12}$ and $r_2=k_s/k_{34}$.
Note that this result holds to all orders in the powers of $k_s$. To compare this result with what we obtained in (\ref{eq_bubbleNL}), we take the $k_s\to 0$ limit, which amounts to keep the $n=0$ term only. It is straightforward to check that the $n=0$ term of the above result agrees exactly with our result (\ref{eq_bubbleNL}). Note that the full result (\ref{eq_fullBubbleNL}) was obtained via the spectral decomposition method. It is thus good to see the agreement between the two results from very different methods.

\subsection{Triangle diagram}
\label{sec_triangle}
Next, we consider the triangle diagram. Contrary to the bubble diagram, there is no available analytical result for the triangle 1-loop diagram. Even worse, it was not clear how we can reliably compute the nonlocal signal for the triangle loop in the squeezed limit. For the CC application, it was usually assumed that one of the three loop lines can be pinched so that one can effectively compute the triangle nonlocal signal with a bubble graph. While such a pinch could be intuitive, the coupling strength of the resulting effective operator after the pinch was undetermined. In the literature, it was normally assumed that this effective coupling strength scales as $\wt\nu^{-2}$, where $\wt\nu$ is the mass parameter of the \emph{pinched} line. This behavior may be well justified by the EFT argument when $\wt\nu\gg 1$. When $\wt\nu\sim 1$ or $\wt\nu <1$, the situation is much less clear.  

\begin{figure}
\centering
\includegraphics[width=0.95\textwidth]{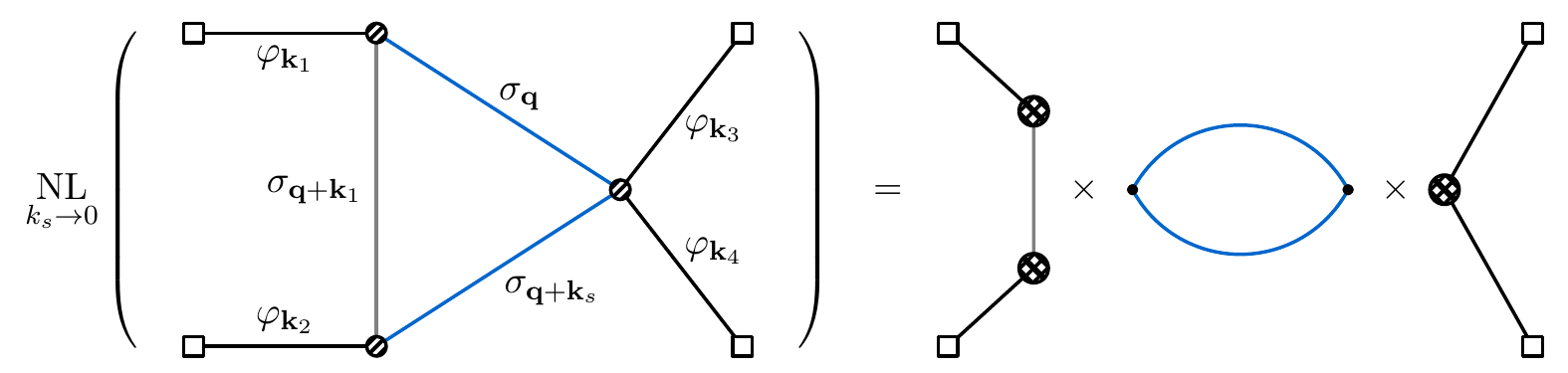} 
\caption{The nonlocal signal in the 1-loop triangle graph.} 
  \label{fig_triangle}
\end{figure}

 Below, we use the factorization theorem to solve this problem. Coupled with our previous results for the partial amplitude which corresponds to the left tree graph, we obtain the triangle nonlocal signal in the $s$-channel squeezed limit valid for all $\wt\nu>0$.

The triangle 1-loop for the 4-point function requires a four-point vertex and a three-point vertex, which we choose to be:
\bge
\label{eq_triangleL}
\Delta \ld  = \FR12 a^3\varphi' \si^2 + \FR14 a^2\varphi'^2 \si^2,
\ede
where the massive field $\si$ has mass parameter $\wt\nu>0$.
Our starting point is again (\ref{eq_4ptFac}), and we need to compute the left and right tree graphs for the triangle graph. See Fig.\ \ref{fig_triangle}. Clearly, the right tree graph is identical to the case of bubble graph computed in (\ref{eq_TRbubble}):
\bge
\label{eq_TRtri}
 \mathcal T_\cc^{(\text{R})} = \mathcal{P}^{(1)}_\cc(\wt\nu)\FR{k_{34}^{-4-2\cc\ii\wt\nu}}{4k_3k_4} .
\ede
The left tree graph $\mathcal{T}_\cc^{\text{(L)}}$ is less trivial, which, according to the general expression 
\begin{align}
\label{eq_TriangleTLeft}
\mathcal T_\cc^{(\text{L})}(k_1,k_2) =&~ \FR{1}{4k_1k_2}\sum_{\aa_1,\aa_2=\pm}(-\aa_1\aa_2) \int_{-\infty}^0 \FR{\di\tau_1}{\tau_1^2}\FR{\di\tau_2}{\tau_2^2} e^{\aa_1 \ii k_1 \tau_1+\aa_2\ii k_2 \tau_2}D^{(\wt\nu)}_{\aa_1\aa_2}(k_1;\tau_1,\tau_2)\n\\
& \times (-\tau_1)^{3/2+\cc\ii\wt\nu}(-\tau_2)^{3/2+\cc\ii\wt\nu}.
\end{align}
At the leading order of $k_s$ in the squeezed limit $k_s\to0$, we should set $k_1=k_2$. The resulting left tree graph $\mathcal T_\cc^{(\text{L})}(k_1,k_1)$ is nothing but a two-point correlator mediated by a single massive scalar $\si$ at the tree level, but its two couplings have unusual complex power dependences on the two time variables. This amplitude has been computed in \cite{Qin:2023ejc} in terms of a two-point seed integral $\wt{\mathcal{I}}_{\wt\nu|\aa_1\aa_2}^{p_1p_2}(1,1)$. In \cite{Qin:2023ejc}, the seed integral $\wt{\mathcal{I}}_{\nu|\aa_1\aa_2}^{p_1p_2}(u_1,u_2)$ is defined as:\footnote{See Eq.\ (10) of \cite{Qin:2023ejc}. For applications to cases involving multiple masses, we include an additional subscript $\wt\nu$ in $\wt{\mathcal{I}}_{\nu|\aa_1\aa_2}^{p_1p_2}(u_1,u_2)$ to highlight the mass parameter of the field mediating the two-point correlator.}
\begin{align} 
\label{eq_Iseed}
  \wt{\mathcal{I}}_{\wt\nu|\aa\bb}^{p_1p_2}(u_1,u_2)\equiv (-\aa\bb)k_s^{5+p_{12}}\int_{-\infty}^0\di\tau_1\di\tau_2(-\tau_1)^{p_1}(-\tau_2)^{p_2}e^{\ii\aa k_{12}\tau_1+\ii\bb k_{34}\tau_2}D_{\aa\bb}^{(\wt\nu)}(k_s;\tau_1,\tau_2).
\end{align}
where $u_{1}\equiv 2k_s/(k_{12}+k_s)$ and $u_2\equiv 2k_s/(k_{34}+k_s)$. Comparing  (\ref{eq_TriangleTLeft}) with (\ref{eq_Iseed}), we get:
\bge
\label{eq_TLeftTri}
 \mathcal T_\cc^{(\text{L})}(k_1,k_2) = \mathcal{P}^{(2)}_\cc (\wt\nu) \FR{k_{12}^{-4-2\cc\ii\wt\nu} }{4k_1k_2},
\ede
where we have introduced the \emph{pinch coefficient of second order} $\mathcal{P}^{(2)}_\cc (\wt\nu)$, which is defined by:
\bge
\label{eq_Pinch2}
\mathcal{P}^{(2)}_\cc (\wt\nu)\equiv 2^{4+2\cc\ii\wt\nu}\sum_{\aa_1,\aa_2=\pm} \wt{\mathcal I}_{\wt\nu|\aa_1\aa_2}^{-1/2+\cc\ii\wt\nu,-1/2+\cc\ii\wt\nu} (1,1).
\ede
The seed integral $\wt{\mathcal{I}}_{\wt\nu|\aa_1\aa_2}^{p_1p_2}(1,1)$ with both arguments being unity was computed in \cite{Qin:2023ejc} with the following results:
\begin{align}
\label{eq_H2ptResultPP}
  &\wt{\mathcal I}_{\wt\nu|\pm\pm}^{p_1p_2}(1,1)
  =\FR{\pm\ii e^{\mp\ii {p}_{12}\pi/2}e^{-\pi\wt\nu}}{2^{5+p_{12}}}\Gamma\left[\bgm\fr52+p_1-\ii\wt\nu,\fr52+p_1+\ii\wt\nu, \fr52+p_2-\ii\wt\nu,\fr52+p_2+\ii\wt\nu\\
3+p_1,3+p_2\edm\right] \n\\
   &-\FR{e^{\mp \ii p_{12}\pi/2}}{2^{5+p_{12}}}\Gamma\Big[5+p_{12},\fr52+p_1\pm\ii\wt\nu,\fr52+p_2\pm\ii\wt\nu\Big]  
  {}_3\wt{\mathrm{F}}_2 
\left[\bgm  5+p_{12},\fr12\pm\ii\wt\nu,1\\ \fr72+p_1\pm\ii\wt\nu,\fr72+p_2\pm\ii\wt\nu\edm\middle|1\right],\\
\label{eq_H2ptResultPM}
 & \wt{\mathcal I}_{\wt\nu|\pm\mp}^{p_1p_2}(1,1)
  =\FR{e^{\mp\ii\bar{p}_{12}\pi/2}}{2^{5+p_{12}} }\Gamma\left[\bgm \fr52+p_1-\ii\wt\nu,\fr52+p_1+\ii\wt\nu,\fr52+p_2-\ii\wt\nu,\fr52+p_2+\ii\wt\nu\\
3+p_1,3+p_2\edm\right].
\end{align}
With these explicit expressions, the summation in (\ref{eq_Pinch2}) can be directly done, and the result is:
\bge
\label{eq_P2}
  \mathcal{P}^{(2)}_\cc (\wt\nu)=-\FR{8(2+\ii\cc\wt\nu)\Gamma(2+2\ii\cc\wt\nu)\sin(\pi\ii\cc \wt\nu)}{3+2\ii\cc\wt\nu}.
\ede
Now we put together the left graph $\mathcal{T}_\cc^{(L)}$ in (\ref{eq_TLeftTri}), the right graph $\mathcal{T}_\cc^{(R)}$ in (\ref{eq_TRtri}), and the bubble signal $\mathcal{B}_\cc$ in (\ref{eq_bubbleNu}), as illustrated in Fig.\ \ref{fig_triangle}, and get the expression for the nonlocal CC signal in the triangle graph in the squeezed limit $k_s\to 0$:
\begin{keyeqn}
\begin{align}
\label{eq_triangleNL}
  \lim_{k_s\to 0} \Big[\mathcal T_\text{triangle}(\{\mb k\})\Big]_{\text{NL}} 
  =&~ \FR{k_s^3}{(4\pi)^{7/2}k_1k_2k_3k_4k_{12}^4k_{34}^4}\Big(\FR{k_s^2}{4k_{12}k_{34}}\Big)^{2\ii\wt\nu} (2+\ii\wt\nu)\sinh^2(\pi\wt\nu) \n\\
  &\times \Gamma\Big[2+2\ii\wt\nu,-\FR32-2\ii\wt\nu\Big]\Gamma^2\Big[\FR32+\ii\wt\nu,-\ii\wt\nu\Big] + \text{c.c.}.
\end{align}
\end{keyeqn}
We note that this result holds for all $\wt\nu>0$. 

\subsection{Box diagram}

\begin{figure}
\centering
\includegraphics[width=0.95\textwidth]{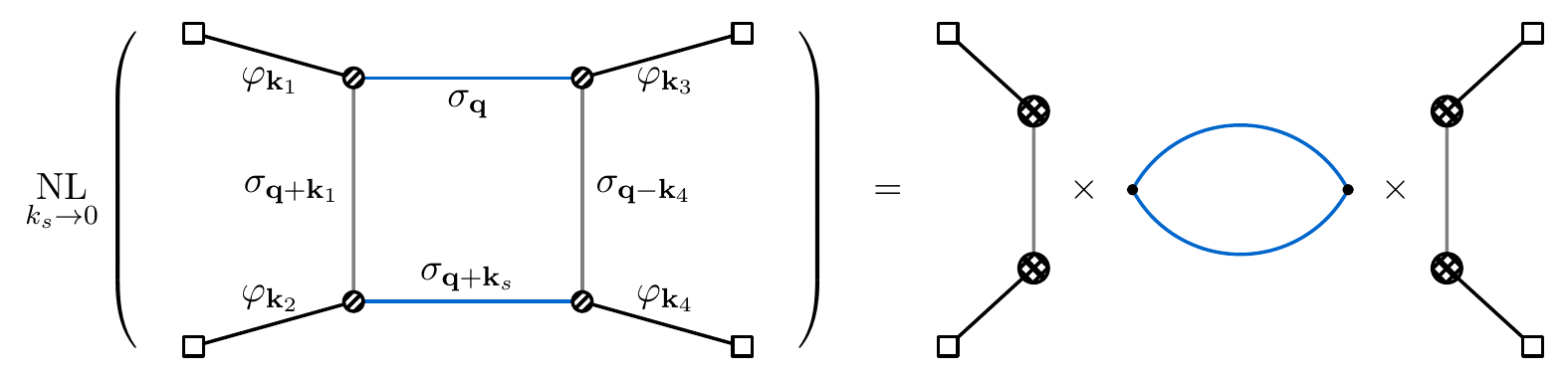} 
\caption{The nonlocal signal in the 1-loop box graph.} 
  \label{fig_box}
\end{figure}

Finally, we consider the box diagram with the following interaction, as shown in Fig.\ \ref{fig_box}:
\bge
\Delta \ld = \FR12 a^3 \varphi' \si^2.
\ede
This time, both the left and the right tree graphs have exactly the same structure with the left tree graph of the triangle graph computed in the last subsection. Therefore, we can directly quote the previous result and write:
 \begin{align}
 &\mathcal T_\cc^{(\text{L})}(k_1,k_2) = \mathcal{P}^{(2)}_\cc (\wt\nu) \FR{k_{12}^{-4-2\cc\ii\wt\nu} }{4k_1k_2},
 &&\mathcal T_\cc^{(\text{R})}(k_3,k_4) = \mathcal{P}^{(2)}_\cc (\wt\nu) \FR{k_{34}^{-4-2\cc\ii\wt\nu} }{4k_3k_4},
\end{align}
where $\mathcal{P}_\cc^{(2)}(\wt\nu)$ is given in (\ref{eq_P2}). Then, we can directly substitute this result into the general expression (\ref{eq_4ptFac}). Together with the bubble signal $\mathcal{B}_\cc$ in (\ref{eq_bubbleNu}), we get the nonlocal signal of the box diagram in the limit $k_s\to 0$:
\begin{keyeqn}
\begin{align}
\label{eq_boxNL}
  \lim_{k_s\to 0} \Big[\mathcal T_\text{box}(\{\mb k\})\Big]_{\text{NL}} = &- \FR{k_s^3}{(4\pi)^{7/2}k_1k_2k_3k_4k_{12}^4k_{34}^4} \Big(\FR{k_s^2}{4k_{12}k_{34}}\Big)^{2\ii\wt\nu}\FR{(2+\ii\wt\nu)^4}{(3+2\ii\wt\nu)^2}\sinh^2(\pi\wt\nu)\n\\
  &\times \Gamma\Big[3+2\ii\wt\nu,-\FR32-2\ii\wt\nu\Big]\Gamma^2\Big[\FR32+\ii\wt\nu,-2-\ii\wt\nu\Big] + \text{c.c.}.
\end{align}
\end{keyeqn}

\subsection{Summary: pinched operator and phases} 

In this section, we have computed the nonlocal signals from all 1PI 1-loop 4-point graphs, including the bubble, the triangle, and the box diagrams. By applying the 1-loop factorization theorem, we have obtained the exact and analytical expressions for the nonlocal CC signals in the $s$-channel squeezed limit $k_s\to 0$ for all these graphs. 

\begin{figure}
\[
\parbox{0.45\textwidth}{\includegraphics[width=0.45\textwidth]{fd_triangle}}
\To
\parbox{0.45\textwidth}{\includegraphics[width=0.45\textwidth]{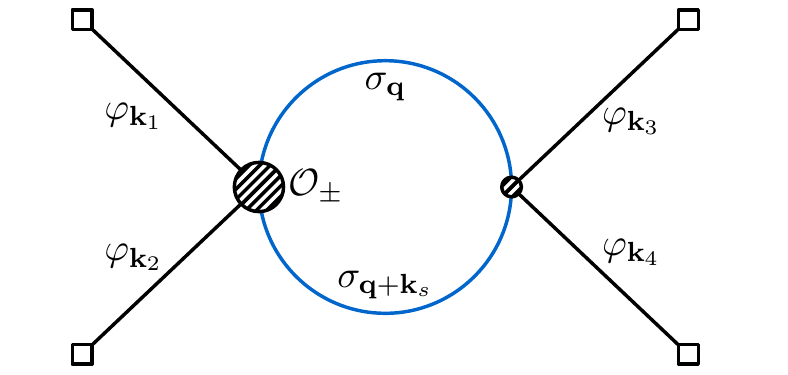}}
\]
\caption{The triangle diagram and its pinched version.} 
  \label{fig_pinched_triangle}
\end{figure}
\paragraph{Pinched operator.}
Our result shows that, so far as the nonlocal CC signal is concerned, it is legitimate to pinch the hard lines in the loop graph, so that the oscillatory signals are essentially generated by an effective bubble diagram. We illustrate this procedure for the triangle diagram in Fig.\ \ref{fig_pinched_triangle}. We emphasize that this pinch procedure does not require the pinched mode to be heavy, although we do expect that our result recovers the EFT intuition when $\wt\nu\gg 1$. That is, the coupling constant of the pinched operator $\mathcal O_\pm$ should scale as $1/\wt\nu^2$. To explain this point more explicitly, we introduce the following effective \emph{pinched operators}:
\bge
\label{eq_O0}
\mathcal O_\pm = \mathcal O_\pm^{(0)} + \mathcal O_\pm^{(1)} + \cdots.
\ede
This pinched operator $\mathcal{O}_\pm$ is defined such that the bubble graph with one insertion of $\mathcal{O}_\pm$, as shown in the right graph of Fig.\ \ref{fig_pinched_triangle}, gives rise to a  nonlocal signal in the $s$-channel soft limit $k_s\to 0$ identical to the signal from the triangle graph, namely the left graph of Fig.\ \ref{fig_pinched_triangle}.  For later convenience, we have rewritten the pinched operator $\mathcal{O}_\pm$ as a summation of $n$-th order pinched operator $\mathcal{O}_\pm^{(n)}$ with $n=0,1,\cdots$. An $n$'th order pinched operator generates a nonlocal signal starting from order $k_s^{3+2n}$. We will have more to say about the higher order couplings in Sec.\ \ref{sec_trianglefull}. For now, let us focus on the leading operator $\mathcal{O}_\pm^{(0)}$. By comparing the two pinch coefficients defined in (\ref{eq_P1}) and (\ref{eq_P2}), we can identify the leading pinched operator $\mathcal{O}_\pm^{(0)}$ to be:
\bge
\mathcal O_\pm^{(0)} = \FR14 \mathcal C^{(0)}_\pm(\wt\nu) a^2\varphi'^2\si^2,
\ede
where the pinched coupling constant $\mathcal C_\pm^{(0)}$ is given by:
\begin{align}
\label{eq_C2}
  \mathcal{C}_\pm^{(0)}(\wt\nu)\equiv  \FR{2\mathcal{P}^{(2)}_\pm (\wt\nu)}{\mathcal{P}^{(1)}_\pm (\wt\nu)}= -\FR{\pm\ii\wt\nu+2}{(\pm\ii\wt\nu+3/2)^2(\pm\ii\wt\nu+1)}.
\end{align}
Here we include a factor of $2$, since the symmetry factors for the triangle and the bubble diagrams are $1$ and $2$, respectively.
As we can see, the pinched operator provides a convenient way to generate nonlocal signals in diagrams with more complicated topologies. For instance, the nonlocal signal in the box graph can be obtained by computing a bubble graph with both of its vertices replaced by the pinched operator.\footnote{For the nonlocal signal of the box diagram in Fig.\ \ref{fig_box}, one can start from the nonlocal signal of the bubble diagram \eqref{eq_bubbleNL}, replace the two $\mathcal P_\pm^{(1)}(\wt\nu)$ by $\mathcal P_\pm^{(2)}(\wt\nu)$, and add a factor $2$ since the symmetry factor for box diagram is $1$, and the result will be \eqref{eq_boxNL}. Furthermore, there is another box graph with ($\mb k_3\leftrightarrow\mb k_4$) giving identical nonlocal signal in the $s$-channel squeezed limit. As a result, the total $s$-channel nonlocal signal is given by \eqref{eq_boxNL} multiplied by 2, and it can be equivalently obtained by computing a bubble graph with two pinched operators \eqref{eq_C2}.} In this section, we are only working at the leading order in the $k_s$ expansion, and we only need a single pinched operator $\mathcal{O}_\pm^{(0)}$. At higher orders in $k_s$, the situation can be more complicated and the choice of pinched operators is not unique. We can simply make convenience choices at each order, as we shall show in the next section.

\begin{figure}
\centering
\includegraphics[width=0.5\textwidth]{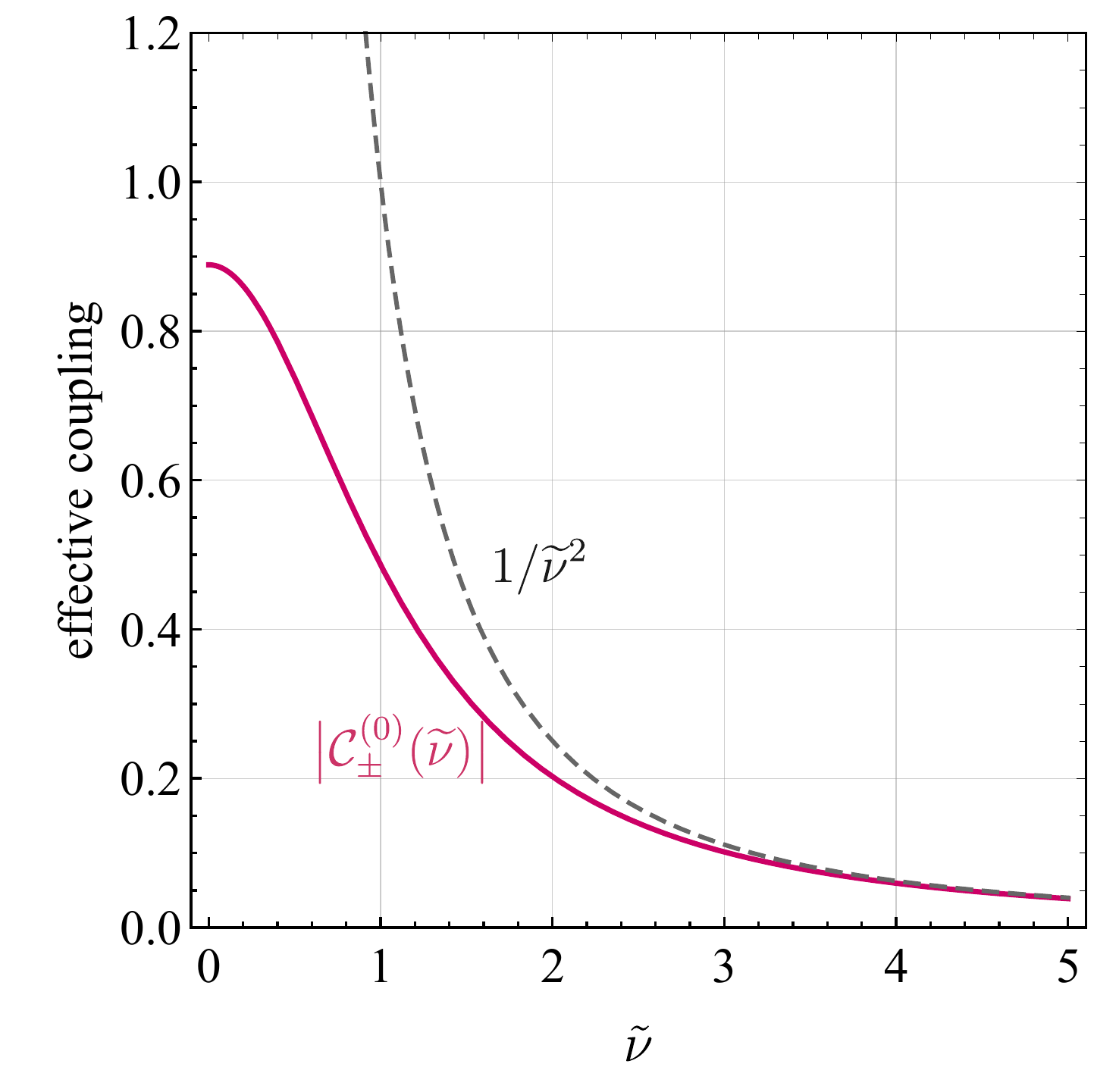}
\caption{The size of the effective coupling from pinching the hard line as a function of the mass parameter $\wt\nu$ of the loop mode. The magenta curve shows the exact result in (\ref{eq_C2}), and the black dashed curve shows the EFT approximation.} 
  \label{fig_eff_coup}
\end{figure}

\paragraph{Signal phase.} A peculiar feature of the effective pinched coupling $\mathcal{C}_\pm^{(0)}$ in (\ref{eq_C2}) is that it is complex, and is dependent on the choices of the two scaling dimensions $2\Delta_\pm=3\pm2\ii\wt\nu$ of the soft loop. This is quite natural from the boundary viewpoint, since we have two sets of boundary operators $\si_\pm$ with scaling dimensions $\Delta_\pm =3/2\pm\ii\wt\nu$, and each set independently generates its own nonlocal signal. Phenomenologically, the complex coupling implies that there will be nontrivial phase shift in the nonlocal CC signals in the triangle or box graph relative to the bubble graph. This phase shift is physical and measurable, as has been emphasized in \cite{Qin:2022lva}. Therefore, the phase shift gives us a chance to tell apart the bubble, triangle, and box diagrams even in the deep squeezed limit. 

\begin{figure}
\centering
\includegraphics[width=0.5\textwidth]{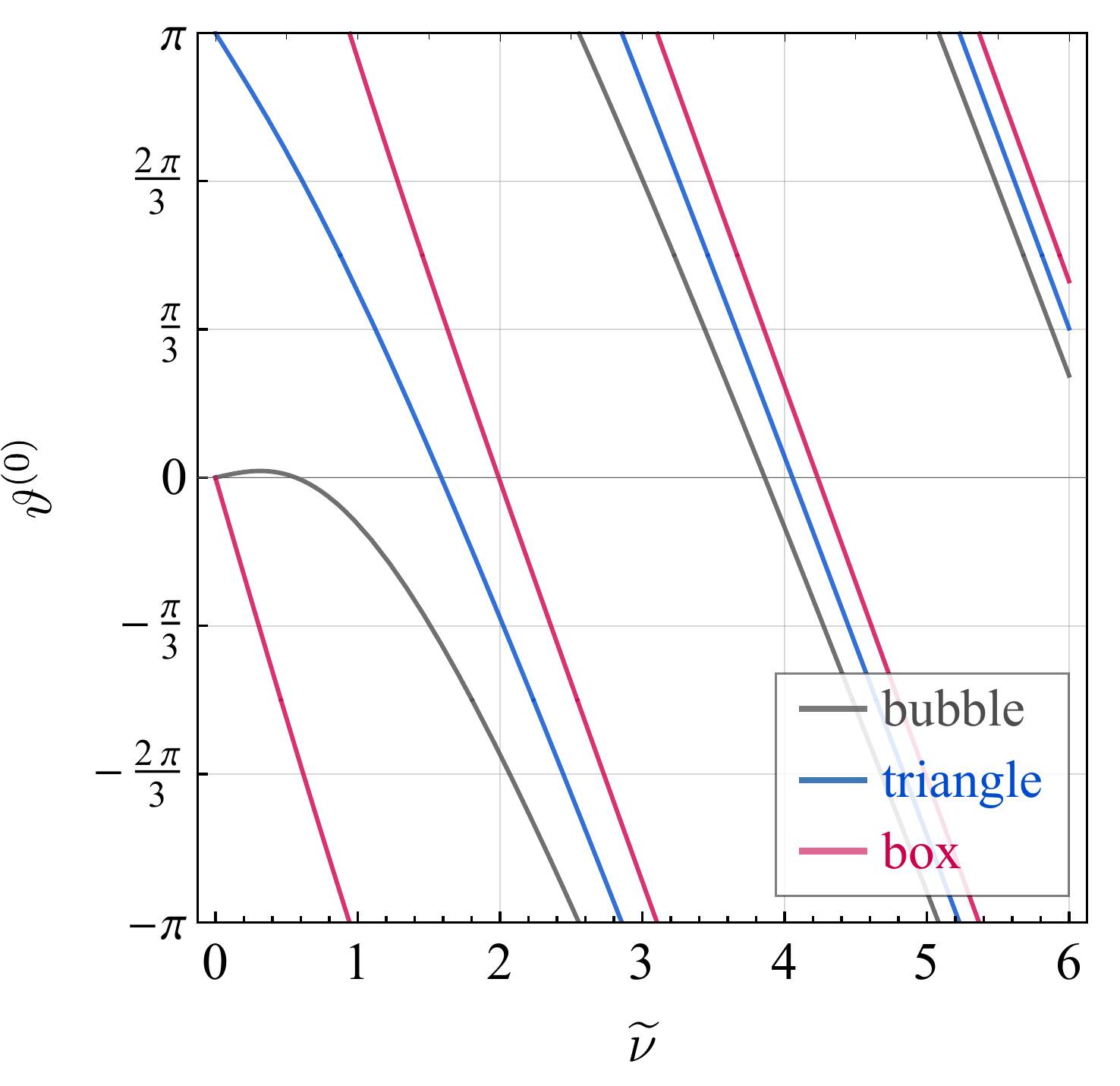}
\caption{The phases of nonlocal CC signals in the bubble, triangle, and box diagrams as functions of the mass parameter $\wt\nu$.} 
  \label{fig_phase}
\end{figure}

Specifically, we can rewrite the nonlocal CC signals in all three diagrams as:
\begin{align}
\label{eq_SqueezedNL}
  \lim_{k_s\to 0}\Big[\mathcal{T}_{I}(\{\mb k\})\Big]_\text{NL}=&~\mathcal{A}_I^{(0)}\FR{k_s^3}{k_1k_2k_3k_4k_{12}^4k_{34}^4}\Big(\FR{k_s^2}{k_{12}k_{34}}\Big)^{2\ii\wt\nu} + \text{c.c.}\n\\
  =&~| \mathcal{A}^{(0)}_I |\FR{k_s^3}{k_1k_2k_3k_4k_{12}^4k_{34}^4} \cos\bigg[2\wt\nu\log\Big(\FR{k_s^2}{k_{12}k_{34}}\Big)+\vartheta_I^{(0)}\bigg]
\end{align}
where $I\in\{\text{bubble},\text{triangle},\text{box}\}$, $\mathcal{A}_I^{(0)}$ denotes the (complex) amplitude of the signal, and $\vartheta^{(0)}_I \equiv \text{Arg} \mathcal A_I^{(0)}$ denotes the phase shift of the signal relative to the folded point $k_s=k_{12}=k_{34}$. Here we use a superscript $(0)$ to highlight that this is the leading result in the squeezed limit $k_s\to0$. Corrections of higher orders in $k_s$ will be considered in the next section. Then, from the three expressions for the nonlocal signals, (\ref{eq_bubbleNL}), (\ref{eq_triangleNL}), and (\ref{eq_boxNL}), we get the phases for the nonlocal signals from these three graphs:
\begin{align}
\vartheta^{(0)}_\text{bubble}=&~\text{Arg}\bigg\{-4^{-2\ii\wt\nu} (3+2\ii\wt\nu)\Gamma\Big[4+2\ii\wt\nu,-\FR32-2\ii\wt\nu\Big]\Gamma^2\Big[\FR32+\ii\wt\nu,-\ii\wt\nu\Big]\bigg\},\\
\vartheta^{(0)}_\text{triangle}=&~\text{Arg}\bigg\{4^{-2\ii\wt\nu}(2+\ii\wt\nu) \Gamma\Big[2+2\ii\wt\nu,-\FR32-2\ii\wt\nu\Big]\Gamma^2\Big[\FR32+\ii\wt\nu,-\ii\wt\nu\Big]\bigg\},\\
\vartheta^{(0)}_\text{box}=&~\text{Arg}\bigg\{-4^{-2\ii\wt\nu}\FR{(2+\ii\wt\nu)^4}{(3+2\ii\wt\nu)^2} \Gamma\Big[3+2\ii\wt\nu,-\FR32-2\ii\wt\nu\Big]\Gamma^2\Big[\FR32+\ii\wt\nu,-2-\ii\wt\nu\Big]\bigg\}.
\end{align}
We plot these phases as functions of the mass parameter $\wt\nu$ in Fig.\ \ref{fig_phase}. The differences among these phases are evident from the plot, and thus one can in principle use the phase information to distinguish the signals from all different 1-loop diagrams. We should note that the phase also depends on particle species and interaction types in a complicated way. Here we are only considering a particular type of coupling with a particular choice of loop particles, but our method here directly applies to more general cases. Further discussions about the use of signal phases can be found in \cite{Qin:2022lva}.

\section{Nonlocal Signal for Finite Momentum Ratios}
\label{sec_full4pt}

Up to this point in this work, we have only considered the explicit expressions for the 1-loop nonlocal signals in a given soft limit, such as in the factorization theorem stated in (\ref{eq_SoftThm}). For practical applications of CC physics, we have to measure the CC signals for a range of configurations where $P_N$ is nonzero but is parametrically smaller than all other independent external momenta. For such configurations, the result of (\ref{eq_SoftThm}) will be corrected by a factor of $[1+\order{P_N^2}]$. Among other things, such corrections introduce a momentum-ratio dependence in the signal phase $\vartheta^{(0)}$ defined in (\ref{eq_SqueezedNL}). As emphasized in \cite{Qin:2022lva}, such running of signal phase can provide important physical information about the particles and interactions in a loop process. Therefore, it is phenomenologically useful to push our factorization results to higher orders in the soft momentum.

In this section, we demonstrate how to go beyond the leading order results of nonlocal CC signals by reconsidering all 1PI diagrams studied in Sec.\ \ref{sec_4pt}. We will calculate explicitly the $\order{k_s^2}$ corrections to the nonlocal CC signals of 1-loop bubble, triangle, and box diagrams with four external points. To achieve this goal, we can no longer quote the factorization theorem \eqref{eq_SoftThm} directly, since it is a statement in the leading order in the squeezed limit. However, the proof in Sec.\ \ref{sec_proof} can be easily generalized to higher orders. So, before working on specific examples, we describe how to obtain the higher order corrections in the general settings of Sec.\ \ref{sec_cuttingrule}.

Recall that the nonlocal signals in the $P_N\to 0$ limit are contributed by the signal poles \eqref{eq_NonlocalPoles}. The choice $n_V=\bar n_1=n_N=\bar n_{N+1}=0$ gives the leading result, while other choices give corrections of higher orders in $P_N$. As emphasized in Sec.\ \ref{sec_cuttingrule}, the cutting rule presented in (\ref{eq_Cut}) holds to all orders in the expansion of $P_N$. Therefore, we can still separate the 1-loop nonlocal signal into three parts at higher orders of $P_N$. With these points in mind, let us summarize the sources of corrections to the leading-order result \eqref{eq_SoftThm}:
\begin{enumerate}
\item The contributions of higher order positive/negative poles in \eqref{eq_NonlocalPoles}. From the boundary OPE point of view, these contributions correspond to the higher order terms in the late-time expansion \eqref{eq_latetime1}-\eqref{eq_latetime4} of the four mode functions in the two soft lines.

\item The higher-order terms in the loop momentum integral expansion, namely the $\order{P_N}$ part in the first line of \eqref{eq_LoopMomIntResult}. From the boundary viewpoint, they correspond to the higher terms of the OPEs \eqref{eq_OPE1} and \eqref{eq_OPE2}. 

\item The last one is a little subtle. In the left tree graph (\ref{eq_Tleft}) and the right tree graph (\ref{eq_Tright}), the sum of all external momentum should be $\sum\{\mb k^{(\text{L})}\}=-\sum\{\mb k^{(\text{R})}\}=\mb P_N$. At the leading order, we can set $\mb P_N = \mb 0$ and thus turn off the momentum flow $\mb P_N$ carried by the bubble signal. At higher orders, the soft yet nonzero momentum $\mb P_N$ could introduce more complicated kinematic dependence in the nonlocal signal. Take the 4-point triangle diagram in Fig.\ \ref{fig_triangle} as an example. In this case, we have $N=2$ and $\mb k_1+\mb k_2\equiv \mb k_s\to \mb 0$. At the leading order, the left tree graph depends on only one scalar variable $k_{12}$ since the constraint $\mb k_1+\mb k_2=\mb 0$ implies $k_1=k_2=k_{12}/2$. However, for nonzero $\mb k_s$, the left tree graph can depend on three independent variable $k_{12}$, $\bar k_{12}\equiv k_1-k_2$, and $k_s$. Now, the triangle inequality requires $|\bar k_{12}|\leq k_s$. Therefore, the new dependences on $k_s$ and $\bar k_{12}$ must be at least of order $\order{k_s}$.
\end{enumerate}
Taking all above contributions into account, we can write down an expression which generalizes the factorization theorem (\ref{eq_SoftThm}) to all orders in $P_N$:
\begin{align}
\label{eq_SoftThm2}
\mathcal{T}\big(\{\mb k\}\big)=&\sum_m \int_{-\ii\infty}^{\ii\infty} \FR{\di s_V}{2\pi\ii}\FR{\di \bar s_1}{2\pi\ii}\FR{\di s_N}{2\pi\ii}\FR{\di \bar s_{N+1}}{2\pi\ii}\mathcal{T}_m^\text{(L)}\Big(\{\mb k^\text{(L)}\},\{s\}\Big)\mathcal{T}_m^\text{(R)}\Big(\{\mb k^\text{(R)}\},\{s\}\Big) \mathcal{B}_m\Big(P_N,\{s\}\Big)\n\\
& + \text{terms analytic in $P_N$ as $P_N\to0$},
\end{align}
where $\{s\}\equiv \{s_V,\bar s_1,s_N,\bar s_{N+1}\}$.
Here the subscript $m$ denotes the each term in the Taylor expansion of the loop integral, and the integral over Mellin variables should be finished by summing residues at all nonlocal poles \eqref{eq_NonlocalPoles}, which are poles of the factor \eqref{eq_NonlocalGammas} contained in the $m$-th order ``bubble signal" $\mathcal B_m$.

Specialized to the 1PI 1-loop 4-point functions studied in Sec.\ \ref{sec_4pt}, we again consider the $s$-channel soft limit $k_s\to 0$ and parameterize the full nonlocal signal in the following form:
\bge
\Big[\mathcal{T}_{I}(\{\mb k\})\Big]_\text{NL}=\FR{k_s^3}{k_1k_2k_3k_4k_{12}^4k_{34}^4}\bigg[ \mathcal{A}_I(\{\mb k\}) \Big(\FR{k_s^2}{k_{12}k_{34}}\Big)^{2\ii\wt\nu}+\text{c.c.}\bigg],
\ede
where again $I\in\{\text{bubble},\text{triangle},\text{box}\}$. $\mathcal A_I$ is a dimensionless function of external momenta and can be expanded around the squeezed limit $k_s\to 0$, whose leading term reproduces the expression \eqref{eq_SqueezedNL}:
\bge
\lim_{k_s\to 0} \mathcal A_I(\{\mb k\}) = \mathcal A_I^{(0)}.
\ede
We find it is convenient to choose the following set of independent momentum magnitudes to express $\mathcal A_I$:
\bge
\mathcal A_I(\{\mb k\}) = \mathcal A_I(k_{12},k_{34},\bar k_{12},\bar k_{34},k_s,k_t),
\ede
where $\bar k_{12} \equiv k_1 - k_2$ and $\bar k_{34} \equiv k_3-k_4$. For the particles and couplings we are choosing in this work, there is no $k_t$ dependence in the $s$-channel exchange diagrams. Also, the triangle inequalities from the momentum conservations require that $|\bar k_{12}|,|\bar k_{34}|\leq k_s$. Thus, when we treat $k_s$ as a small expansion parameter, $\bar k_{12}$ and $\bar k_{34}$ are automatically small as well. Furthermore, because of the exchange symmetries $k_1\leftrightarrow k_2$ and $k_3\leftrightarrow k_4$, the complex amplitude $\mathcal A_I$ depends on $\bar k_{12}$ and $\bar k_{34}$ only through even powers such as $\bar k_{12}^2$ and $\bar k_{34}^2$. With all these remarks made, in the following, we shall set $k_1=k_2$ and $k_3=k_4$ to remove the dependences on $\bar k_{12}$ and $\bar k_{34}$, just to keep the presentation simple.

\subsection{Bubble diagram}
Higher order corrections to the bubble diagram in Fig.\ \ref{fig_bubble} are simple:
We did not assume $\mb k_s=\mb 0$ when calculating left and right tree graphs in \eqref{eq_TLbubble} and \eqref{eq_TRbubble}. Also, there is no $\mathcal O(k_s)$ terms in the loop integral \eqref{eq_LoopMomIntResult}, so the higher order corrections come only from the higher order signal poles in (\ref{eq_NonlocalPoles}).
Therefore, we only need to sum over residues at all the positive and negative poles in \eqref{eq_NonlocalPoles}. The final result is a series of momentum ratios $k_s/k_{12}$ and $k_s/k_{34}$, and has been worked out in our previous work \cite{Qin:2022lva}, which is equivalent to \eqref{eq_fullBubbleNL}. Here we simply quote the first correction, which can be obtained from \eqref{eq_fullBubbleNL} by keeping $n=0$ term in the summation and expanding the two $_{2}\text{F}_1$ functions to linear order in $r_1^2$ and $r_2^2$:
\bge
\label{eq_BubbleFull}
\mathcal A_{\text{bubble}} = \mathcal A_{\text{bubble}}^{(0)} \bigg[1+\al(\wt\nu) \FR{k_s^2}{k^2_{12}}+\al(\wt\nu) \FR{k_s^2}{k_{34}^2} + \mathcal O(k_s^4)\bigg],
\ede
where $\mathcal{A}_{\text{bubble}}^{(0)}$ is the complex amplitude of the leading-order nonlocal signal of the bubble diagram, which is given in (\ref{eq_bubbleNL}), and the coefficient $\al(\wt\nu)$ is defined as:
\bge
\label{eq_alpha}
\al(\wt\nu) \equiv \FR{(2+\ii\wt\nu)(5+2\ii\wt\nu)}{5+4\ii\wt\nu}.
\ede

\subsection{Triangle diagram}
\label{sec_trianglefull}
Unlike the bubble diagram, all three types of corrections contribute to the higher order terms of triangle diagram. In this subsection, we will study the higher order terms of the triangle diagram in details, and calculate the leading corrections to the squeezed limit result. In principle, we can treat the three types of corrections separately. However, we will take another approach, since the loop integral is simple enough (though more complicated than the bubble diagram) such that we can derive the full nonlocal signal associated with $k_s$. 

\paragraph{Loop integral}
Let us take a closer look at the loop momentum integral (\ref{eq_LoopMomInt}) for the triangle diagram:
\bge
\label{eq_triangleloopInt}
\mathcal L = \int \FR{\di^3\mb q}{(2\pi)^3} |\mb q|^{-2s_{3\bar 1}}|\mb q+\mb k_1|^{-2s_{1\bar 2}}|\mb q+\mb k_s|^{-2s_{2\bar 3}}.
\ede
This integral is calculated in App.\ \ref{app_triangle}, and the nonanalytic part in $k_s$ can be expanded as a double series in $k_s/k_1$ and $1-k_2/k_1$ in the squeezed limit $k_s\to 0$:
\bge
\label{eq_TriangleLoopInt}
\mathcal L = \FR{1}{(4\pi)^{3/2}} k_s^{3-2s_{3\bar12\bar3}}k_2^{-2s_{1\bar2}} \sum_{m,\ell=0}^\infty \mb P_m^\ell(\{s_i\})\Big(\FR{k_s}{k_1}\Big)^{2m}\Big(1-\FR{k_2}{k_1}\Big)^\ell + \text{terms analytic in $k_s$}.
\ede
We can easily rewrite the above expression in terms of $k_{12}$, $\bar k_{12}$ and $k_s$.
However, if we focus on the symmetric configuration $k_1=k_2$, only terms with $\ell=0$ survive and the loop integral is simplified to:
\bge
\mathcal L|_{k_1=k_2} = \FR{1}{(4\pi)^{3/2}} k_s^{3-2s_{3\bar12\bar3}}k_1^{-2s_{1\bar2}} \sum_{m=0}^\infty \mb P_m(\{s_i\})\Big(\FR{k_s}{k_1}\Big)^{2m} + \text{terms analytic in $k_s$},
\ede
where
\bge
\label{eq_Pm}
\mb P_m(\{s_i\}) = f_m(s_{1\bar2},s_{3\bar12\bar3}) \times \FR{(-1)^m}{m!}\Gamma\left[\bgm -m-\fr32+s_{3\bar12\bar3},m+\fr32-s_{3\bar1},m+\fr32-s_{2\bar3}\\2m+3-s_{3\bar12\bar3},s_{3\bar1},s_{2\bar3}\edm\right],
\ede
and $f_m(s_{1\bar2},s_{3\bar12\bar3})$ is a polynomial:
\bge
\label{eq_fm}
f_m(s_{1\bar2},s_{3\bar12\bar3}) = (s_{1\bar2})_m (3-s_{3\bar12\bar3}-s_{1\bar2})_m.
\ede
Similar to the case of bubble loop integral (\ref{eq_BubbleMomInt}), the only left poles in $\mathbf{P}_m(\{s_i\})$ integral come from the factor $\Gamma(-m-3/2-s_{3\bar12\bar3})$. In addition, the factors $1/\Gamma(s_{3\bar1})$ and $1/\Gamma(s_{2\bar3})$ in \eqref{eq_Pm} imply that the cutting rule \eqref{eq_Cut} holds for the two soft propagators at each order, therefore we can simply sum over residues at nonlocal poles \eqref{eq_NonlocalPoles} to complete the integral over $s_{3,\bar1,2,\bar3}$. At the leading order in $k_s$, this is exactly our factorization theorem. We note that, although $f_m(s_{1\bar 2},s_{3\bar12\bar3})$ depends on these 4 Mellin variables, this polynomial does not affect the pole structure, so the nonlocal poles in this case are still given by \eqref{eq_NonlocalPoles}. In fact, given that $f_m(s_{1\bar 2},s_{3\bar12\bar3})$ is dependent on $s_{1\bar2}$, we find it more proper to absorb the factor $f_m(s_{1\bar 2},s_{3\bar12\bar3})$ into the definition of the left tree graph. We will come back to this point below. With this $f_m(s_{1\bar 2},s_{3\bar12\bar3})$ factor removed, we define the $m$-th order bubble signal as:
\begin{align}
\label{eq_Bm}
  \mathcal B_m |_{k_1=k_2} =&~ \FR{(-1)^m k_s^3}{m!(4\pi)^{7/2}}\Big(\FR{k_s}{k_1}\Big)^{2m}
  \int_{-\ii\infty}^{\ii\infty} \FR{\di s_3}{2\pi\ii}\FR{\di s_{\bar1}}{2\pi\ii}\FR{\di s_2}{2\pi\ii}\FR{\di s_{\bar3}}{2\pi\ii}\,
 e^{\ii\pi (s_{\bar12} -s_{\bar 33} )} \Big(\FR{k_s}{2}\Big)^{-2s_{3\bar12\bar3}}\n\\
  &\times \Gamma\left[\bgm -m-\fr32+s_{3\bar12\bar3},m+\fr32-s_{3\bar1},m+\fr32-s_{2\bar3}\\2m+3-s_{3\bar12\bar3},s_{3\bar1},s_{2\bar3}\edm\right] \prod_{i=3,\bar1,2,\bar3} \Gamma\Big[s_i -\FR{\ii\wt\nu}2,s_i +\FR{\ii\wt\nu}2\Big].
\end{align}
We also note that the leading term with $m=0$ recovers the expression \eqref{eq_bubbleNu}.
 
\paragraph{Left and right tree graphs.} Next let us look at the two $m$'th order tree graphs, $\mathcal{T}_{m}^{\text{(L)}}$ and $\mathcal{T}_{m}^{\text{(R)}}$, given in (\ref{eq_SoftThm2}). For the triangle loop, the corresponding two tree graphs are shown on the right hand side of Fig.\ \ref{fig_triangle}. 

Similar to the leading result in Sec.\ \ref{sec_triangle}, the right tree graph only receives a polynomial of conformal time $(-\tau_3)^{-2s_{\bar33}}$:
\begin{align}
\label{eq_TRm}
  \mathcal T^{(\text{R})} =&~ \FR{1}{4k_3k_4}\sum_{\aa_3=\pm}(\ii\aa_3)\int_{-\infty}^0 \di\tau_3\,e^{\aa_3\ii k_{34}\tau_3}
  \times (-\tau_3)^{3/2-2s_{\bar3}}(-\tau_3)^{3/2-2s_3}\n\\
  =&-\FR{\sin(\pi s_{\bar33})}{2k_3k_4} k_{34}^{-4+2s_{\bar33}}\Gamma(4-2s_{\bar33}),
\end{align}
which is in fact independent of $m$, so we omit the subscript. 

On the other hand, the left tree graph $\mathcal{T}_{m}^{\text{(L)}}$, is much less trivial. As we mentioned above, here we choose to include the polynomial $f_m(s_{1\bar 2},s_{3\bar12\bar3})$ in the definition of the left tree graph. We remind the reader that this polynomial comes from the loop momentum integral (\ref{eq_Pm}). Due to this extra polynomial factor, in general, the integral over $s_1$ and $\bar s_2$ will not directly recover the propagator $D^{(\wt\nu)}_{\aa_1\aa_2}(k;\tau_1,\tau_2)$. Explicitly, the expression for the $m$'th order left tree graph now reads:
\begin{align}
\label{eq_TLm}
\mathcal T^{(\text{L})}_m |_{k_1=k_2} =&~ \FR{1}{4k_1^2}\sum_{\aa_1,\aa_2=\pm}(-\aa_1\aa_2)\int_{-\infty}^0 \di\tau_1\di\tau_2\,(-\tau_1)^{-1/2-2s_{\bar1}}(-\tau_2)^{-1/2-2s_2}
e^{\aa_1\ii k_1\tau_1+\aa_2\ii k_1\tau_2}\n\\
&\times \int_{-\ii\infty}^{\ii\infty} \FR{\di s_1}{2\pi\ii}\FR{\di s_{\bar2}}{2\pi\ii}\, f_m(s_{1\bar 2},s_{3\bar12\bar3}) \mathfrak{D}^{(\wt\nu)}_{\aa_1\aa_2}(k_1;\tau_1,\tau_2;s_1,s_{\bar 2}).
\end{align}
where $\mathfrak{D}^{(\wt\nu)}_{\aa_1\aa_2}$ denotes the Mellin integrand of the PMB representation of the propagator $D_{\aa_1\aa_2}^{(\wt\nu)}$. The explicit expressions for $\mathfrak{D}^{(\wt\nu)}_{\aa_1\aa_2}$ can be easily extracted from the PMB representation (\ref{eq_DScalarMB1}):
\begin{align}
\label{eq_frakDpm}
    \mathfrak{D}^{(\wt\nu)}_{\pm\mp}(k;\tau_1,\tau_2;s_1,s_2) =&~ \FR{1}{4\pi}
    e^{\mp\ii\pi(s_1-s_2)}\Big(\FR{k}2\Big)^{-2s_{12}}
    (-\tau_1)^{-2s_1+3/2}(-\tau_2)^{-2s_2+3/2}\n\\
    &\times \Gamma\Big[s_1-\FR{\ii\wt\nu}2,s_1+\FR{\ii\wt\nu}2,s_2-\FR{\ii\wt\nu}2,s_2+\FR{\ii\wt\nu}2\Big],\\
\label{eq_frakDpp}
    \mathfrak{D}^{(\wt\nu)}_{\pm\pm}(k;\tau_1,\tau_2;s_1,s_2) =&~ \mathfrak{D}^{(\wt\nu)}_{\mp\pm}(k;\tau_1,\tau_2;s_1,s_2)\theta(\tau_1-\tau_2) + \mathfrak{D}^{(\wt\nu)}_{\pm\mp}(k;\tau_1,\tau_2;s_1,s_2)\theta(\tau_2-\tau_1).
\end{align}
Thus, in the absence of the polynomial $f_m(s_{1\bar 2},s_{3\bar12\bar3})$, the Mellin integral in the second line of (\ref{eq_TLm}) can be directly done and gives nothing but the massive propagator $D_{\aa_1\aa_2}^{(\wt\nu)}(k_1;\tau_1,\tau_2)$. Now, with the presence of $f_m(s_{1\bar 2},s_{3\bar12\bar3})$, we make use of the following relation, which can be easily verified with the expressions in (\ref{eq_frakDpm}) and (\ref{eq_frakDpp}):
\bge
s_{1\bar 2} \mathfrak{D}^{(\wt\nu)}_{\aa_1\aa_2}(k_1;\tau_1,\tau_2;s_1,s_{\bar2}) = \mathcal D_{\tau_1,\tau_2} \mathfrak{D}^{(\wt\nu)}_{\aa_1\aa_2}(k_1;\tau_1,\tau_2;s_1,s_{\bar2}), \quad \mathcal D_{\tau_1,\tau_2} \equiv \FR12(3- \tau_1\pd_{\tau_1} - \tau_2\pd_{\tau_2}).
\ede
Consequently, the polynomial $f_m(s_{1\bar 2},s_{3\bar12\bar3})$ in the integral (\ref{eq_TLm}) can be replaced by a differential operator $f_m(\mathcal D_{\tau_1,\tau_2},s_{3\bar12\bar3})$. Then, we can commute this operator with the integral over $s_1,s_{\bar2}$. The integral can then be directly done, yielding the original propagator $D_{\aa_1\aa_2}^{(\wt\nu)}$:
\begin{align}
\label{eq_TLmFinal}
\mathcal T^{(\text{L})}_m |_{k_1=k_2} =&~ \FR{1}{4k_1^2}\sum_{\aa_1,\aa_2=\pm}(-\aa_1\aa_2)\int_{-\infty}^0 \di\tau_1\di\tau_2\,(-\tau_1)^{-1/2-2s_{\bar1}}(-\tau_2)^{-1/2-2s_2}
e^{\ii \aa_1k_1\tau_1+\ii \aa_2 k_1\tau_2}\n\\
&\times  f_m(\mathcal D_{\tau_1,\tau_2},s_{3\bar12\bar3}) D^{(\wt\nu)}_{\aa_1\aa_2}(k_1;\tau_1,\tau_2).
\end{align}
Having finished the Mellin integral, the last step is to perform an integration by parts and move the $\tau_{1,2}$-derivatives in $\mathcal{D}_{\tau_1,\tau_2}$ to the front of exponential factor $e^{\ii \aa_1k_1\tau_1+\ii \aa_2 k_1\tau_2}$. The resulting expression can then be expressed as a combination of seed integrals $\wt{\mathcal{I}}_{\wt\nu|\aa\bb}^{p_1p_2}$ defined in (\ref{eq_Iseed}).

Now, with the expressions for the $m$'th order bubble signal $\mathcal B_m$ in \eqref{eq_Bm}, the right tree graph $\mathcal{T}^{\text{(R)}}$ in \eqref{eq_TRm}, and the left tree graph $\mathcal T^{(\text{L})}_m$ in \eqref{eq_TLmFinal}, we can calculate the nonlocal signals in \eqref{eq_SoftThm2} to all orders in the $s$-channel momentum $k_s$. Here we are  only considering the symmetric configuration $k_1=k_2$. For the asymmetric case, we can start from \eqref{eq_TriangleLoopInt} and follow the same procedure. 

Below, we  do this calculation explicitly to find the first nontrivial correction to the leading order nonlocal signal in the triangle diagram in Fig.\ \ref{fig_triangle}. As was made clear in the discussion above \eqref{eq_SoftThm2}, these corrections come from three different sources. First, we have the contributions from the higher order nonlocal poles beyond $n_3=\bar n_1=n_2=\bar n_3=0$. Second, we have the contributions of higher powers of $k_s$ from the loop momentum integral, which correspond to $m=1$ terms in (\ref{eq_SoftThm2}). Third, there are contributions from asymmetric configuration which are of order $\bar k_{12}^2$. For simplicity, we shall take the freedom of choosing external momenta and set $k_1=k_2$, so that the third contributions of $\order{\bar k_{12}^2}$ do not appear. 

\paragraph{Corrections from higher order nonlocal poles.} To compute the leading corrections from the higher order nonlocal poles, we can set $m=0$ in (\ref{eq_SoftThm2}). Then, from (\ref{eq_fm}), we get $f_m(s_{1\bar2},s_{3\bar12\bar3})=1$. In this case, the left tree graph is simplified to:
\begin{align}
\mathcal T^{(\text{L})}_0 |_{k_1=k_2} =&~ \FR{1}{4k_1^2}\sum_{\aa_1,\aa_2=\pm}(-\aa_1\aa_2)\int_{-\infty}^0 \di\tau_1\di\tau_2\,(-\tau_1)^{-1/2-2s_{\bar1}}(-\tau_2)^{-1/2-2s_2}
e^{\ii\aa_1 k_1\tau_1+\ii\aa_2 k_1\tau_2}\n\\
&\times D^{(\wt\nu)}_{\aa_1\aa_2}(k_1;\tau_1,\tau_2)\n\\
=&~\FR{1}{4k_1k_2}\Big(\FR{k_{12}}2\Big)^{-4+2s_{\bar12}} \sum_{\aa_1,\aa_2=\pm} \mathcal I_{\aa_1\aa_2}^{(\wt\nu)-1/2-2s_{\bar1},-1/2-2s_2}(1,1),
\end{align}
where the seed integral $\wt{\mathcal{I}}_{\wt\nu|\aa\bb}^{p_1p_2}$ is defined in (\ref{eq_Iseed}), and we have symmetrized expression with respect to $k_1\leftrightarrow k_2$. On the other hand, the bubble signal \eqref{eq_Bm} with $m=0$ is:
\begin{align}
  \mathcal B_0 |_{k_1=k_2} =&~ \FR{k_s^3}{(4\pi)^{7/2}}
  \int_{-\ii\infty}^{\ii\infty} \FR{\di s_3}{2\pi\ii}\FR{\di s_{\bar1}}{2\pi\ii}\FR{\di s_2}{2\pi\ii}\FR{\di s_{\bar3}}{2\pi\ii}\,
   e^{\ii\pi (s_{\bar12} -s_{\bar 33} )}\Big(\FR{k_s}{2}\Big)^{-2s_{3\bar12\bar3}}\n\\
  &\times \Gamma\left[\bgm -\fr32+s_{3\bar12\bar3},\fr32-s_{3\bar1},\fr32-s_{2\bar3}\\3-s_{3\bar12\bar3},s_{3\bar1},s_{2\bar3}\edm\right]\times \prod_{i=3,\bar1,2,\bar3} \Gamma\Big[s_i -\FR{\ii\wt\nu}2,s_i +\FR{\ii\wt\nu}2\Big].
\end{align}
Recall that the nonlocal poles for the triangle diagram are:
\bge
s_3=-n_3-\cc\FR{\ii\wt\nu}{2},\quad
   \bar s_1=-\bar n_1-\cc\FR{\ii\wt\nu}{2}, \quad
   s_2=-n_2-\cc\FR{\ii\wt\nu}{2},\quad
   \bar s_3=-\bar n_3-\cc\FR{\ii\wt\nu}{2}.
\ede
For the leading correction we can set one of the four integers $n_3$, $\bar n_1$, $n_2$, $\bar n_3$ to be $1$, and the other three to be $0$. Summing up the four choices, the result is: 
\bge
\label{eq_TriangleCorrection1}
\mathcal A_{\text{triangle}}^{(0)} \bigg[1+\be(\wt\nu) \FR{k_s^2}{k^2_{12}}+\al(\wt\nu) \FR{k_s^2}{k_{34}^2} \bigg] + \text{c.c.},
\ede
where $\mathcal A_{\text{triangle}}^{(0)}$ is the complex amplitude of the nonlocal signal in the triangle 1-loop graph, which we calculated in Sec.\ \ref{sec_triangle} with the result given in (\ref{eq_triangleNL}). The coefficient $\al(\wt\nu)$ was defined in \eqref{eq_alpha}. In the current example, the piece proportional to $\al(\wt\nu)$ is contributed by the sum of the residues of two choices of signal poles: either $n_3=1$ with all other $n$'s being 0, or $\bar n_3=1$ with all other $n$'s being 0. Likewise, the coefficient $\be(\wt\nu)$ is defined as:
\begin{align}
\label{eq_beta}
\be(\wt\nu) =& -\FR{\sum_{\aa_1,\aa_2=\pm} [ \mathcal I_{\aa_1\aa_2}^{(\wt\nu)3/2+\ii\wt\nu,-1/2+\ii\wt\nu}(1,1)+ \mathcal I_{\aa_1\aa_2}^{(\wt\nu)-1/2+\ii\wt\nu,3/2+\ii\wt\nu}(1,1)]}{(5+4\ii\wt\nu)\sum_{\aa_1,\aa_2=\pm}  \mathcal I_{\aa_1\aa_2}^{(\wt\nu)-1/2+\ii\wt\nu,-1/2+\ii\wt\nu}(1,1)}\n\\
=&~ \FR{(1+\ii\wt\nu)(3+\ii\wt\nu)(3+2\ii\wt\nu)}{(2+\ii\wt\nu)(5+4\ii\wt\nu)},
\end{align}
and it is contributed by the signal poles with either $\bar n_1=1$ with all other $n$'s being zero, or $n_2=1$ with all other $n$'s being zero.

The result in (\ref{eq_TriangleCorrection1}) summarizes the contributions from the higher order nonlocal poles of $m=0$ term in (\ref{eq_SoftThm2}). As explained, this part should be combined with the contributions from the leading poles (with all $n$'s being 0) in the $m=1$ term of (\ref{eq_SoftThm2}), which we shall compute next.

\paragraph{Correction from the higher order terms in the loop momentum integral.}
Now, let us consider the contributions from $m=1$ term in (\ref{eq_SoftThm2}) specialized to the triangle diagram. In this case, the bubble signal \eqref{eq_Bm} becomes:
\begin{align}
  \mathcal B_1 |_{k_1=k_2} =&~ \FR{(-1) k_s^3}{(4\pi)^{7/2}}\Big(\FR{2k_s}{k_{12}}\Big)^{2}
  \int_{-\ii\infty}^{\ii\infty} \FR{\di s_3}{2\pi\ii}\FR{\di s_{\bar1}}{2\pi\ii}\FR{\di s_2}{2\pi\ii}\FR{\di s_{\bar3}}{2\pi\ii}\,
 e^{\ii\pi (s_{\bar12} -s_{\bar 33} )} \Big(\FR{k_s}{2}\Big)^{-2s_{3\bar12\bar3}}\n\\
  &\times \Gamma\left[\bgm -\fr52+s_{3\bar12\bar3},\fr52-s_{3\bar1},\fr52-s_{2\bar3}\\5-s_{3\bar12\bar3},s_{3\bar1},s_{2\bar3}\edm\right]\times \prod_{i=3,\bar1,2,\bar3} \Gamma\Big[s_i -\FR{\ii\wt\nu}2,s_i +\FR{\ii\wt\nu}2\Big],
\end{align}
where we have substituted $k_s/k_1$ by $2k_s/k_{12}$.

On the other hand, taking $m=1$ in (\ref{eq_fm}), we get $f_1(s_{1\bar2},s_{3\bar12\bar3}) = (s_{1\bar2})(3-s_{3\bar12\bar3}-s_{1\bar2})$. Therefore, the left tree graph $\mathcal T^{(\text{L})}_1$ with symmetric configuration $k_1=k_2$ becomes:
\begin{align}
\mathcal T^{(\text{L})}_1 |_{k_1=k_2} =&~ \FR{1}{4k_1^2}\sum_{\aa_1,\aa_2=\pm}(-\aa_1\aa_2)\int_{-\infty}^0 \di\tau_1\di\tau_2\,(-\tau_1)^{-1/2-2s_{\bar1}}(-\tau_2)^{-1/2-2s_2}
e^{\ii\aa_1 k_1\tau_1+\ii\aa_2 k_1\tau_2}\n\\
&\times  f_1(\mathcal D_{\tau_1,\tau_2},s_{3\bar12\bar3}) D^{(\wt\nu)}_{\aa_1\aa_2}(k_1;\tau_1,\tau_2).
\end{align}
After integrating by parts, we obtain:
\begin{align}
\mathcal T_1^{(\text{L})} |_{k_1=k_2} =&~ \FR{1}{4k_1k_2} \Big(\FR{k_{12}}2\Big)^{-4+2s_{\bar12}}\sum_{\aa_1,\aa_2=\pm} \bigg\{
(2-s_{\bar12})(1-s_{\bar33}) \mathcal I^{(\wt\nu)-1/2-2s_{\bar1},-1/2-2s_2}_{\aa_1\aa_2}(1,1)\n\\
&  +\FR\ii4(3-2s_{\bar12}+2s_{\bar33}) \Big[\aa_1 \mathcal I^{(\wt\nu)1/2-2s_{\bar1},-1/2-2s_2}_{\aa_1\aa_2}(1,1) +\aa_2\mathcal I^{(\wt\nu)-1/2-2s_{\bar1},1/2-2s_2}_{\aa_1\aa_2}(1,1)\Big]\n\\
&+\FR14 \Big[\mathcal I^{(\wt\nu)3/2-2s_{\bar1},-1/2-2s_2}_{\aa_1\aa_2}(1,1)+\mathcal I^{(\wt\nu)-1/2-2s_{\bar1},3/2-2s_2}_{\aa_1\aa_2}(1,1)\Big]\n\\
&+\FR12\aa_1\aa_2 \mathcal I^{(\wt\nu)1/2-2s_{\bar1},1/2-2s_2}_{\aa_1\aa_2}(1,1)
\bigg\}.
\end{align}
Now we can finish the four Mellin integrals over $(s_3,\bar s_1,s_2,\bar s_3)$ by taking residues at the leading poles, namely
\bge
s_3=-\cc\FR{\ii\wt\nu}{2},\quad
   \bar s_1=-\cc\FR{\ii\wt\nu}{2}, \quad
   s_2=-\cc\FR{\ii\wt\nu}{2},\quad
   \bar s_3=-\cc\FR{\ii\wt\nu}{2},
\ede
and the resulting correction is:
\bge
\mathcal A_{\text{triangle}}^{(0)} \bigg[1+\ga(\wt\nu) \FR{k_s^2}{k^2_{12}}\bigg] + \text{c.c.},
\ede
where the coefficient $\ga(\wt\nu)$ is defined as:
\begin{align}
\label{eq_TriangleCorrection2}
\ga(\wt\nu) =&~ \FR{3+2\ii\wt\nu}{(2+\ii\wt\nu)(5+4\ii\wt\nu)}
\times \FR{\mathcal T_1^{(\text{L})} |_{k_1=k_2}}{\mathcal T_0^{(\text{L})} |_{k_1=k_2}}\Big|_{s_3=\bar s_1= s_2=\bar s_3=-\ii\wt\nu/2}\n\\
=&~ \FR{2(1+\ii\wt\nu)(3+\ii\wt\nu)^2(3+2\ii\wt\nu)}{(2+\ii\wt\nu)^2(5+2\ii\wt\nu)(5+4\ii\wt\nu)}.
\end{align}
On the right hand side of the first equality, the first factor comes from the change of the bubble signal $\mathcal B_1/\mathcal B_0$ at leading positive poles.
Combining the corrections \eqref{eq_TriangleCorrection1} and \eqref{eq_TriangleCorrection2}, we obtain the leading correction:
\bge
\label{eq_TriangleNLO}
\mathcal A_{\text{triangle}}|_{k_1=k_2} = \mathcal A_{\text{triangle}}^{(0)} \bigg[1+\Big(\be(\wt\nu)+\ga(\wt\nu)\Big)\FR{k_s^2}{k^2_{12}} + \al(\wt\nu) \FR{k_s^2}{k^2_{34}} + \mathcal O(k_s^4)\bigg].
\ede
As explained before, when $k_1\neq k_2$, there will be another correction of order $\bar k_{12}^2$. Therefore, we can write down the correction for general configuration as:
\bge
\label{eq_TriangleFull}
\mathcal A_{\text{triangle}} = \mathcal A_{\text{triangle}}^{(0)} \bigg[1+\Big(\be(\wt\nu)+\ga(\wt\nu)\Big)\FR{k_s^2}{k^2_{12}} + \al(\wt\nu) \FR{k_s^2}{k^2_{34}} + \mathcal O(\bar k_{12}^2,k_s^4)\bigg].
\ede
\paragraph{Pinched operator to the next order.} Now, we use the $\order{k_s^2}$ correction of the nonlocal signal (\ref{eq_TriangleNLO}) from the triangle diagram to find the first order pinched operator $\mathcal{O}_\pm^{(1)}$ defined in (\ref{eq_O0}). For this purpose, let us consider the pinched diagram on the right hand side of Fig.\ \ref{fig_pinched_triangle}. If we use the pinched operator at the leading order $\mathcal{O}_\pm^{(0)}$ for the left vertex,  use the original four-point vertex, namely the second term in (\ref{eq_triangleL}), for the right vertex, and compute the nonlocal signal to the second order in $k_s$, we will get the following result for the complex amplitude:
\bge
\label{eq_triLO}
\mathcal A_{\text{triangle}}^{(0)} \bigg[1+\al(\wt\nu)\FR{k_s^2}{k^2_{12}} + \al(\wt\nu) \FR{k_s^2}{k^2_{34}} + \mathcal O(k_s^4)\bigg],
\ede
This differs from the triangle nonlocal signal at $\order{k_s^2}$, showing that we should include a first order pinched operator $\mathcal O_\pm^{(1)}$ which takes account of the difference between (\ref{eq_triLO}) and (\ref{eq_TriangleNLO}):
\bge
\label{eq_ExtraNL}
\mathcal A_{\text{triangle}}^{(0)}\Big[\be(\wt\nu)+\ga(\wt\nu)-\al(\wt\nu)\Big]\FR{k_s^2}{k^2_{12}}.
\ede
Since the only requirement is that $\mathcal O_\pm^{(1)}$ generates a nonlocal signal as above, the form of $\mathcal O_\pm^{(1)}$ is not uniquely fixed. For example, we can make the following choice, which turns out to be convenient:
\bge
\label{eq_O1}
\mathcal O_\pm^{(1)} = \FR14 \mathcal C_\pm^{(1)}(\wt\nu) \varphi'^2 \pd_i^2 \si^2,
\ede
where the two additional derivatives on $\si$ just produces the right scaling of the signal. Now, to find the coupling constant $\mathcal C^{(1)}_\pm$, we should calculate the bubble diagram generated by the operator \eqref{eq_O1} at the left vertex (together with the original operator $a^2\varphi'^2\si^2/4$ at the right vertex), to the leading order. The calculation is very similar to the bubble case in Sec.\ \ref{sec_bubble}: The $\pd_i^2$ gives $-k_s^2$, and there is an extra $(-\tau_1)^2$ in the left tree graph integral \eqref{eq_TLbubble}, which is fixed by the dilatation symmetry of the problem. As a result, the factor $\Gamma(4+2\cc\ii\wt\nu)$ in (\ref{eq_P1}) should be replaced by $-\Gamma(6+2\cc\ii\wt\nu) k_{12}^{-2}$. 
Therefore, the operator $\mathcal O_\pm^{(1)}$ generates the following nonlocal signal:
\bge
\label{eq_DerivativeBubble}
\mathcal C_+^{(1)}\mathcal A^{(0)}_{\text{bubble}}  \FR{\Gamma(6+2\ii\wt\nu)}{\Gamma(4+2\ii\wt\nu)} \FR{k_s^2}{k_{12}^2} + \mathcal O(k_s^4) = \FR{\mathcal C_+^{(1)}}{\mathcal C_+^{(0)}}\mathcal A^{(0)}_{\text{triangle}}\times (4+2\ii\wt\nu)(5+2\ii\wt\nu) \FR{k_s^2}{k_{12}^2} + \mathcal O(k_s^4).
\ede
Now, by comparing \eqref{eq_DerivativeBubble} and \eqref{eq_ExtraNL}, we can determine the coupling $\mathcal C^{(1)}_\pm(\wt\nu)$:
\bge
\mathcal C^{(1)}_+(\wt\nu) = \FR{\be(\wt\nu)+\ga(\wt\nu)-\al(\wt\nu)}{(4+2\ii\wt\nu)(5+2\ii\wt\nu)}\mathcal C_+^{(0)}(\wt\nu), \qquad  \mathcal C^{(1)}_-(\wt\nu) = \mathcal C^{(1)}_+(-\wt\nu),
\ede
and the effective pinched operator to the next order:
\bge
\mathcal O_\pm = \FR14 \mathcal C_\pm^{(0)}(\wt\nu) a^2\varphi'^2\si^2 + \FR14 \mathcal C_\pm^{(1)}(\wt\nu)\varphi'^2\pd_i^2\si^2 + \cdots.
\ede

\subsection{Box diagram}

Given all the results for the triangle diagram, we do not have to compute the box signal from scratch. Thanks to the pinch procedure discussed previously, we can generate the nonlocal signal of a box diagram by the nonlocal signal of a bubble diagram, with two pinched operators $\mathcal O_\pm \sim \mathcal O_\pm^{(0)}+\mathcal O_\pm^{(1)}$ inserted, and the correction to the box diagram can be simply read out from \eqref{eq_TriangleFull} as the following,
\bge
\mathcal A_{\text{box}} = \mathcal A_{\text{box}}^{(0)} \bigg[1+\Big(\be(\wt\nu)+\ga(\wt\nu)\Big)\FR{k_s^2}{k^2_{12}} + \Big(\be(\wt\nu)+\ga(\wt\nu)\Big) \FR{k_s^2}{k^2_{34}} + \mathcal O(\bar k_{12}^2,\bar k_{34}^2,k_s^4)\bigg].
\ede
where the coefficients $\be(\wt\nu)$ and $\ga(\wt\nu)$ are given in (\ref{eq_beta}) and (\ref{eq_TriangleCorrection2}), respectively.

\section{Conclusions and Outlook}
\label{sec_conclusion}

Inflation correlators are on the one hand observables in Cosmological Collider physics, and on the other hand dS counterparts of scattering amplitudes in Minkowski spacetime. A better understanding of these objects is thus crucial for both CC phenomenology and theoretical study of dS QFTs. A special example in this regard is the nonlocal signal in massive inflation correlators. They generate characteristic oscillatory signals in CC observables, and also produce branch cut on the complex plane of external momenta. For this reason, the theoretical question of the analytic structure of massive inflation correlators is closely related to the phenomenological question of identifying all nonlocal signals in a given process. 

In this work, we make a first step towards understanding the nonlocal signal in general inflation correlators. By working with arbitrary 1PI 1-loop processes, we formulated a factorization theorem \eqref{eq_SoftThm}, and proved it using the method of partial Mellin-Barnes representation. With the factorization theorem, we can not only identify all the nonlocal signals in a given 1PI 1-loop graph, but also get its explicit expression. The theorem shows that, at the leading order in the soft momentum, this expression factorizes into three parts: the left tree graph \eqref{eq_Tleft}, the right tree graph \eqref{eq_Tright}, and the bubble signal \eqref{eq_BubbleSignal}. Our proof of the theorem made it clear that the nonlocal signal could only appear in the configuration where two of the internal momenta become soft simultaneously. As a byproduct of the proof, we found that the time orderings between the endpoints of the two soft lines do not contribute to the nonlocal signal, so we can simply ``cut" these two soft lines when computing the nonlocal signal, which is a natural generalization of the known tree-level cutting rules to the 1-loop order. Furthermore, we can also ``pinch" an endpoint of a soft propagator with an endpoint of the other soft propagator, to get an effective pinched operator. We also show that the factorization theorem cam be intuitively understood from a boundary OPE point of view: The two soft lines are effectively closed to the future boundary of dS, so that the corresponding four massive bulk operators can be pushed to the boundary. As a result, the two soft propagators are automatically cut, and we can do the OPE and pinch the remaining bulk operators.

With the help of the factorization theorem, we calculated the leading nonlocal signals of all possible 1-loop 4-point correlators, including the bubble, triangle, and box diagrams. As guaranteed by the theorem, the loop diagrams are factorized to simpler tree graphs, together with a universal bubble signal that contains the nonanalyticity. For 4-point correlators we considered, the tree graphs from the factorizations are 2-point functions of inflaton, either contact or with an exchange of a massive scalar. The contact 2-point correlators are easy to calculate, and the closed-form expressions for the single-exchange case have been derived in our earlier work \cite{Qin:2023ejc}. It is then straightforward to construct nonlocal signals for loop diagrams by these building blocks. Our results are intuitive: The coupling constant in the effective pinched operator scales as $1/m^2$ at large mass, which meets the expectation from the EFT viewpoint. We also emphasize that both the factorization theorem and the cutting rule still hold at higher orders, and we carefully calculated the leading corrections to those nonlocal signals for illustration. These corrections can be also regarded as the contributions of subleading effective pinched operators.

There are many open questions to be explored along this direction. 

First, in this work, we have only considered the simple case with  massive scalar fields running in the loop. It would be useful to generalize our analysis to include spinning fields and also to include dS boost breaking effects. This generalization is particularly relevant to CC phenomenology. The reason is that the nonlocal signal generated by a single dS covariant scalar of mass $m\gtrsim H$ suffers from the Boltzmann suppression $e^{-\pi m}$. For a 1-loop process where we have two soft lines contributing nonlocal signals simultaneously, there would be double suppression $\propto e^{-2\pi m}$. Therefore, seeing dS covariant loop signal is challenging for cosmological observations. While this problem can be partially circumvented in a dS covariant way by considering nonminimal scenarios such as curvaton mechanism or modulated reheating \cite{Lu:2019tjj,Kumar:2019ebj}, a natural solution within the minimal inflation setup is to consider dS boost breaking dispersions and interactions. As has been emphasized in the literature \cite{Chen:2018xck,Hook:2019zxa,Hook:2019vcn,Wang:2019gbi,Wang:2020ioa,Wang:2021qez,Tong:2022cdz,Qin:2022lva}, in this case, one can make use of the boost-breaking helical chemical potential to remove the Boltzmann suppression. As a result, the nonlocal signals in such scenarios can be naturally large even for loop process. Although, due to the loss of full dS isometry, the calculation of inflation correlators or just (local or nonlocal) signals becomes much more difficult in this case, our main tool used in this work, namely the PMB representation, does not explicitly rely on the existence of full dS isometries. Our previous works \cite{Qin:2022lva,Qin:2022fbv} also showed that the PMB representation is particularly useful to study dS boost breaking correlators. Therefore, we expect that our results in this paper can be directly generalized to those boost breaking cases.

Second, as we mentioned in the Introduction, besides the nonlocal signal, there is another class of nonanalyticity of inflation correlators, called the local signal, which shows up as branch cuts in the soft limit of partial energy sum instead of partial momentum sum. The local signal is also an observable of CC physics. It would be interesting to do a similar analysis to identify all local signals in a 1-loop graph and find their explicit expressions in the corresponding soft limits. It would also be interesting to explore the loop counterparts of other types of nonanalyticities in a tree correlator, such as the total energy pole and the partial energy poles. 

Finally, it would be interesting to generalize our result to arbitrary loop order. We leave all these open questions for future studies.

\paragraph{Acknowledgments.} We thank Hayden Lee and Yi Wang for useful discussions. This work is supported by the National Key R\&D Program of China (2021YFC2203100), NSFC under Grant No.\ 12275146, an Open Research Fund of the Key Laboratory of Particle Astrophysics and Cosmology, Ministry of Education of China, and a Tsinghua University Initiative Scientific Research Program.

\newpage
\begin{appendix}

\section*{Appendix}

\section{Mathematical Functions}
\label{app_math}

Here we collect some mathematical functions and compact notations frequently used in this work.

First, we denote the products of Euler $\Gamma$ functions in the following compact way:
\begin{align}
  \Gamma\left[ z_1,\cdots,z_m \right]
  \equiv&~ \Gamma(z_1)\cdots \Gamma(z_m) ,\\
  \Gamma\left[\bgm z_1,\cdots,z_m \\w_1,\cdots, w_n\edm\right]
  \equiv&~\FR{\Gamma(z_1)\cdots \Gamma(z_m)}{\Gamma(w_1)\cdots \Gamma(w_n)}.
\end{align}
Second, the Pochhammer symbol is often used:
\begin{align}
\label{eq_pochhammer}
  (z)_n\equiv\Gamma\left[\bgm z+n \\ z\edm\right].
\end{align}
Third, the generalized hypergeometric function $ {}_p\mathrm{F}_q$ is defined by the following series when the series is convergent, and is defined by analytic continuation outside the convergence reigion:
\begin{align}
\label{eq_HGF}
  {}_p\mathrm{F}_q\left[\bgm a_1,\cdots,a_p \\ b_1,\cdots,b_q \edm  \middle| z \right]=\sum_{n=0}^\infty\FR{(a_1)_n\cdots (a_p)_n}{(b_1)_n\cdots (b_q)_n}\FR{z^n}{n!}.
\end{align}
We sometimes also use the following dressed hypergeometric function:
\begin{equation}
\label{eq_DressedF}
    {}_p\mathcal{F}_q\left[\begin{matrix}
        a_1, \cdots, a_p \\
        b_1, \cdots, b_q
    \end{matrix}\middle|z\right]=\Gamma\left[\begin{matrix}
        a_1, \cdots, a_p \\
        b_1, \cdots, b_q
    \end{matrix}\right]{}_p\mathrm{F}_q\left[\begin{matrix}
        a_1, \cdots, a_p \\
        b_1, \cdots, b_q
    \end{matrix}\middle|z\right].
\end{equation}
Finally, the PMB representation of the inflation correlator in this work draws heavily use of the Mellin-Barnes representation of the Hankel functions:
\bge
  \text{H}_{\nu}^{(j)}(az)=\int_{-\ii\infty}^{\ii\infty}\FR{\di s}{2\pi\ii}\FR{(az/2)^{-2s}}{\pi}e^{(-1)^{j+1}(2s-\nu-1)\pi\ii/2}\Gamma\Big[s-\FR{\nu}{2},s+\FR{\nu}{2}\Big].~~~~(j=1,2)
\ede

\section{Loop Momentum Integrals}
\label{app_loop}
Under the partial MB representation, the loop integral is resolved to the form of \eqref{eq_LoopMomInt}:
\bge
  \mathcal{L}=\int\FR{\di^3\mb q}{(2\pi)^3}
 |\mb q|^{-2s_{V\bar1}}|\mb q+\mb P_1|^{-2s_{1\bar 2}}|\mb q+\mb P_2|^{-2s_{2\bar 3}}\cdots|\mb q+\mb P_{V-1}|^{-2s_{(V-1)\ob V}}.
\ede
If we stick to the leading order in the squeezed limit $P_N\to 0$, this integral is simplified to the bubble integral, as shown in Sec.\ \ref{sec_proof}. However, if we want to consider the higher order corrections, which is the case in Sec.\ \ref{sec_full4pt}, we shall treat this loop integral more carefully, taking higher order corrections into account. In some simple cases, including the bubble and the triangle diagrams, we can explicitly calculate (the nonlocal part of) the loop integral to arbitrary order, as we will show below.

\subsection{Bubble loop integral}
\label{app_bubble}

In the calculation of the bubble signal in Sec.\ \ref{sec_proof}, we encounter the bubble loop integral in the form of \eqref{eq_BubbleMomInt}:
\bge
\label{eq_bubbleloop}
\mathcal L_{\text{bubble}} = \int \FR{\di^3\mb q}{(2\pi)^3}\, |\mb q|^{-2a} |\mb q+\mb k_s|^{-2b}.
\ede
To compute this integral, we use the standard method of Feynman parameterization:
\bge
\label{eq_bubbleFP}
|\mb q|^{-2a} | \mb q+\mb k_s|^{-2b} = \Gamma\left[\bgm a+b\\a,b\edm\right]
\int_0^1 \di\xi_1\di\xi_2\,\de(1-\xi_1-\xi_2) \FR{\xi_1^{a-1}\xi_2^{b-1}}{(\xi_1|\mb q|^2+\xi_2|\mb q+\mb k_s|^2)^{a+b}}.
\ede
Now we can make a shift to the variable $\mb q$ by introducing new variable $\mb p$ such that:
\bge
\mb q = \mb p -\xi_2 \mb k_s.
\ede
Together with the constraint $\xi_1+\xi_2=1$, the denominator of the integrand in \eqref{eq_bubbleFP} becomes:
\bge
\label{eq_bubbleShift}
(\xi_1|\mb q|^2+\xi_2|\mb q+\mb k_s|^2)^{a+b} = (|\mb p|^2 + \Delta)^{a+b}, \qquad \Delta = \xi_1\xi_2 k_s^2,
\ede
so we can easily perform the integral over $\mb p$:
\bge
\label{eq_bubbleIntp}
\int \FR{\di^3\mb p}{(2\pi)^3} \FR{1}{(|\mb p|^2 + \Delta)^{a+b}} = \FR{1}{(4\pi)^{3/2}}\Gamma\left[\bgm a+b-\fr32\\a+b \edm\right]\Delta^{3/2-a-b}.
\ede
Then insert \eqref{eq_bubbleFP}, \eqref{eq_bubbleShift} and \eqref{eq_bubbleIntp} into the original integral \eqref{eq_bubbleloop}, we obtain:
\bge
\mathcal L_{\text{bubble}} = \FR{k_s^{3-2a-2b}}{(4\pi)^{3/2}}\Gamma\left[\bgm a+b-\fr32\\a,b\edm\right]
\int_0^1 \di\xi_1\di\xi_2\,\de(1-\xi_1-\xi_2) \xi_1^{1/2-b}\xi_2^{1/2-a}.
\ede
The integral over Feynman parameters can be calculated explicitly:
\bge
\int_0^1 \di\xi_1\di\xi_2\,\de(1-\xi_1-\xi_2) \xi_1^{\al}\xi_2^{\be} = \Gamma\left[\bgm 1+\al,1+\be\\2+\al+\be\edm \right],
\ede
so finally we obtain:
\bge
\mathcal L_{\text{bubble}} = \FR{k_s^{3-2a-2b}}{(4\pi)^{3/2}}
\Gamma\left[\bgm a+b-\fr32,\fr32-a,\fr32-b\\3-a-b,a,b\edm\right].
\ede

\subsection{Triangle loop integral}
\label{app_triangle}

Next we consider a more nontrivial case, namely the loop integral of a triangle diagram \eqref{eq_triangleloopInt} which we considered in Sec.\ \ref{sec_trianglefull}:
\bge
\label{eq_triangleloop}
\mathcal L_{\text{triangle}} = \int \FR{\di^3\mb q}{(2\pi)^3} |\mb q|^{-2a}|\mb q+\mb k_s|^{-2b}|\mb q+\mb k_1|^{-2c}.
\ede
Similarly, we apply Feynman parameterization:
\begin{align}
\label{eq_triangleFP}
|\mb q|^{-2a}|\mb q+\mb k_s|^{-2b}|\mb q+\mb k_1|^{-2c} =&~ \Gamma\left[\bgm a+b+c\\a,b,c\edm\right]
\int_0^1 \di\xi_1\di\xi_2\di\xi_3\,\de(1-\xi_1-\xi_2-\xi_3)\n\\
&\times \FR{\xi_1^{a-1}\xi_2^{b-1}\xi_3^{c-1}}{(\xi_1|\mb q|^2+\xi_2|\mb q+\mb k_s|^2+\xi_3|\mb q+\mb k_1|^2)^{a+b+c}}.
\end{align}
After the following change of variable:
\bge
\mb q = \mb p- \xi_2\mb k_s - \xi_3 \mb k_1,
\ede
the denominator of the integrand in \eqref{eq_triangleFP} becomes:
\bge
\label{eq_triangleShift}
(\xi_1|\mb q|^2+\xi_2|\mb q+\mb k_s|^2+\xi_3|\mb q+\mb k_1|^2)^{a+b+c} = |\mb p|^2 + \Delta,\qquad
\Delta = \xi_1\xi_2 k_s^2 + \xi_1\xi_3k_1^2 + \xi_2\xi_3 k_2^2,
\ede
where we have used the constraints $\xi_1+\xi_2+\xi_3=1$ and $\mb k_1\cdot \mb k_s = (k_s^2+k_1^2-k_2^2)/2$. Then the integral over $\mb p$ can be again calculated:
\bge
\label{eq_triangleIntp}
\int \FR{\di^3\mb p}{(2\pi)^3} \FR{1}{(|\mb p|^2 + \Delta)^{a+b+c}} = \FR{1}{(4\pi)^{3/2}}\Gamma\left[\bgm a+b+c-\fr32\\a+b+c \edm\right]\Delta^{3/2-a-b-c}.
\ede
Insert \eqref{eq_triangleFP}, \eqref{eq_triangleShift} and \eqref{eq_triangleIntp} into \eqref{eq_triangleloop}, we obtain:
\bge
\label{eq_IntermediateInt}
\mathcal L_{\text{triangle}} = \FR{1}{(4\pi)^{3/2}}\Gamma\left[\bgm a+b+c-\fr32\\a,b,c \edm\right]
\int_0^1\di\xi_1\di\xi_2\di\xi_3\,\de(1-\xi_1-\xi_2-\xi_3) \xi_1^{a-1}\xi_2^{b-1}\xi_3^{c-1} \Delta^{3/2-a-b-c}.
\ede

\paragraph{Symmetric configuration}
As we can see, the triangle loop integral depends on the shape of the triangle formed by the three momenta $\mb k_1$, $\mb k_2$ and $\mb k_s$, and this dependence is manifest in the expression of $\Delta$ in \eqref{eq_triangleShift}. Let us consider a special configuration, namely we fix the shape of the triangle by by setting $k_1=k_2$. The result, though much simpler than the general case, gives the leading correction of the integral \eqref{eq_triangleloop} in arbitrary shapes (so long as $\mb k_1$ and $\mb k_2$ are not anti-parallel).

Now that $k_1=k_2$, we obtain
\bge
\label{eq_triangleSymDelta}
\Delta|_{k_1=k_2} = \xi_1\xi_2 k_s^2 + \xi_3(1-\xi_3)k_1^2.
\ede
The integral \eqref{eq_IntermediateInt} is still difficult. However, we only focus on the nonlocal part of this integral, so we can make use of the MB representation to resolve the $k_s$-dependence in the factor $\Delta^{3/2-a-b-c}$. With the help of the following identity:
\bge
\label{eq_MBz}
\FR{1}{(X_0+X_1+\cdots+X_n)^\lam}= \FR{1}{\Gamma(\lam)}\int_{-\ii\infty}^{\ii\infty} \prod_{\ell=1}^n \Big[ \FR{\di z_\ell}{2\pi\ii}X_\ell^{z_\ell}\Big] X_0^{-\lam-z_1-\cdots-z_n}
\Gamma(\lam+z_1+\cdots+z_n)\prod_{\ell=1}^n \Gamma(-z_\ell),
\ede
we can rewrite $\Delta^{3/2-a-b-c}$ as:
\bge
\Delta^{3/2-a-b-c}|_{k_1=k_2} = \int_{-\ii\infty}^{\ii\infty} \FR{\di z}{2\pi\ii}\, [\xi_3(1-\xi_3)k_1^2]^z [\xi_1\xi_2 k_s^2]^{3/2-a-b-c-z}
\times \Gamma\left[\bgm a+b+c-\fr32+z,-z\\ a+b+c-\fr32 \edm\right],
\ede
and therefore,
\begin{align}
\mathcal L_{\text{triangle}}|_{k_1=k_2} =&~ \FR{k_s^{3-2a-2b-2c}}{(4\pi)^{3/2}}\int_{-\ii\infty}^{\ii\infty} \FR{\di z}{2\pi\ii}\, \Big(\FR{k_s}{k_1}\Big)^{-2z} \Gamma\left[\bgm a+b+c-\fr32+z,-z\\a,b,c\edm\right]\n\\
&\times \int_0^1\di\xi_1\di\xi_2\di\xi_3\,\de(1-\xi_1-\xi_2-\xi_3) \xi_1^{1/2-b-c-z}\xi_2^{1/2-a-c-z}\xi_3^{c-1+z}(1-\xi_3)^z.
\end{align}
The integral over Feynman parameters in the second line can be calculated explicitly:
\bge
\int_0^1 \di\xi_1\di\xi_2\di\xi_3\, \de(1-\xi_1-\xi_2-\xi_3) \xi_1^\al\xi_2^\be\xi_3^\ga(1-\xi_3)^\de = \Gamma \bgb 1+\al,1+\be,1+\ga,2+\al+\be+\de\\2+\al+\be,3+\al+\be+\ga+\de\edb,
\ede
so we obtain:
\begin{align}
\mathcal L_{\text{triangle}}|_{k_1=k_2} =&~ \FR{k_s^{3-2a-2b-2c}}{(4\pi)^{3/2}}\int_{-\ii\infty}^{\ii\infty} \FR{\di z}{2\pi\ii}\, \Big(\FR{k_s}{k_1}\Big)^{-2z} \Gamma\left[\bgm a+b+c-\fr32+z,-z\\a,b,c\edm\right]\n\\
&\times \bgb \fr32-a-c-z, \fr32-b-c-z, c+z, 3-a-b-2c-z \\ 3-a-b-2c-2z,3-a-b-c\edb.
\end{align}
The last step is to integrate over Mellin variable $z$, using the residue theorem. Since $k_s<k_1$, we should close the contour from left and sum over residues at the following two sets of left poles:
\bge
z = -m + \FR32 -a-b-c, \qquad z = -m-c,
\ede
which come from the factors $\Gamma(a+b+c-3/2+z)$ and $\Gamma(c+z)$, respectively. Furthermore, we find that the first set of poles contributes to residues proportional to $k_s^{2m}$, which is analytic as $k_s\to 0$. Therefore, we are only concerned with contribution of the second set of poles:
\begin{align}
\label{eq_triangleLoopSym}
\mathcal L_{\text{triangle}}|_{k_1=k_2} =&~ \FR{k_s^{3-2a-2b}k_1^{-2c}}{(4\pi)^{3/2}}\sum_{m=0}^\infty \FR{(-1)^m}{m!}\Big(\FR{k_s}{k_1}\Big)^{2m} \Gamma\left[\bgm -m+a+b-\fr32,m+c\\a,b,c\edm\right]\n\\
&\times \bgb m+\fr32-a, m+\fr32-b, m+3-a-b-c \\ 2m+3-a-b,3-a-b-c\edb + \text{terms analytic in $k_s$}\n\\
=&~\FR{k_s^{3-2a-2b}k_1^{-2c}}{(4\pi)^{3/2}}\sum_{m=0}^\infty \FR{(-1)^m}{m!}\Big(\FR{k_s}{k_1}\Big)^{2m} (3-a-b-c)_m(c)_m\n\\
&\times \bgb -m+a+b-\fr32,m+\fr32-a,m+\fr32-b\\ 2m+3-a-b,a,b\edb+ \text{terms analytic in $k_s$}.
\end{align}

\paragraph{General configuration}
Now we consider the most general case, with $\Delta$ given in \eqref{eq_triangleShift}. Again we can use the MB representation \eqref{eq_MBz} and express $\Delta^{3/2-a-b-c}$ in \eqref{eq_IntermediateInt} as the following:
\begin{align}
\Delta^{3/2-a-b-c} =& \int_{-\ii\infty}^{\ii\infty} \FR{\di z_1}{2\pi\ii}\FR{\di z_2}{2\pi\ii}\,(\xi_1\xi_3k_1^2)^{z_1}(\xi_2\xi_3k_2^2)^{z_2}(\xi_1\xi_2k_s^2)^{3/2-a-b-c-z_1-z_2}\n\\
&\times \Gamma\bgb a+b+c-\fr32+z_1+z_2,-z_1,-z_2\\a+b+c-\fr32 \edb.
\end{align}
Then we can integrate out Feynman parameters using:
\bge
\int_0^1 \di\xi_1\di\xi_2\di\xi_3\,\de(1-\xi_1-\xi_2-\xi_3) \xi_1^\al\xi_2^\be\xi_3^\ga = \Gamma\bgb 1+\al,1+\be,1+\ga\\3+\al+\be+\ga\edb,
\ede
and the integral \eqref{eq_IntermediateInt} becomes:
\begin{align}
\mathcal L_{\text{triangle}} =&~ \FR{k_s^{3-2a-2b-2c}}{(4\pi)^{3/2}}\int_{-\ii\infty}^{\ii\infty} \FR{\di z_1}{2\pi\ii}\FR{\di z_2}{2\pi\ii}\,\Big(\FR{k_s}{k_1}\Big)^{-2z_1}\Big(\FR{k_s}{k_2}\Big)^{-2z_2} \Gamma\bgb a+b+c-\fr32+z_1+z_2,-z_1,-z_2\\a,b,c \edb\n\\
&\times \Gamma\bgb \fr32-a-c-z_1,\fr32-b-c-z_2,c+z_1+z_2\\ 3-a-b-c\edb.
\end{align}
Now we should integrate out $z_1$ and $z_2$, respectively. Assume $k_1>k_2$, and let us integrate out $z_2$ first. Since $k_2<k_s$, we should again close the contour from left and sum over residues at left poles:
\bge
z_2 = -m_2 -a-b-c+\FR32-z_1, \qquad z_2 = -m_2 -c-z_1,
\ede
where the first set contributes to analytic terms proportional to $k_s^{2m_2}$, so we only consider the second set, which gives:
\begin{align}
\mathcal L_{\text{triangle}} =&~ \FR{k_s^{3-2a-2b}k_2^{-2c}}{(4\pi)^{3/2}}\sum_{m_2=0}^\infty \FR{(-1)^{m_2}}{m_2!}\Big(\FR{k_s}{k_2}\Big)^{2m_2} \int_{-\ii\infty}^{\ii\infty} \FR{\di z_1}{2\pi\ii}\,\Big(\FR{k_2}{k_1}\Big)^{-2z_1}\n\\
&\times \Gamma\bgb -m_2+a+b-\fr32,-z_1,m_2+c+z_1\\a,b,c \edb
\times \Gamma\bgb \fr32-a-c-z_1,m_2+\fr32-b+z_1\\ 3-a-b-c\edb\n\\
&+ \text{terms analytic in $k_s$}.
\end{align}
Next we should integrate out $z_1$. Since we have assumed $k_1>k_2$, we should still sum over residues at left poles:
\bge
z_1 = -m_1 -m_2-c, \qquad z_1 = -m_1 - m_2 -\FR32+b.
\ede
This time we can complete the summation of $m_1$ to hypergeometric functions with argument $k_2^2/k_1^2$:
\begin{align}
\label{eq_Long}
\mathcal L_{\text{triangle}} =&~ \FR{k_s^{3-2a-2b}k_2^{-2c}}{(4\pi)^{3/2}}
\sum_{m_2=0}^{\infty} \FR{(-1)^{m_2}}{m_2!}
\Big(\FR{k_s}{k_1}\Big)^{2m_2}
\n\\
&\times \bigg\{\Big(\FR{k_2}{k_1}\Big)^{2c}\Gamma\left[
\bgm -m_2-\fr32+a+b,m_2+c,m_2+\fr32-a,\fr32-b-c\\ a,b,c,3-a-b-c\edm
\right]\n\\
&\times {}_2\mathrm F_1\left[\bgm m_2+c,m_2+\fr32-a\\ -\fr12+b+c\edm\middle| \FR{k_2^2}{k_1^2}\right]\n\\
&+\Big(\FR{k_2}{k_1}\Big)^{3-2b}
\Gamma\left[
\bgm -m_2-\fr32+a+b,m_2+\fr32-b,-\fr32+b+c,m_2+3-a-b-c\\ a,b,c,3-a-b-c\edm
\right]\n\\
&\times {}_2\mathrm F_1\left[\bgm m_2+\fr32-b,m_2+3-a-b-c\\ \fr52-b-c\edm\middle| \FR{k_2^2}{k_1^2}\right]
\bigg\} + \text{terms analytic in $k_s$}.
\end{align}
Finally, we can expand the above the result around $k_2 \sim k_1$. Let us write:
\bge
\mathcal L_{\text{triangle}} = \FR{k_s^{3-2a-2b}k_2^{-2c}}{(4\pi)^{3/2}}\sum_{m,\ell=0}^\infty \mb P_m^\ell(a,b,c)\Big(\FR{k_s}{k_1}\Big)^{2m}\Big(1-\FR{k_2}{k_1}\Big)^\ell + \text{terms analytic in $k_s$}.
\ede
This is exactly the right hand side of \eqref{eq_TriangleLoopInt}. The coefficients $\mb P_m^\ell$ can be determined by matching each term in the Taylor expansion of \eqref{eq_Long}. Below we list the expressions of $\mb P_m^\ell$ for $\ell=0,1,2$ and arbitrary $m$:
\begin{align}
\mb P_m^0(a,b,c) = &~ \FR{(-1)^m}{m!}(c)_m (3-a-b-c)_m \Gamma\left[\bgm -m+a+b-\fr32,m+\fr32-a,m+\fr32-b\\2m+3-a-b,a,b\edm\right],\\
\mb P_m^1(a,b,c) = &~  \FR{m(2m+3-2a) -(2m+3-2b)c}{2m+3-a-b}\times \mb P_m^0(a,b,c)
,\\
\mb P_m^2(a,b,c) = &~ \Big[m(2m+3-2a)\big(2m^2+(5-2a)m+1-a+b\big)\n\\
& -(2m+3-2b)\big(4m^2+(10-4a)m+7-3a-b\big)c\n\\
&+(2m+3-2b)(2m+5-2b)c^2\Big]\Big/\Big[2(2m+3-a-b)_2\Big]\times \mb P_m^0(a,b,c).
\end{align}
One can also check that terms with $\ell=0$ recover the result \eqref{eq_triangleLoopSym} in the symmetric case.

\section{One-Loop Four-Point Functions with Unequal Masses}
\label{app_diffmass}
In this appendix, we consider the leading results of all possible 1-loop 4-point functions, which has been considered in Sec.\ \ref{sec_4pt}, but now with unequal masses for all the loop modes. For convenience, we always set the mass parameters of the two soft lines to be $\wt\nu_1$ and $\wt\nu_2$, respectively.
\subsection{Bubble diagram}
First we consider the bubble diagram with the following interaction:
\bge
\Delta \ld = \FR12  a^2\varphi'^2 \si^{(1)}\si^{(2)},
\ede
where $\si^{(1)}$ and $\si^{(2)}$ are distinct massive scalar fields, with mass parameters $\wt\nu_1$ and $\wt\nu_2$, respectively. Following our factorization theorem \eqref{eq_SoftThm},
the bubble signal is given by \eqref{eq_BubbleSignal}:
\bge
\label{eq_bubblesignalDiff}
\mathcal B_{\cc\dd} = \FR{k_s^3}{(4\pi)^{7/2}}\Gamma\bgb-\cc\ii\wt\nu_1-\dd\ii\wt\nu_2-\fr32,\fr32+\cc\ii\wt\nu_1,\fr32+\dd\ii\wt\nu_2,-\cc\ii\wt\nu_1,-\dd\ii\wt\nu_2\\ 3+\cc\ii\wt\nu_1+\dd\ii\wt\nu_2\edb \Big(\FR {k_s}2\Big)^{2\ii(\cc\wt\nu_1+\dd\wt\nu_2)}.
\ede
The left and the right tree graphs are simple:
\begin{align}
\label{eq_bubbleTLDiff}
\mathcal T_{\cc\dd}^{(\text{L})} =&~ \FR{1}{4k_1k_2}\sum_{\aa_1=\pm}(\ii\aa_1)\int_{-\infty}^0 \di\tau_1\,e^{\aa_1\ii k_{12}\tau_1}
   (-\tau_1)^{3/2+\cc\ii \wt\nu_1}(-\tau_1)^{3/2+\cc\ii \wt\nu_2} \n\\
  =&~ \mathcal{P}^{(1)}_{\cc\dd}(\wt\nu_1,\wt\nu_2)\FR{k_{12}^{-4-\cc\ii\wt\nu_1-\dd\ii\wt\nu_2}}{4k_1k_2} ,\\
  \label{eq_bubbleTRDiff}
  \mathcal T_{\cc\dd}^{(\text{R})} =&~ \FR{1}{4k_3k_4}\sum_{\aa_2=\pm}(\ii\aa_2)\int_{-\infty}^0 \di\tau_2\,e^{\aa_2\ii k_{34}\tau_2}
  (-\tau_2)^{3/2+\cc\ii \wt\nu_1}(-\tau_2)^{3/2+\dd\ii \wt\nu_2} \n\\
  =&~ \mathcal{P}^{(1)}_{\cc\dd}(\wt\nu_1,\wt\nu_2)\FR{k_{34}^{-4-\cc\ii\wt\nu_1-\dd\ii\wt\nu_2}}{4k_3k_4} ,
\end{align}
where the first order pinch coefficient now becomes:
\bge
\mathcal P^{(1)}_{\cc\dd}(\wt\nu_1,\wt\nu_2) \equiv 2 \sin\Big[\ii\FR\pi2 (\cc\wt\nu_1+\dd\wt\nu_2) \Big]\Gamma(4+\cc\ii\wt\nu_1+\dd\ii\wt\nu_2).
\ede
Therefore, the leading nonlocal signal is:
\bge
  \lim_{k_s\to 0} \Big[\mathcal T_\text{bubble}(\{\mb k\})\Big]_{\text{NL}} 
  =   \FR{k_s^3}{k_1k_2k_3k_4k_{12}^4k_{34}^4} 
  \bigg[ \mathcal A_{\text{bubble},+}^{(0)}\Big(\FR{k_s^2}{k_{12}k_{34}}\Big)^{\ii\wt\nu_+} 
  + \mathcal A_{\text{bubble},-}^{(0)}\Big(\FR{k_s^2}{k_{12}k_{34}}\Big)^{\ii\wt\nu_-}  \bigg]
 +\text{c.c.},
\ede
where
\bge
\mathcal A_{\text{bubble},\pm}^{(0)} = -\FR{4^{-1-\ii\wt\nu_\pm}}{(4\pi)^{7/2}}\sinh^2\Big(\FR{\pi\wt\nu_\pm}2\Big)(3+\ii\wt\nu_\pm)\Gamma\Big[4+\ii\wt\nu_\pm, \FR32+\ii\wt\nu_1,\FR32\pm\ii\wt\nu_2,-\ii\wt\nu_1,\mp\ii\wt\nu_2,-\FR32-\ii\wt\nu_\pm\Big],
\ede
and
\bge
\wt\nu_\pm \equiv \wt\nu_1 \pm \wt\nu_2,
\ede
are the positive and negative frequency of the signal, respectively.
\subsection{Triangle diagram}
As the second example, we consider the triangle diagram with the interaction:
\bge
\Delta\ld = a^3 \varphi' \si^{(1)}\si^{(3)} + a^3 \varphi' \si^{(2)}\si^{(3)} + \FR12a^2\varphi'^2\si^{(1)}\si^{(2)}.
\ede
The bubble signal and the right tree graph are the same as the bubble case, given by \eqref{eq_bubblesignalDiff} and \eqref{eq_bubbleTRDiff}, respectively. 
The left tree graph is:
\begin{align}
\label{eq_triangleTLDiff}
\mathcal T_{\cc\dd}^{(\text{L})} =&~ \FR{1}{4k_1k_2}\sum_{\aa_1,\aa_2=\pm}(-\aa_1\aa_2) \int_{-\infty}^0 \FR{\di\tau_1}{\tau_1^2}\FR{\di\tau_2}{\tau_2^2} e^{\aa_1 \ii k_1 \tau_1+\aa_2\ii k_2 \tau_2}D^{(\wt\nu_3)}_{\aa_1\aa_2}(k_1;\tau_1,\tau_2)\n\\
& \times (-\tau_1)^{3/2+\cc\ii\wt\nu_1}(-\tau_2)^{3/2+\cc\ii\wt\nu_2}\n\\
  =&~ \mathcal{P}^{(2)}_{\cc\dd}(\wt\nu_1,\wt\nu_2;\wt\nu_3)\FR{k_{12}^{-4-\cc\ii\wt\nu_1-\dd\ii\wt\nu_2}}{4k_1k_2},
\end{align}
where the second order pinch coefficient can be expressed by the following combination of seed integrals:
\bge
\mathcal{P}^{(2)}_{\cc\dd} (\wt\nu_1,\wt\nu_2;\wt\nu_3)\equiv 2^{4+\cc\ii\wt\nu_1+\dd\ii\wt\nu_2}\sum_{\aa_1,\aa_2=\pm} \wt{\mathcal I}_{\wt\nu_3|\aa_1\aa_2}^{-1/2+\cc\ii\wt\nu_1,-1/2+\dd\ii\wt\nu_2} (1,1).
\ede
Therefore, the leading nonlocal signal for the triangle diagram is:
\bge
  \lim_{k_s\to 0} \Big[\mathcal T_\text{triangle}(\{\mb k\})\Big]_{\text{NL}} 
  =   \FR{k_s^3}{k_1k_2k_3k_4k_{12}^4k_{34}^4} 
  \bigg[ \mathcal A_{\text{triangle},+}^{(0)}\Big(\FR{k_s^2}{k_{12}k_{34}}\Big)^{\ii\wt\nu_+} 
  + \mathcal A_{\text{triangle},-}^{(0)}\Big(\FR{k_s^2}{k_{12}k_{34}}\Big)^{\ii\wt\nu_-}  \bigg]
 +\text{c.c.},
\ede
where
\begin{align}
\mathcal A_{\text{triangle},\pm}^{(0)} =&~ \FR{\ii 2^{1-\ii\wt\nu_\pm}}{(4\pi)^{7/2}}\sinh\Big(\FR{\pi\wt\nu_\pm}2\Big)(3+\ii\wt\nu_\pm)\Gamma\Big[\FR32+\ii\wt\nu_1,\FR32\pm\ii\wt\nu_2,-\ii\wt\nu_1,\mp\ii\wt\nu_2,-\FR32-\ii\wt\nu_\pm\Big]\n\\
&\times\sum_{\aa_1,\aa_2=\pm} \wt{\mathcal I}_{\wt\nu_3|\aa_1\aa_2}^{-1/2+\ii\wt\nu_1,-1/2\pm\ii\wt\nu_2} (1,1).
\end{align}
We can also define the effective pinched operator:
\bge
\mathcal O_{\cc\dd} = \mathcal O^{(0)}_{\cc\dd} + \cdots,\qquad
\mathcal O_{\cc\dd}^{(0)} = \FR12 \mathcal C_{\cc\dd}^{(0)}(\wt\nu_1,\wt\nu_2;\wt\nu_3) a^2\varphi'^2\si^{(1)}\si^{(2)},
\ede
and the pinched coupling is given by:
\bge
\mathcal C_{\cc\dd}^{(0)}(\wt\nu_1,\wt\nu_2;\wt\nu_3) \equiv \FR{\mathcal{P}^{(2)}_{\cc\dd}(\wt\nu_1,\wt\nu_2;\wt\nu_3)}{\mathcal{P}^{(1)}_{\cc\dd}(\wt\nu_1,\wt\nu_2;\wt\nu_3)}.
\ede
Although the expression for $\mathcal C_{\cc\dd}^{(0)}$ involves generalized hypergeometric functions and thus is complicated,
we find it scales as $1/\wt\nu_3^2$ at large $\wt\nu_3$ by fitting the asymptotic behavior when $\wt\nu_3\to\infty$ and $\wt\nu_{1,2}$ fixed.
This behavior meets our expectation:
\bge
\lim_{\wt\nu_3\to\infty}\mathcal A_{\text{triangle},\pm}^{(0)} \sim \wt\nu_3^{-2} \mathcal A_{\text{bubble},\pm}^{(0)}.
\ede

\subsection{Box diagram}
Finally we consider the box diagram with the interaction:
\bge
\Delta\ld = a^3\varphi' \si^{(1)}\si^{(3)} + a^3\varphi' \si^{(2)}\si^{(3)} +  a^3\varphi' \si^{(1)}\si^{(4)} + a^3\varphi' \si^{(2)}\si^{(4)}.
\ede
Now the bubble signal is still \eqref{eq_bubblesignalDiff}. The left tree graph is given in \eqref{eq_triangleTLDiff}, and the right tree graph is similar, given by:
\begin{align}
\mathcal T_{\cc\dd}^{(\text{R})} = &~\mathcal{P}^{(2)}_{\cc\dd}(\wt\nu_1,\wt\nu_2;\wt\nu_4)\FR{k_{34}^{-4-\cc\ii\wt\nu_1-\dd\ii\wt\nu_2}}{4k_3k_4},
\end{align}
and thus the nonlocal signal is:
\bge
  \lim_{k_s\to 0} \Big[\mathcal T_\text{box}(\{\mb k\})\Big]_{\text{NL}} 
  =   \FR{k_s^3}{k_1k_2k_3k_4k_{12}^4k_{34}^4} 
  \bigg[ \mathcal A_{\text{box},+}^{(0)}\Big(\FR{k_s^2}{k_{12}k_{34}}\Big)^{\ii\wt\nu_+} 
  + \mathcal A_{\text{box},-}^{(0)}\Big(\FR{k_s^2}{k_{12}k_{34}}\Big)^{\ii\wt\nu_-}  \bigg]
 +\text{c.c.},
\ede
where
\begin{align}
\mathcal A_{\text{box},\pm}^{(0)} =&~ \FR{16}{(4\pi)^{7/2}}\Gamma\bgb\FR32+\ii\wt\nu_1,\FR32\pm\ii\wt\nu_2,-\ii\wt\nu_1,\mp\ii\wt\nu_2,-\FR32-\ii\wt\nu_\pm\\ 3+\ii\wt\nu_\pm\edb\n\\
&\times\sum_{\aa_1,\aa_2=\pm} \wt{\mathcal I}_{\wt\nu_3|\aa_1\aa_2}^{-1/2+\ii\wt\nu_1,-1/2\pm\ii\wt\nu_2} (1,1) \times\sum_{\aa_3,\aa_4=\pm} \wt{\mathcal I}_{\wt\nu_4|\aa_3\aa_4}^{-1/2+\ii\wt\nu_1,-1/2\pm\ii\wt\nu_2} (1,1),
\end{align}
and the large mass behaviors are:
\begin{align}
\lim_{\wt\nu_4 \to \infty} \mathcal A_{\text{box},\pm}^{(0)} \sim&~ \wt\nu_4^{-2} \mathcal A_{\text{triangle},\pm}^{(0)},\\
\lim_{\wt\nu_3,\wt\nu_4 \to \infty} \mathcal A_{\text{box},\pm}^{(0)} \sim&~ \wt\nu_3^{-2} \wt\nu_4^{-2} \mathcal A_{\text{bubble},\pm}^{(0)}.
\end{align}
\end{appendix}

\newpage
\bibliography{CosmoCollider} 
\bibliographystyle{utphys}

\end{document}